\documentclass[journal]{IEEEtran}
\usepackage{cite}
   \usepackage{graphicx}
  \usepackage{subfigure}
   \usepackage{epstopdf}
   \usepackage{float}
   \usepackage{paralist}
   \usepackage{url}
   
\usepackage{rotating,array}
\usepackage{graphicx} 

\ifCLASSINFOpdf
\else
\fi
\usepackage{amsmath,amssymb,amsthm}
\usepackage{algorithm}
\usepackage{algpseudocode}
\newenvironment{seqnarray*}{\small\begin{eqnarray*}}{\end{eqnarray*}}
\newenvironment{sequation*}{\small\begin{equation*}}{\end{equation*}}

\begin{document}
%
\title{Restricted Linearized Augmented Lagrangian Method for Euler's Elastica Model}
%
%
%
\author{Yinghui Zhang, Xiaojuan Deng, Jun Zhang, Hongwei Li$^*$
\thanks{This work is supported in part by the National Natural Science Foundation of China under Grant 61771324 and 61401289.}
\thanks{\emph{Asterisk indicates corresponding author.}}
\thanks{Yinghui Zhang, Xiaojuan Deng, and Hongwei Li are with School of Mathematical Sciences, Capital Normal University, Beijing, 100048, China, and are also with Beijing Advanced Innovation Center for Imaging Technology, Capital Normal University, Beijing, 100048, China.}
\thanks{Jun Zhang is with Jiangxi Province Key Laboratory of Water Information Cooperative Sensing and Intelligent Processing, Nanchang Institute of Technology, Nanchang 330099, Jiangxi, China.}
\thanks{E-mail: hongwei.li91@cnu.edu.cn.}
\thanks{}}
%
%

\markboth{Journal of \LaTeX\ Class Files,~Vol.~xx, No.~xx,xx-xxxx}%
{Shell \MakeLowercase{\textit{et al.}}: Bare Demo of IEEEtran.cls for IEEE Journals}
%



\maketitle

\begin{abstract}
Euler's elastica model has been extensively studied and applied to image processing tasks. However, due to the high nonlinearity and nonconvexity of the involved curvature term, conventional algorithms suffer from slow convergence and high computational cost. Various fast algorithms have been proposed, among which, the augmented Lagrangian based ones are very popular in the community. However, parameter tuning might be very challenging for these methods. In this paper, a simple cutting-off strategy is introduced into the augmented Lagrangian based algorithms for minimizing the Euler's elastica energy, which leads to easy parameter tuning and fast convergence. The cutting-off strategy is based on an observation of inconsistency inside the augmented Lagrangian based algorithms. 
When the weighting parameter of the curvature term goes to zero, the energy functional boils down to the ROF model. So, a natural requirement is that its augmented Lagrangian based algorithms should also approach the augmented Lagrangian based algorithms formulated directly for solving the ROF model from the very beginning. Unfortunately, this is not the case for certain existing augmented Lagrangian based algorithms.   The proposed cutting-off strategy helps to decouple the tricky dependence between the auxiliary splitting variables, so as to remove the observed inconsistency.  Numerical experiments suggest that the proposed algorithm enjoys easier parameter-tuning, faster convergence and even higher quality of image restorations. 
\end{abstract}

\begin{IEEEkeywords}
Euler's elastica, augmented Lagrangian, image denoising
\end{IEEEkeywords}

%
\IEEEpeerreviewmaketitle

\section{Introduction}\label{sec:intro}
%
%

%
%
\IEEEPARstart{T}{he} classical image denoising problem is that,
given a noisy image $f: \Omega \rightarrow \mathbb{R}$, where $\Omega$ is a bounded open subset of $\mathbb{R}^{2}$, a clean image $u$ is to be estimated from
$$f = u + n, $$
where $n$ is the noise. One of the most popular and successful variational methods for solving this denoising problem is developed by Rudin, 
  Osher, and Fatemi(ROF) \cite{rudin1992nonlinear}. It is defined as the following minimization problem
  \begin{equation}\label{eq:ROF}
\min\limits_u \, \int_{\mathrm{\Omega}}|\nabla u| + \frac{\lambda}{2}\int_{\mathrm{\Omega}}(f-u)^2,
  \end{equation}
where $\lambda > 0$ is a scalar weighting parameter. The ROF model is effective on preserving edges while removing noise thanks to the total variation term in the energy functional. Several fast algorithms have been proposed for solving the ROF model 
\cite{chambolle2004algorithm,goldstein2009split,wu2010augmented,wahlberg2012admm}. 

Despite its appealing properties, the ROF model yields the staircase effect and fails to preserve image contrasts. To overcome these drawbacks, some high order variational models have been proposed \cite{mumford1994elastica,Masnou1998,chan2000high,lysaker2003noise,brito2010multigrid,schonlieb2011unconditionally,zhu2013augmented}. Among them, Euler's elastica energy functional has a wide application in image processing, such as image inpainting \cite{brito2010fast,yashtini2016fast}, segmentation \cite{tai2017simple,bae2017augmented}, restoration \cite{zhu2013augmented,yashtini2015alternating} and reconstruction \cite{Zhang2017,zhang2017eulerelastica}. For image restoration, the Euler's elastica model is defined as
\begin{equation}
\label{ee}
\min\limits_{u}\int_\mathrm{\Omega}\left(a + b\kappa^2|\nabla u|\right) d\mathrm{\Omega}+{\frac{\lambda}{2}}\int_\mathrm{\Omega} (u-f)^2d\mathrm{\Omega},
\end{equation}
where $a,b,\kappa=\nabla\cdot\left(\frac{\nabla u}{|\nabla u|}\right)$ are the weighting parameters for the total variational term and the curvature term respectively. 
The Euler's elastica functional was first introduced by Mumford, Nitzberg and Shiota \cite{mumford1993filtering} for segmenting images with occlusions.  The conventional algorithms  \cite{shen2003euler,ni2008variational,brito2010fast} for solving the Euler's elastica model seek effective discretization schemes for solving the model's Euler-Lagrange equation. However, due to the nonlinear and nonconvex curvature term, the involved computations are of high complexity  and time-consuming.

 Recently, several fast numerical algorithms   \cite{Bae2011,duan2014two,chen2016augmented,2018arXiv181107091D}  have been proposed based on the augmented Lagrangian and other techniques, such as primal-dual, operator splitting or linearization techniques to convert the original problem into a few simple and easily-solved subproblems. Tai, Hahn and Chung  \cite{tai2011fast} reformulated  (\ref{ee}) as a constrained optimization problem by using the idea of augmented Lagrangian. Then, an alternating minimization method reduces the original problem into a series of subproblems that either have closed-form solutions or can be dealt with by the Fast Fourier Transform(FFT). Another related work can be found in   \cite{zhang2016new}, where Jianping Zhang and Ke Chen proposed an augmented Lagrangian-based primal dual approach, such that the number of auxiliary variables was reduced. To propose a simpler method with fewer parameters, Maryam Yashtini and Sung Ha Kang  \cite{yashtini2016fast} tried to relax the curvature term and solve the associated subproblems  by FFT and a shrinkage operator.  Besides, they also tried to consider the Euler's elastica energy minimization problem as a weighted ROF problem, so as to develop fast algorithms. However, the application of FFT is limited to periodic boundary conditions, which is not desirable for many image processing models. To relieve this concern, Tai  \cite{hahn2011fast} proposed another fast algorithm for Euler's elastica model by replacing the FFT with an inexpensive arithmetic operation. In the same spirit, Duan et al. \cite{Duan2013} introduced one new variable to decouple the mean curvature term. By using one sweep of the Gauss-Seidel iteration to replace the FFT operation, Duan's method further reduced the needed computation cost. Inspired by this, Zhang et al.  \cite{zhang2017fast} adopted the linearization technique to approximate the quadratic terms in the Euler-Lagrange equation, such that all subproblems have closed-form solutions. 

Despite the achieved promising results, the fast solvers developed so far for the Euler's elastica model (\ref{ee}) might suffer from tricky parameter tuning due to the mutual dependence between the auxiliary variables.
In this paper, we develop a cutting-off strategy for the augmented Lagrangian based algorithms for solving (\ref{ee}), which enjoys  easier parameter tuning, faster convergence, and even higher quality of restorations. 
The proposed algorithm is based on a key observation of inconsistency inside the conventional augmented Lagrangian approach for  
solving (\ref{ee}). When $b\rightarrow 0$, the Euler's elastica model (\ref{ee})  approaches the following ROF model (\ref{eq:rofa}) 
\begin{equation}
\label{eq:rofa}
\min\limits_{u}\int_\mathrm{\Omega} a |\nabla u| d\mathrm{\Omega}+{\frac{\lambda}{2}}\int_\mathrm{\Omega} (u-f)^2d\mathrm{\Omega}.
\end{equation}
Nevertheless, the augmented Lagrangian method for solving 
(\ref{ee}) doesn't reduce to the augmented Lagrangian method for solving (\ref{eq:rofa}). This will be explained and verified in the following sections. By introducing  a cutting-off strategy, the restricted augmented Lagrangian method developed in this paper aims to remove such an inconsistency. 
Besides, we will show that a  technique often used by conventional augmented Lagrangian based algorithms for simplifying  the computations involving the auxiliary variable $\boldsymbol{n}$, is actually problematic. The proposed method also deals with this issue.

The remainder of this paper is organized as follows. In Section 2, the proposed method is described in detail. Numerical experiments to illustrate the effectiveness and efficiency of the proposed method are presented in Section 3. Conclusions and remarks are given in Section 4.

\section{The Restricted Augmented Lagrangian method for minimizing the Euler's elastica energy}\label{sec:RALM}

\subsection{Conventional Augmented Lagrangian Approaches}

  Tai et al. \cite{tai2011fast} applied the augmented Lagrangian method  \cite{wu2010augmented} to solve the Euler's elastica model. This was done by introducing four auxiliary variables $\boldsymbol{p}$, $\boldsymbol{m} $, $v$ and $\boldsymbol{n}$ to relax the energy functional (\ref{ee}) as 
\begin{equation}
\begin{aligned}
\label{THC}
&\min\limits_{v,\boldsymbol{p},\boldsymbol{m},\boldsymbol{n}}\int_{\mathrm{\Omega}}\left(a+b(\nabla \cdot\boldsymbol{n})^2\right)|\boldsymbol{p}|+{\frac{\lambda}{2}}\int_{\mathrm{\Omega}} (v-u_0)^2\\
&s.t. \, \boldsymbol{p}=\nabla u,\ \boldsymbol{n} = \boldsymbol{m},\ v=u, \ |\boldsymbol{p}| = \boldsymbol{m} \cdot \boldsymbol{p},\ |\boldsymbol{m}|\le 1.
\end{aligned} 
\end{equation}

Instead of introducing four auxiliary variables, Duan et al. \cite{Duan2013} developed an  algorithm that needs just three auxiliary variables, and avoids the use of FFT, which looses the periodic boundary condition restriction.  This algorithm starts by reformulating (\ref{ee}) as
\begin{equation}
\begin{aligned}
\label{DWH}
&\min\limits_{u,\boldsymbol{p},\boldsymbol{n},h}\int_{\mathrm{\Omega}}\left(a+bh^2\right)|\boldsymbol{p}|+{\frac{\lambda}{2}}\int_{\mathrm{\Omega}} (u-f)^2\\
&~~~s.t. \, \boldsymbol{p}=\nabla u,\  \boldsymbol{p} = |\boldsymbol{p}|\boldsymbol{n},\ h=\nabla \cdot\boldsymbol{n}.
\end{aligned}
\end{equation}
Then, the augmented Lagrangian formulation for (\ref{DWH}) reads
\begin{equation}
\begin{aligned}
\label{ALDWH}
 \mathcal{L}(u, \boldsymbol{p}, \boldsymbol{n}, h&; \boldsymbol{\lambda}_{1}, \boldsymbol{\lambda}_{2}, {\lambda}_{3})=
 \int_{\mathrm{\mathrm{\Omega}}}(a+bh^2)|\boldsymbol{p}|+ {\frac{\lambda}{2}}\int_{\mathrm{\Omega}} (u-f)^2\\
&+\int_{\mathrm{\Omega}}\boldsymbol{\lambda}_{1}\cdot (\boldsymbol{p} - |\boldsymbol{p}|\boldsymbol{n})+\frac{r_1}{2}\int_{\mathrm{\Omega}}|\boldsymbol{p} - |\boldsymbol{p}|\boldsymbol{n}|^2\\
&+\int_{\mathrm{\Omega}}\boldsymbol{\lambda}_{2}\cdot(\boldsymbol{p}-\nabla u)+\frac{r_2}{2}\int_{\mathrm{\Omega}}|\boldsymbol{p}-\nabla u|^2\\
&+\int_{\mathrm{\Omega}}{\lambda}_{3}(h-\nabla \cdot \boldsymbol{n})+\frac{r_3}{2}\int_{\mathrm{\Omega}}(h-\nabla \cdot \boldsymbol{n})^2,
\end{aligned}
\end{equation}
where,  $\boldsymbol{\lambda}_{1}, \boldsymbol{\lambda}_{2} \in \mathbb{R}^2$, and ${\lambda}_{3} \in \mathbb{R}$ are Lagrangian multipliers corresponding to the constraints $\boldsymbol{p} = |\boldsymbol{p}|\boldsymbol{n}$, $\boldsymbol{p}=\nabla u$ and $h=\nabla \cdot \boldsymbol{n}$, respectively, and for $ i=1,2,3$, $r_i$ and $\lambda$ are positive scalar parameters. The above energy functional is then minimized by the alternating minimization method, i.e. given the previous iterate ($u^k,\boldsymbol{p}^k,\boldsymbol{n}^k,h^k;\boldsymbol{\lambda}_{1}^k,\boldsymbol{\lambda}_{2}^k,{\lambda}_{3}^k
$), the next iterate is computed by solving the following subproblems (\ref{pre_u-sub})-(\ref{pre_h-sub}) and updating the Lagrangian multipliers through (\ref{lambda_1})-(\ref{lambda_3}). 
\begin{equation}
\begin{aligned}
\label{pre_u-sub}
u^{k+1}& =\arg\min\limits_{u} {\frac{\lambda}{2}} \int_{\mathrm{\Omega}} (u-f)^2
+\int_{\mathrm{\Omega}}\boldsymbol{\lambda}_{2}^k\cdot(\boldsymbol{p}^k-\nabla u)\\
&+\frac{r_2}{2}\int_{\mathrm{\Omega}}|\boldsymbol{p}^k-\nabla u|^2,
\end{aligned}
\end{equation}
\begin{equation}
\begin{aligned}
\label{pre_p-sub}
\boldsymbol{p}^{k+1}&=\arg\min\limits_{\boldsymbol{p}} \int_{\mathrm{\Omega}} \left(a+b\left(h^k\right)^2\right)|\boldsymbol{p}|\\
&+\int_{\mathrm{\Omega}}\boldsymbol{\lambda}_{2}^k\cdot(\boldsymbol{p}-\nabla u^{k+1})
+\frac{r_2}{2}\int_{\mathrm{\Omega}}|\boldsymbol{p}-\nabla u^{k+1}|^2\\
&+\int_{\mathrm{\Omega}}\boldsymbol{\lambda}_{1}^{k}\cdot\left(\boldsymbol{p}-\boldsymbol{n}^k|\boldsymbol{p}|\right)
+\frac{r_1}{2}\int_{\mathrm{\Omega}}\left|\boldsymbol{p}-\boldsymbol{n}^k|\boldsymbol{p}|\right|^2,
\end{aligned}
\end{equation}
\begin{equation}
\label{pre_n-sub}
\begin{aligned}
\boldsymbol{n}^{k+1}&=\arg\min\limits_{\boldsymbol{n}}  \int_{\mathrm{\Omega}}\boldsymbol{\lambda}_{1}^{k}\cdot\left(\boldsymbol{p}^{k+1}-|\boldsymbol{p}^{k+1}|\boldsymbol{n}\right)\\
&+\frac{r_1}{2}\int_{\mathrm{\Omega}}\left|\boldsymbol{p}^{k+1}-|\boldsymbol{p}^{k+1}|\boldsymbol{n}\right|^2\\
&+\int_{\mathrm{\Omega}}\lambda_3^k(h^k-\nabla \cdot \boldsymbol{n})+\frac{r_3}{2}\int_{\mathrm{\Omega}}(h^k-\nabla \cdot \boldsymbol{n})^2,
\end{aligned}
\end{equation}
\begin{equation}
\begin{aligned}
\label{pre_h-sub}
h^{k+1}&=\arg\min\limits_{h}  \int_{\mathrm{\Omega}}\left (a+b\left(h\right)^2\right)|\boldsymbol{p}^{k+1}|\\
&+\int_{\mathrm{\Omega}}{\lambda}_{3}^k(h-\nabla \cdot \boldsymbol{n}^{k+1}) 
+\frac{r_3}{2}\int_{\mathrm{\Omega}}(h-\nabla \cdot \boldsymbol{n}^{k+1})^2,
\end{aligned}
\end{equation}
\begin{equation}
\label{lambda_1}
\boldsymbol{\lambda}_{1}^{k+1}=\boldsymbol{\lambda}_{1}^{k}+r_1(\boldsymbol{p}^{k+1}-|\boldsymbol{p}^{k+1}|\boldsymbol{n}^{k+1}),
\end{equation}
\begin{equation}
\boldsymbol{\lambda}_{2}^{k+1}=\boldsymbol{\lambda}_{2}^{k}+r_2(\boldsymbol{p}^{k+1}-\nabla u^{k+1}),
\end{equation}
\begin{equation}
\label{lambda_3}
{\lambda}_{3}^{k+1}={\lambda}_{3}^{k}+r_3(h^{k+1}-\nabla \cdot \boldsymbol{n}^{k+1}).
\end{equation}
The $\boldsymbol{p}$ subproblem can be solved by applying a shrinkage operator. The $u$ and  $\boldsymbol{n}$ subproblems don't possess closed-form solutions, so one sweep of Gauss-Seidel iteration is  employed to compute approximate ones. 

 To further reduce computational complexities, Zhang et al. proposed a linearized augmented Lagrangian method  \cite{zhang2017fast}  for Euler's elastica model. Due to the use of the linearization technique, all the associated subproblems  have closed-form solutions. In terms of image quality, Zhang's algorithm achieves similar results  as Tai's method if parameters are chosen properly.

For the conventional algorithms mentioned above, we have the following two key observations.
\vskip .2cm
\par\noindent {\bf Observation 1:}
The constraint associated with the variable $\boldsymbol{n}$ is $\boldsymbol{p} = |\boldsymbol{p}|\boldsymbol{n}$, which is a relaxation to $\boldsymbol{n} = \boldsymbol{p}/|\boldsymbol{p}|$.
This is an essential technique to reduce the computational complexity for the optimization algorithms, which has been extensively utilized in the literature \cite{tai2011fast,Duan2013,yashtini2016fast,zhang2017fast}.
This technique seems quite reasonable since the two equations are mathematically equivalent to each other when $\boldsymbol{p}\neq 0$, i.e.
\[
   \boldsymbol{n} = \dfrac{\boldsymbol{p}}{|\boldsymbol{p}|} \ \Leftrightarrow \quad \boldsymbol{p} = |\boldsymbol{p}|\boldsymbol{n}.  
\]
When involved into the optimization process, however, they could demonstrate different behaviors, i.e. the two terms
$$\int_{\mathrm{\Omega}}\left|\boldsymbol{n} - \frac{\boldsymbol{p}}{|\boldsymbol{p}|}\right|^2, \quad  \int_{\mathrm{\Omega}}|\boldsymbol{p}-|\boldsymbol{p}|\boldsymbol{n}|^2 $$
are not equivalent to each other when they perform as penalty terms for
$\boldsymbol{n}$ in the optimization procedure.
 For instance, when $\boldsymbol{p}\to \boldsymbol{0}, \boldsymbol{p}\neq \boldsymbol{0}$,  the second term will lose constraint over $\boldsymbol{n}$, while the first term  requires that
 $|\boldsymbol{n}| = 1, $ regardless of the value of $\boldsymbol{p}$.  This observation indicates that replacing $\boldsymbol{n} = \boldsymbol{p}/|\boldsymbol{p}|$ by
 $\boldsymbol{p} = |\boldsymbol{p}|\boldsymbol{n}$  might be problematic. More discussions will be given later.
 
 \vskip .2cm
 \par\noindent {\bf Observation 2:}
 For the Euler's elastica model, either in the form (\ref{THC}) or (\ref{DWH}),  when $b \to 0$, it reduces to the ROF model, and only one auxiliary variable  $\boldsymbol{p}$ and one multiplier $\boldsymbol{\lambda}$ are needed to be introduced. 
So, a consistent augmented Lagrangian method for the Euler's elastica model should correspondingly reduce to the augmented Lagrangian method for solving the ROF model. However,  when applying the alternating optimization technique, the subproblem for the optimization with respect to $\boldsymbol{p}$ obviously depends on  $\boldsymbol{n}$, regardless of the value of parameter $b$. That's to say, the augmented Lagrangian method for the Euler's elastica model doesn't immediately reduce to the  augmented Lagrangian method for the ROF model when $b=0$. This is an internal inconsistency for applying the augmented Lagrangian approach directly for solving the Euler's elastica model, which we think is the main reason responsible for the tricky coupling between the parameters and slow convergence. More discussions will be given later. 
 
\subsection{The Restricted Augmented Lagrangian Method}
  
 In light of {\bf Observation 1} and {\bf Observation 2}, in this subsection we will develop an efficient augmented Lagrangian based  algorithm for solving the Euler's elastica model.  The proposed algorithm consists of three key ingredients. The first one accounts for {\bf Observation 1}, and the original constraint
 $\boldsymbol{n} = \boldsymbol{p}/|\boldsymbol{p}|$ is used, instead of the relaxed version 
  $\boldsymbol{p} = |\boldsymbol{p}|\boldsymbol{n}$.
The second one accounts for {\bf Observation 2}, and 
a simple cutting-off strategy is employed to recover the favourable property that the augmented Lagrangian methods for solving the Euler's elastica model should reduce to the augmented Lagrangian methods for solving the ROF model when $b=0$.  The third ingredient is the linearization technique utilized to reduce the computational complexity, as suggested in \cite{zhang2017fast}.
  
Let's first reformulate the Euler's elastica model as the following constraint minimization problem
 
\begin{equation}
\begin{aligned}
\label{REE}
&\min\limits_{u,\boldsymbol{p},\boldsymbol{n},h}\int_{\mathrm{\Omega}}\left(a+bh^2\right)|\boldsymbol{p}|+{\frac{\lambda}{2}}\int_{\mathrm{\Omega}} (u-f)^2\\
&~~~s.t.\quad \boldsymbol{p}=\nabla u,\quad   \boldsymbol{n} = \dfrac{\boldsymbol{p}}{|\boldsymbol{p}|},\quad h=\nabla \cdot\boldsymbol{n},
\end{aligned}
\end{equation}
where the constraint for $\boldsymbol{n}$ takes its original form, rather than the relaxed  one. Then,
the augmented Lagrangian energy functional related to this minimization problem reads
\begin{equation}
\begin{aligned}
\label{RAL}
\mathcal{L}(u, \boldsymbol{p}, \boldsymbol{n}, h&; \boldsymbol{\lambda}_{1}, \boldsymbol{\lambda}_{2}, {\lambda}_{3})= 
\int_{\mathrm{\mathrm{\Omega}}}(a+bh^2)|\boldsymbol{p}| + {\frac{\lambda}{2}}\int_{\mathrm{\Omega}} (u-f)^2\\
&+\int_{\mathrm{\Omega}}\boldsymbol{\lambda}_{1}\cdot (\boldsymbol{n} - \dfrac{\boldsymbol{p}}{|\boldsymbol{p}|})+\frac{r_1}{2}\int_{\mathrm{\Omega}}|\boldsymbol{n} - \dfrac{\boldsymbol{p}}{|\boldsymbol{p}|}|^2\\
&+\int_{\mathrm{\Omega}}\boldsymbol{\lambda}_{2}\cdot(\boldsymbol{p}-\nabla u)+\frac{r_2}{2}\int_{\mathrm{\Omega}}|\boldsymbol{p}-\nabla u|^2\\
&+\int_{\mathrm{\Omega}}{\lambda}_{3}(h-\nabla \cdot \boldsymbol{n})+\frac{r_3}{2}\int_{\mathrm{\Omega}}(h-\nabla \cdot \boldsymbol{n})^2,
\end{aligned}
\end{equation}
where,  $\boldsymbol{\lambda}_{1}, \boldsymbol{\lambda}_{2} \in \mathbb{R}^2$, and ${\lambda}_{3} \in  \mathbb{R}$ are Lagrangian multipliers corresponding to the constraints $\boldsymbol{n} = \boldsymbol{p}/|\boldsymbol{p}|$, $\boldsymbol{p}=\nabla u$ and $h=\nabla \cdot \boldsymbol{n}$, respectively, and  $r_i$, $i=1,2,3$ and $\lambda$ are positive scalars as weighting parameters. It should be noted that, to avoid the degenerate case of $\boldsymbol{p} = 0$, it's a common practice to replace  $|\boldsymbol{p}|$ by $|\boldsymbol{p}|_\epsilon = |\boldsymbol{p}| + \epsilon$, where $\epsilon$ is a small positive constant. In this paper, we fix $\epsilon = 1e-4$ in all the experiments.

As usually done in the literature, we adopt the alternating minimization method to minimize (\ref{RAL}), just as described by (\ref{pre_u-sub})-(\ref{lambda_3}). However, the minimization subproblem with respect to $\boldsymbol{p}$ will be modified to eliminate
its connections to the variable $\boldsymbol{n}$. This is fulfilled by simply cutting-off the terms involving $\boldsymbol{n}$ in (\ref{pre_p-sub}), which results in the  following four sequential subproblems and three updating equations for the Lagrangian multipliers

\begin{equation}
\begin{aligned}
\label{u-sub}
u^{k+1} = &\arg\min\limits_{u} {\frac{\lambda}{2}} \int_{\mathrm{\Omega}} (u-f)^2
+\int_{\mathrm{\Omega}}\boldsymbol{\lambda}_{2}^k\cdot(\boldsymbol{p}^k-\nabla u)\\
&+\frac{r_2}{2}\int_{\mathrm{\Omega}}|\boldsymbol{p}^k-\nabla u|^2,
\end{aligned}
\end{equation}
\begin{equation}
\begin{aligned}
\label{p-sub}
\boldsymbol{p}^{k+1}=&\arg\min\limits_{\boldsymbol{p}} \int_{\mathrm{\Omega}} \left(a+b\left(h^k\right)^2\right)|\boldsymbol{p}|
+\int_{\mathrm{\Omega}}\boldsymbol{\lambda}_{2}^k\cdot(\boldsymbol{p}-\nabla u^{k+1})\\
&+\frac{r_2}{2}\int_{\mathrm{\Omega}}|\boldsymbol{p}-\nabla u^{k+1}|^2,
\end{aligned}
\end{equation}
\begin{equation}
\label{n-sub}
\begin{aligned}
\boldsymbol{n}^{k+1}  =&\arg\min\limits_{\boldsymbol{n}}\int_{\mathrm{\Omega}}\boldsymbol{\lambda}_{1}^{k}\cdot\left(\boldsymbol{n}-\frac{\boldsymbol{p}^{k+1}}{|\boldsymbol{p}^{k+1}|_\epsilon}\right)
+\frac{r_1}{2}\int_{\mathrm{\Omega}}\left|\boldsymbol{n}-\frac{\boldsymbol{p}^{k+1}}{|\boldsymbol{p}^{k+1}|_\epsilon}\right|^2\\
&+\int_{\mathrm{\Omega}}\lambda_3^k(h^k-\nabla \cdot \boldsymbol{n})+\frac{r_3}{2}\int_{\mathrm{\Omega}}(h^k-\nabla \cdot \boldsymbol{n})^2,
\end{aligned}
\end{equation}
\begin{equation}
\begin{aligned}
\label{h-sub}
h^{k+1}  =&\arg\min\limits_{h}\int_{\mathrm{\Omega}}\left (a+b\left(h\right)^2\right)|\boldsymbol{p}^{k+1}| 
+\int_{\mathrm{\Omega}}{\lambda}_{3}^k(h-\nabla \cdot \boldsymbol{n}^{k+1})\\
&+\frac{r_3}{2}\int_{\mathrm{\Omega}}(h-\nabla \cdot \boldsymbol{n}^{k+1})^2,
\end{aligned}
\end{equation}
\begin{equation}
\boldsymbol{\lambda}_{1}^{k+1}=\boldsymbol{\lambda}_{1}^{k}+r_1(\boldsymbol{n}^{k+1}-\frac{\boldsymbol{p}^{k+1}}{|\boldsymbol{p}^{k+1}|_\epsilon}),
\end{equation}
\begin{equation}
\boldsymbol{\lambda}_{2}^{k+1}=\boldsymbol{\lambda}_{2}^{k}+r_2(\boldsymbol{p}^{k+1}-\nabla u^{k+1}),
\end{equation}
\begin{equation}
{\lambda}_{3}^{k+1}={\lambda}_{3}^{k}+r_3(h^{k+1}-\nabla \cdot \boldsymbol{n}^{k+1}).
\end{equation}

Please note that,  when $b = 0$,  the couplings between the variable $\boldsymbol{n}$ and $\boldsymbol{p}$ will be removed, such that the subproblems for $u^{k+1}$ and
$\boldsymbol{p}^{k+1}$  form a closed iteration cycle, which coincides exactly with the augmented Lagrangian method for solving the ROF model. 

There is an intuitive explanation for the rationality of applying the cutting-off strategy. There is an ordered dependence between the auxiliary variables 
$\boldsymbol{p}$, $\boldsymbol{n}$ and $h$, which is usually ignored in the literature. The variable $\boldsymbol{p}$ can be thought of as an essential variable, while $\boldsymbol{n}$ is defined as $\boldsymbol{p}$/$|\boldsymbol{p}|$ to represent some normal field, and $h$ is introduced to denote $\nabla \cdot\boldsymbol{n}$.  The role of the cutting-off strategy mentioned above is to restrict the information flowing from $\boldsymbol{n}$ to $\boldsymbol{p}$ when solving the $\boldsymbol{p}$ subproblem, which is reasonable since $\boldsymbol{p}$ should not depend on $\boldsymbol{n}$.  

From now on, we will name our method as RALM (restricted augmented Lagrangian method) for brevity.  

\subsection {The Solutions of Minimization Subproblems}
\label{solutions}

In \cite{zhang2017fast}, Zhang et al. proposed a linearization technique to efficiently solve the subproblems derived from the augmented Lagrangian formulations. It has been shown that the linearization technique could achieve comparable results as the method proposed in \cite{Duan2013}, with much reduced computational complexity. So, the linearization technique is also adopted here to develop fast solvers for our subproblems, i.e. the $u$-subproblem and $\boldsymbol{n}$-subproblem. For easy reference, the linearization procedure \cite{zhang2017fast} is described below.


\vskip .2cm
 \par\noindent
\textbf{For the $u$-subproblem}
Due to the nonlinearity of $|\boldsymbol{p}^k-\nabla u|^2$, it's difficult to figure out an efficient solver for (\ref{u-sub}). So, we approximate this term as follows:
\begin{equation*}
\begin{aligned}
&\frac{r_2}{2}\int_{\mathrm{\Omega}}|\boldsymbol{p}^k-\nabla u|^2
\approx \frac{r_2}{2}\int_{\mathrm{\Omega}}|\boldsymbol{p}^k-\nabla u^k|^2\\
&+r_2 \int_{\mathrm{\Omega}}\langle \nabla(\boldsymbol{p}^k-\nabla u^k),u-u^k\rangle 
+\frac{1}{2\delta_1}\int_{\mathrm{\Omega}}(u-u^k)^2,
\end{aligned}
\end{equation*}
where $\delta_1 > 0$ is a constant, which denotes the step-size for updating $u$. By applying the above approximation, the  Euler-Lagrange equation of (\ref{u-sub}) can then be simplified to 
\begin{equation}
\label{eq_u}
\frac{1}{\delta_1}(u-u^k)+\lambda u=\lambda f-\nabla\cdot(r_2\boldsymbol{p}^k+\boldsymbol{\lambda}_{2}^k)+r_2\Delta u^k,
\end{equation}
which admits a closed-form solution
\begin{equation}
\label{u_solution}
u^{k+1}=\frac{u^k+\delta_1g_1}{1+\delta_1\lambda},
\end{equation}
where $g_1$ denotes the right-hand side of (\ref{eq_u}).

\vskip .2cm
 \par\noindent
\textbf{For the $\boldsymbol{p}$-subproblem}
The $\boldsymbol{p}$-subproblem can be solved by the soft thresholding operator defined in  
\cite{goldstein2009split},  i.e. the solution of (\ref{p-sub}) can be explicitly expressed as 
\begin{equation}
\boldsymbol{p}^{k+1}= shrinkage\left(\frac{r_2\nabla u^{k+1}-\boldsymbol{\lambda}_{2}^k}{r_2},\frac{c}{r_2}\right),
\end{equation}
where
$$shrinkage(x,\alpha) = sign(x) * max \left(|x| - \alpha, 0 \right),$$ 
and the constant $c=\left(a+b\left(h^k\right)^2\right)$.

\vskip .2cm
 \par\noindent
\textbf{For the $\boldsymbol{n}$-subproblem}
Note that the constraint for $\boldsymbol{n}$ has changed back to its original form, i.e.
$$\boldsymbol{p}=|\boldsymbol{p}|\boldsymbol{n} \quad \rightarrow \quad   \boldsymbol{n} = \dfrac{\boldsymbol{p}}{|\boldsymbol{p}|}.$$
This is the main difference from the derivation  described in \cite{zhang2017fast}.
To handle the singularity of the operator $\nabla(\nabla\cdot)$, a quadratic penalty term shall be added. The $\boldsymbol{n}$-subproblem can then be reformulated as
\begin{equation*}
\begin{aligned}
\label{rn-sub}
&\boldsymbol{n}^{k+1}=\arg\min\limits_{\boldsymbol{n}}  \int_{\mathrm{\Omega}}\boldsymbol{\lambda}_{1}^{k}\cdot(\boldsymbol{n}-\frac{\boldsymbol{p}^{k+1}}{|\boldsymbol{p}^{k+1}|_\epsilon})
+\frac{r_1}{2}\int_{\mathrm{\Omega}}\left|\boldsymbol{n}-\frac{\boldsymbol{p}^{k+1}}{|\boldsymbol{p}^{k+1}|_\epsilon}\right|^2 \\
&+\int_{\mathrm{\Omega}}\lambda_3(h^k-\nabla \cdot \boldsymbol{n})
+\frac{r_3}{2}\int_{\mathrm{\Omega}}(h^k-\nabla \cdot \boldsymbol{n})^2
+\frac{\gamma}{2}\int_{\mathrm{\Omega}}|\boldsymbol{n}-\boldsymbol{n}^k|^2,
\end{aligned}
\end{equation*}
where $\gamma > 0$ is a penalty parameter.
As in solving the $u$-subproblem, we use the linearization technique to approximate the  penalty term $\int_{\mathrm{\Omega}}\frac{r_3}{2}(h^k-\nabla \cdot \boldsymbol{n})^2$ as
\begin{equation*}
\begin{aligned}
&\frac{r_3}{2}\int_{\mathrm{\Omega}}(h^k-\nabla\cdot\boldsymbol{n})^2
\approx \frac{r_3}{2}\int_{\mathrm{\Omega}}(h^k-\nabla\cdot\boldsymbol{n}^k)\\
&+r_3\int_{\mathrm{\Omega}}\langle \nabla(h^k-\nabla\cdot\boldsymbol{n}^k),\boldsymbol{n}-\boldsymbol{n}^k)\rangle 
+\frac{1}{2\delta_2}\int_{\mathrm{\Omega}}|\boldsymbol{n}-\boldsymbol{n}^k|^2,
\end{aligned}
\end{equation*}
where $\delta_2 > 0$ is a constant. By applying the above approximation, the Euler-Lagrange equation of (\ref{n-sub}) can be simplified as
\begin{equation}
\begin{aligned}
\label{equ_n}
\frac{\boldsymbol{n}-\boldsymbol{n}^k}{\delta_2}+\left(\gamma+r_1\right)\boldsymbol{n}=&\gamma\boldsymbol{n}^k+r_1\frac{\boldsymbol{p}^{k+1}}{|\boldsymbol{p}^{k+1}|_\epsilon}
-\boldsymbol{\lambda}_{1}^{k}-r_3\nabla h^k\\
& -\nabla{\lambda}_{3}^k+r_3\nabla(\nabla\cdot\boldsymbol{n}^k),
\end{aligned}
\end{equation}
which admits the following closed-form solution
\begin{equation}
\boldsymbol{n}^{k+1}=\frac{\boldsymbol{n}^k+\delta_2g_2}{1+\delta_2(\gamma+r_1)},
\end{equation}
where $g_2$ denotes the right-hand side of (\ref{equ_n}).

\vskip .2cm
 \par\noindent
\textbf{For the $h$-subproblem}
 The corresponding  Euler-Lagrange equation for (\ref{h-sub}) can be simply derived as
\begin{equation}
2b|\boldsymbol{p}^{k+1}| h^{k+1} + {\lambda}_{3}^k + r_3(h^{k+1}-\nabla\cdot\boldsymbol{n}^{k+1})=0,
\end{equation}
from which, we obtain 
\begin{equation}
h^{k+1}=\frac{r_3\nabla\cdot\boldsymbol{n}^{k+1}-{\lambda}_{3}^k}{2b|\boldsymbol{p}^{k+1}|+r_3}.
\end{equation}
The above solution procedures can be summarized as  
\textbf{Algorithm \ref{alg:RALM}}. 
For easy reference, the algorithm proposed in \cite{zhang2017fast} will be named as LALM (linearized augmented Lagrangian method). To validate the advantage of the constraint $\boldsymbol{n} = \boldsymbol{p}/|\boldsymbol{p}|$ over 
  $\boldsymbol{p} = |\boldsymbol{p}|\boldsymbol{n}$, we  also need adapt RALM to use the latter constraint. Such an algorithm will be named LALMn, which is summarized as \textbf{Algorithm \ref{alg:LALM_N}}.

\begin{algorithm}[] %
	\caption{Restricted Augmented Lagrangian Method ({\bf RALM}) for Euler's elastica model} %
        \label{alg:RALM} 
        \begin{algorithmic} 
        \State {Input}: Given an observed image $f, a, b, \lambda, \gamma, r_1, $ $ r_2, r_3, \delta_1, \delta_2$
        \State {Output}: the restored image $u$
       \State Set $u^0$ = $f$, $\boldsymbol{p}^0$ = $\boldsymbol{n}^0$ = $\boldsymbol{\lambda}_{1}^{0}$ = $\boldsymbol{\lambda}_{2}^0$  =  $\boldsymbol{0}$, $\lambda_3^0$ = $h^0$ = $0$ , Set $k = 0$
           \Repeat
			\State 
\State $u^{k+1}=\frac{u^k+\delta_1g_1}{1+\delta_1\lambda}$
\State $\boldsymbol{p}^{k+1}= shrinkage\left(\frac{r_2\nabla u^{k+1}-\boldsymbol{\lambda}_{2}^k}{r_2},\frac{c}{r_2}\right)$
\State $\boldsymbol{n}^{k+1}=\frac{\boldsymbol{n}^k+\delta_2g_2}{1+\delta_2(\gamma+r_1)}$
\State $h^{k+1}=\frac{r_3\nabla\cdot\boldsymbol{n}^{k+1}-{\lambda}_{3}^k}{2b|\boldsymbol{p}^{k+1}|+r_3}$
\State $\boldsymbol{\lambda}_{1}^{k+1}=\boldsymbol{\lambda}_{1}^{k}+r_1(\boldsymbol{n}^{k+1}-\frac{\boldsymbol{p}^{k+1}}{|\boldsymbol{p}^{k+1}|_\epsilon})$
\State $\boldsymbol{\lambda}_{2}^{k+1}=\boldsymbol{\lambda}_{2}^{k}+r_2(\boldsymbol{p}^{k+1}-\nabla u^{k+1})$
\State ${\lambda}_{3}^{k+1}={\lambda}_{3}^{k}+r_3(h^{k+1}-\nabla \cdot \boldsymbol{n}^{k+1})$
            \State $k = k + 1$
            \Until{stopping criteria is met}
		\end{algorithmic}
	\end{algorithm}

 \begin{algorithm}[] 
	\caption{ Augmented Lagrangian Method ({\bf LALMn}) for Euler's elastica model} %
	\label{alg:LALM_N} 
	\begin{algorithmic} 
	 \State {Input}: Given an observed image $f, a, b, \lambda, \gamma, r_1, $ $ r_2, r_3, \delta_1, \delta_2$
	\Repeat
	\State $u^{k+1}=\frac{u^k+\delta_1g_1}{1+\delta_1\lambda}$
		\State $\boldsymbol{p}^{k+1}= shrinkage\left(\frac{\frac{r_1\boldsymbol{n}^k+\boldsymbol{\lambda_1}^k}{|\boldsymbol{p}^k|^2_\epsilon}+r_2\nabla u^{k+1} -\boldsymbol{\lambda}_{2}^k}{\frac{r_1}{|\boldsymbol{p}^k|^2_\epsilon}+r_2},\frac{c}{\frac{r_1}{|\boldsymbol{p}^k|^2_\epsilon}+r_2}\right)$
		\State $\boldsymbol{n}^{k+1}=\frac{\boldsymbol{n}^k+\delta_2g_2}{1+\delta_2(\gamma+r_1)}$
		\State $h^{k+1}=\frac{r_3\nabla\cdot\boldsymbol{n}^{k+1}-{\lambda}_{3}^k}{2b|\boldsymbol{p}^{k+1}|+r_3}$
		\State $\boldsymbol{\lambda}_{1}^{k+1}=\boldsymbol{\lambda}_{1}^{k}+r_1(\boldsymbol{n}^{k+1}-\frac{\boldsymbol{p}^{k+1}}{|\boldsymbol{p}^{k+1}|_\epsilon})$
		\State $\boldsymbol{\lambda}_{2}^{k+1}=\boldsymbol{\lambda}_{2}^{k}+r_2(\boldsymbol{p}^{k+1}-\nabla u^{k+1})$
		\State ${\lambda}_{3}^{k+1}={\lambda}_{3}^{k}+r_3(h^{k+1}-\nabla \cdot \boldsymbol{n}^{k+1})$
		\State $k = k + 1$
		      \Until{stopping criteria is met}
\end{algorithmic}
\end{algorithm}

 \subsection{Numerical discretization}
 
 For numerical implementations, all the computations in {\bf Algorithm 1} and {\bf Algorithm 2} need to be discretized on some grid. To  simplify the expressions, we use same notations to denote both continuous and discrete  quantities like the image $u$, gradient operator $\nabla u$, etc. The actual meaning of these operators should be clear from the context. 
  Suppose the image we are dealing with is size of $M \times N \triangleq J$, and denote the Euclidean space  $X = \mathbb{R}^{M \times N}$ and $Y = X \times X$.  For given $u \in X$ and $\boldsymbol{v} \in Y$, we use the discrete forward (+) and backward  (-) differential operators for the discretization of the gradient $\nabla u$ and the divergence $\nabla \cdot \boldsymbol{v}$ operators, respectively. With Neumann boundary conditions, the above mentioned operators are defined as follows
 \begin{equation*}
 (\nabla u)(i,j) = ((\nabla_x^{+} u(i,j)),(\nabla_y^{+} u(i,j))),  
 \end{equation*}
with
\begin{equation*}
\nabla_x^{+} u(i,j) = \left\{
             \begin{aligned}
             &u(i+1,j)-u(i,j) \ & \mathrm{if}\ i<M,\\
             &0                             &  \mathrm{if} \ i=M,
             \end{aligned}
\right.
\end{equation*}
\begin{equation*}
\nabla_y^{+} u(i,j) = \left\{
             \begin{aligned}
             &u(i,j+1)-u(i,j) \  &  \mathrm{if}\ j<N,\\
             &0                             &  \mathrm{if} \ j=N,
             \end{aligned}
\right.
\end{equation*}
for $i = 1,\cdots, M$, $j = 1,\cdots, N$, and 
 \begin{equation*}
 (\nabla \cdot \boldsymbol{v})(i,j) = \nabla^{-}_x \boldsymbol{v}_1(i,j) + \nabla^{-}_y \boldsymbol{v}_2(i,j),
 \end{equation*}
 where
\begin{equation*}
\nabla_x^{-} \boldsymbol{v}_1(i,j) = \left\{
             \begin{aligned}
             &\boldsymbol{v}_1(i,j) & \mathrm{if}\ i = 1, \\
             &\boldsymbol{v}_1(i,j)-\boldsymbol{v}_1(i-1,j) & \mathrm{if} \ 1 < i < M,\\
             &-\boldsymbol{v}_1(i-1,j)      &  \mathrm{if}\ i = M,
             \end{aligned}
\right.
\end{equation*}
\begin{equation*}
\nabla_y^{-} \boldsymbol{v}_2(i,j) = \left\{
             \begin{aligned}
             &\boldsymbol{v}_2(i,j) &  \mathrm{if} \ j = 1, \\
             &\boldsymbol{v}_2(i,j) - \boldsymbol{v}_2(i,j-1)&  \mathrm{if} \ 1<j<N,\\
             &-\boldsymbol{v}_2(i,j-1)  &  \mathrm{if} \ j=N,
             \end{aligned}
\right.
\end{equation*}
for  $i = 1,\cdots, M$, $j = 1,\cdots, N$.

\section{Experiments}\label{NE}

Various experiments will be performed against LALM \cite{zhang2017fast} in this section to validate the proposed algorithm.  We will show that, compared with LALM, the proposed method RALM enjoys easier parameter tuning,  less iteration numbers, and recovers images with higher quality.  RALM will not be compared direclty with the fast algorithm proposed in \cite{Duan2013}, since it has been compared carefully with LALM in \cite{zhang2017fast}.  

 A  synthesized image consisting of several circular rings will be utilized to carry out experiments to demonstrate the effect of the  constraint adopted by RALM, as well as  how LALM and RALM perform when the parameter $b\to 0$. Experiments on real images will be then performed to further validate the efficiency and effectiveness of the proposed method.

 To compare the results quantitatively, we have computed three quality indices:  PSNR, NRMSE and NMAD, which are deinfed by (\ref{equ:psnr}), (\ref{equ:NRMSE}) and (\ref{equ:NMAD}), respectively. All the gray images are scaled to the range $[0,1]$ and correspondingly the display windows of all the restored images  are also set to $[0,1]$. When a noisy image is needed,  the Gaussian white noise with  variance 0.01 is always applied. In all the following experiments, we use the relative residuals defined by (\ref{re}) as the stopping criterion. To  check the convergence behaviors of the competing methods, we also monitor the PSNR and other two indices: the numerical energy and and the norm of  $\boldsymbol{n}$, defined by (\ref{E}) and (\ref{norm_n}), respectively, to see how they evolve with the iterations.
\begin{equation}
\label{E}
E(k) = \int_\mathrm{\Omega}\left(a + b\kappa^2\right) |\nabla u^{k}| d\mathrm{\Omega}+{\frac{\lambda}{2}}\int_\mathrm{\Omega}(u^{k}-f)^2d\mathrm{\Omega},
\end{equation}
\begin{equation}
\label{equ:psnr}
PSNR = 10log_{10}\left(\frac{MAX^2_{I}}{\frac{1}{J}\sum\limits_{i=1}^{M}\sum\limits_{j=1}^{N}\left(I(i,j)-f(i,j)\right)^2}\right),
\end{equation}
  
\begin{equation}
\label{equ:NRMSE}
NRMSE = \left(\frac{\sum\limits_{i=1}^{M}\sum\limits_{j=1}^{N}(I(i,j)-f(i,j))}{\sum\limits_{i=1}^{M}\sum\limits_{j=1}^{N}(I(i,j)-\frac{1}{J}\sum\limits_{i=1}^{M}\sum\limits_{j=1}^{N}I(i,j))}\right)^{\frac{1}{2}},
\end{equation}
\begin{equation}
\label{equ:NMAD}
NMAD = \frac{\sum\limits_{i=1}^{M}\sum\limits_{j=1}^{N}|I(i,j)-f(i,j)|}{\sum\limits_{i=1}^{M}\sum\limits_{j=1}^{N}|I(i,j)|},
\end{equation}
\begin{equation}
\label{re}
\frac{\parallel u^k-u^{k-1}\parallel_{2}}{\parallel u^{k-1}\parallel_{2}},
\end{equation}

\begin{equation}
\label{norm_n}
\parallel \boldsymbol{n}^k\parallel = \frac{\sum\limits_{i=1}^{M}\sum\limits_{j=1}^{N}\sqrt{\boldsymbol{n}^{k}_1(i,j)^2+\boldsymbol{n}^{k}_2(i,j)^2}}{\Omega}.
\end{equation}
The computing system is a desktop computer that possesses a 2.4GHz dual-core CPU and 16GB memory. The algorithms are implemented in MATLAB.  
 
\subsection{Easier parameter tuning for the proposed method RALM}
\label{EPT}

\begin{figure*}[!htp]
\centering
\begin{tabular}{p{0.03cm}<{\centering}p{3.2cm}<{\centering}p{3.2cm}<{\centering}p{3.2cm}<{\centering}p{3.2cm}<{\centering}}
&\ \ Noise-free&Noisy&LALM&RALM\\
 \begin{sideways}\hspace{1cm}Synthesized\hspace{1cm}\end{sideways}&
\includegraphics[width=3.5cm]{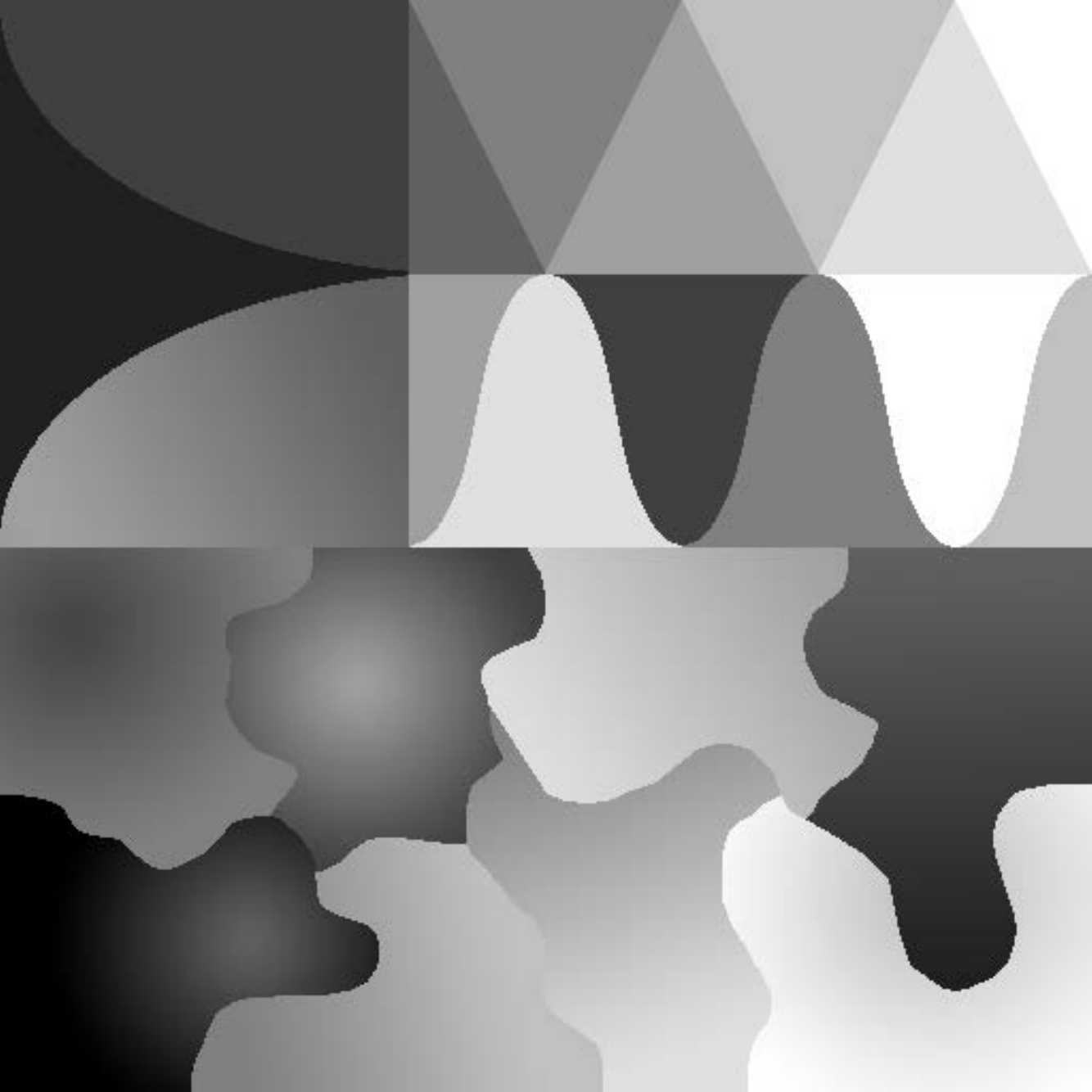}&
\includegraphics[width=3.5cm]{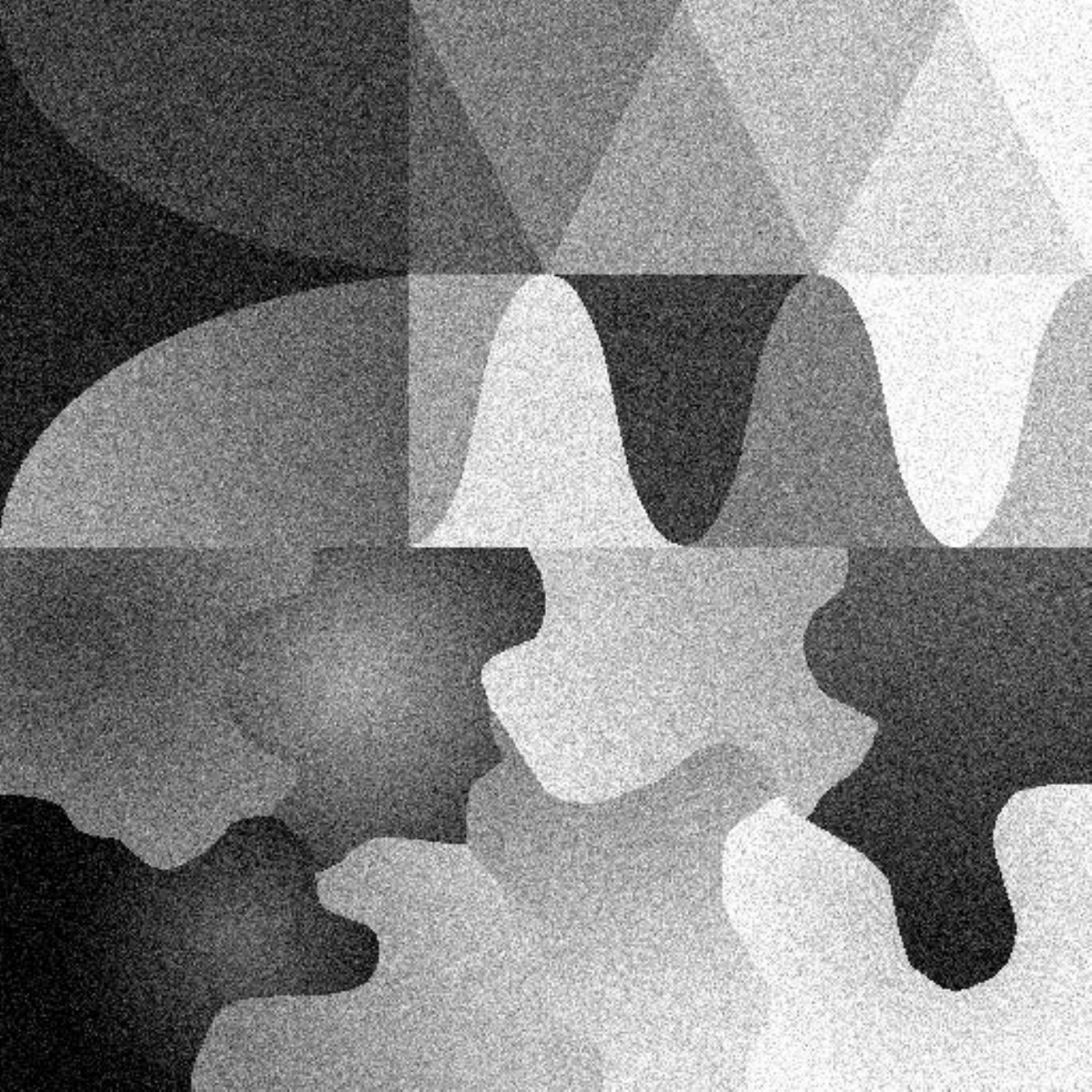}&
\includegraphics[width=3.5cm]{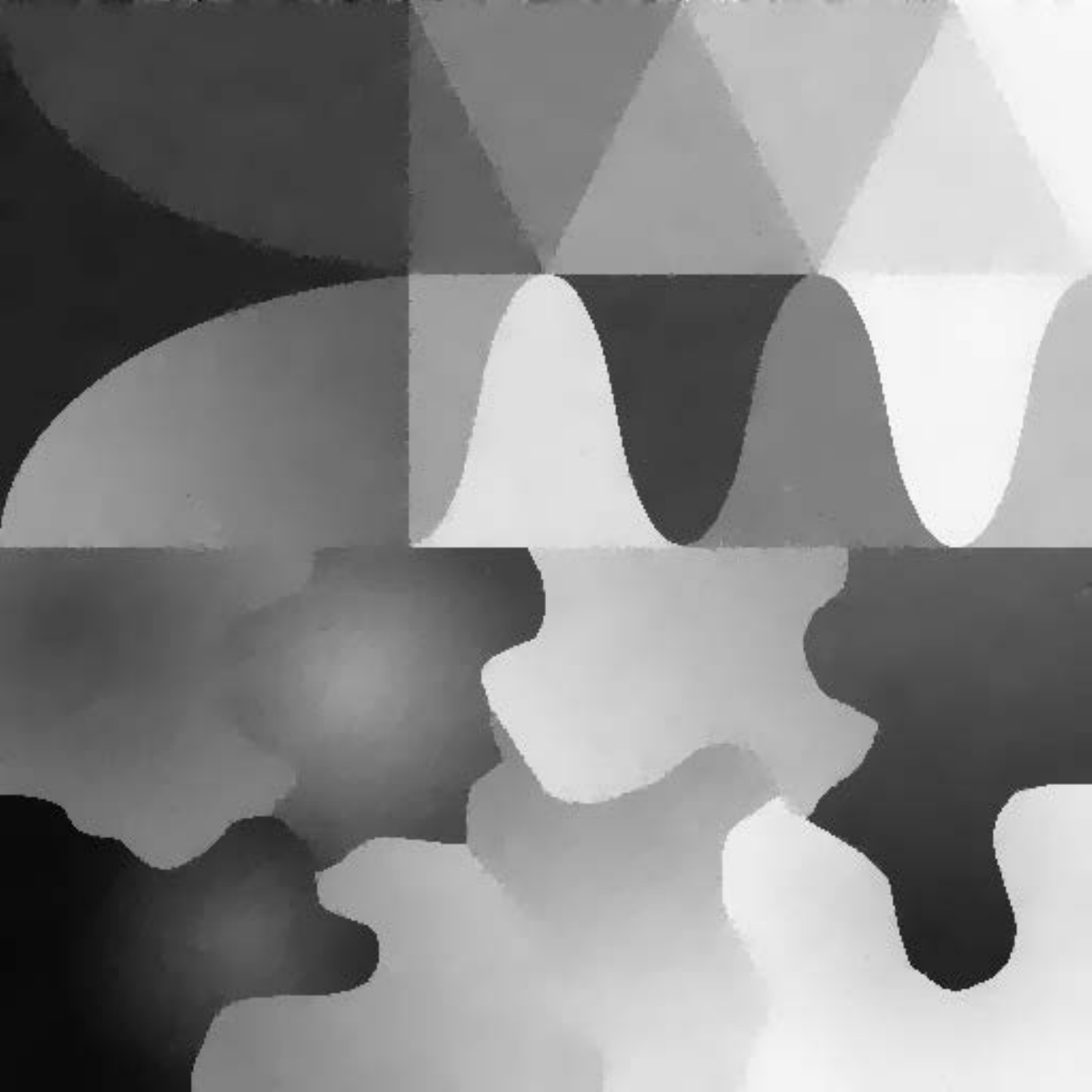}&
\includegraphics[width=3.5cm]{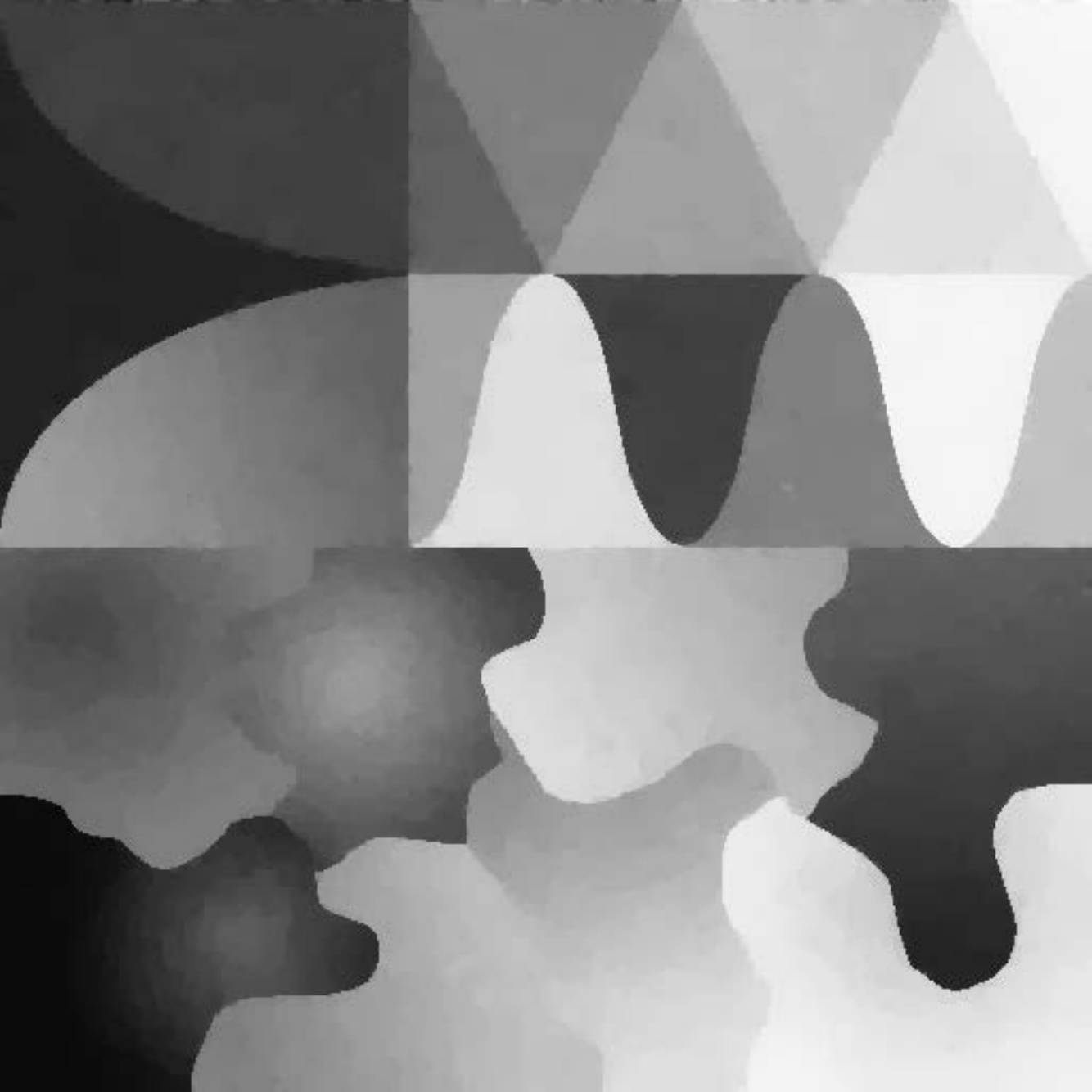}\\
\begin{sideways}\hspace{1.1cm}Pirate\hspace{1.5cm}\end{sideways}&
\includegraphics[width=3.5cm]{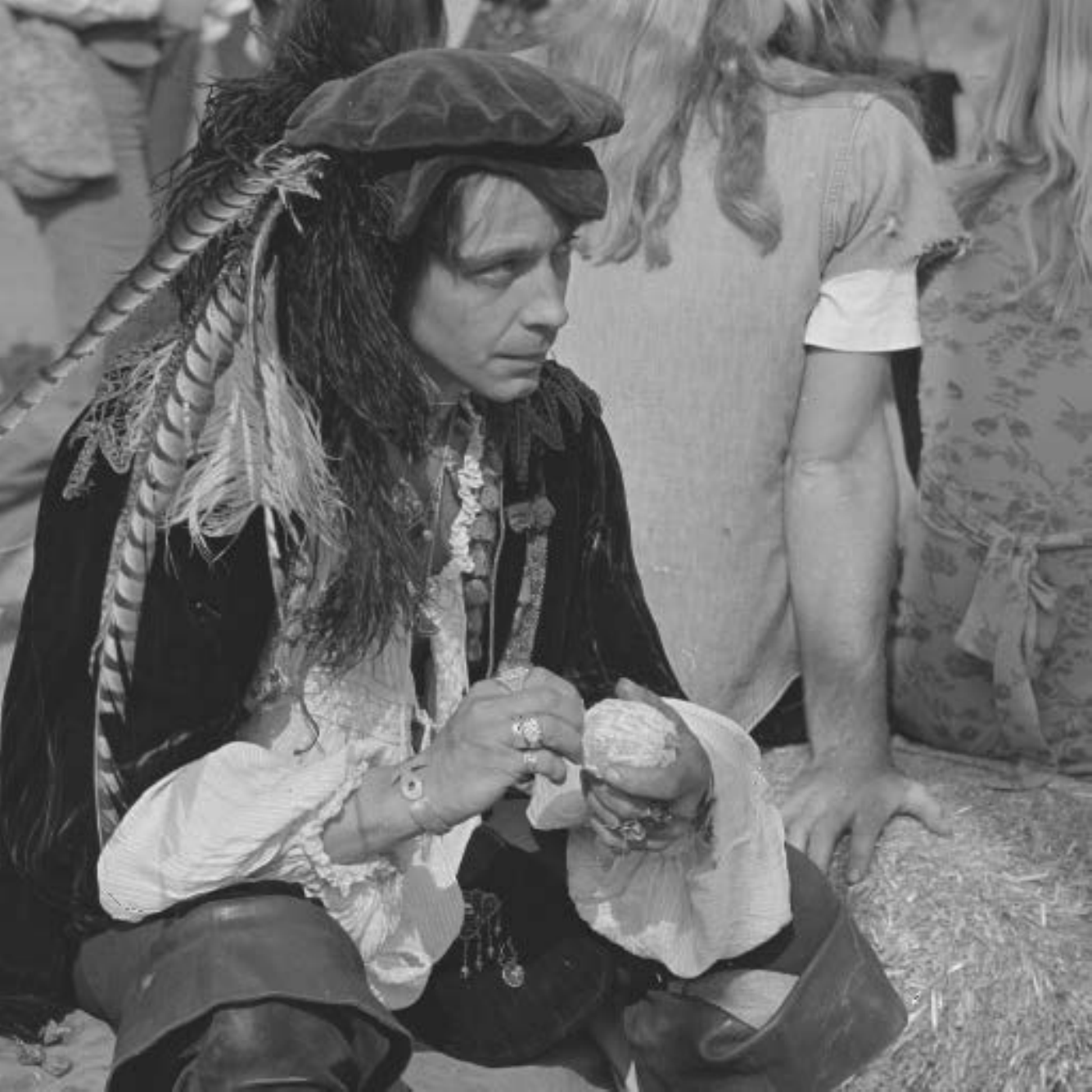}&
\includegraphics[width=3.5cm]{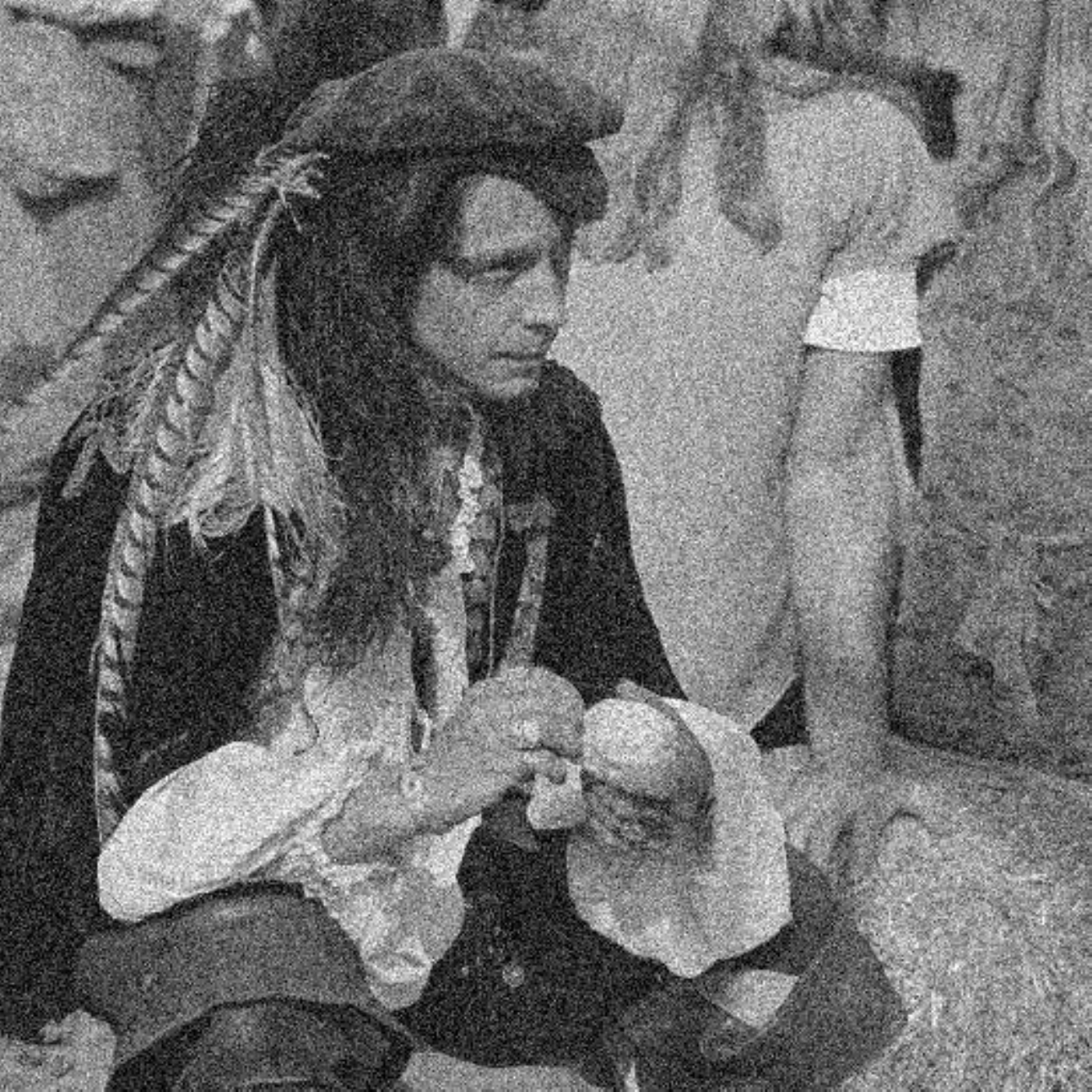}&
\includegraphics[width=3.5cm]{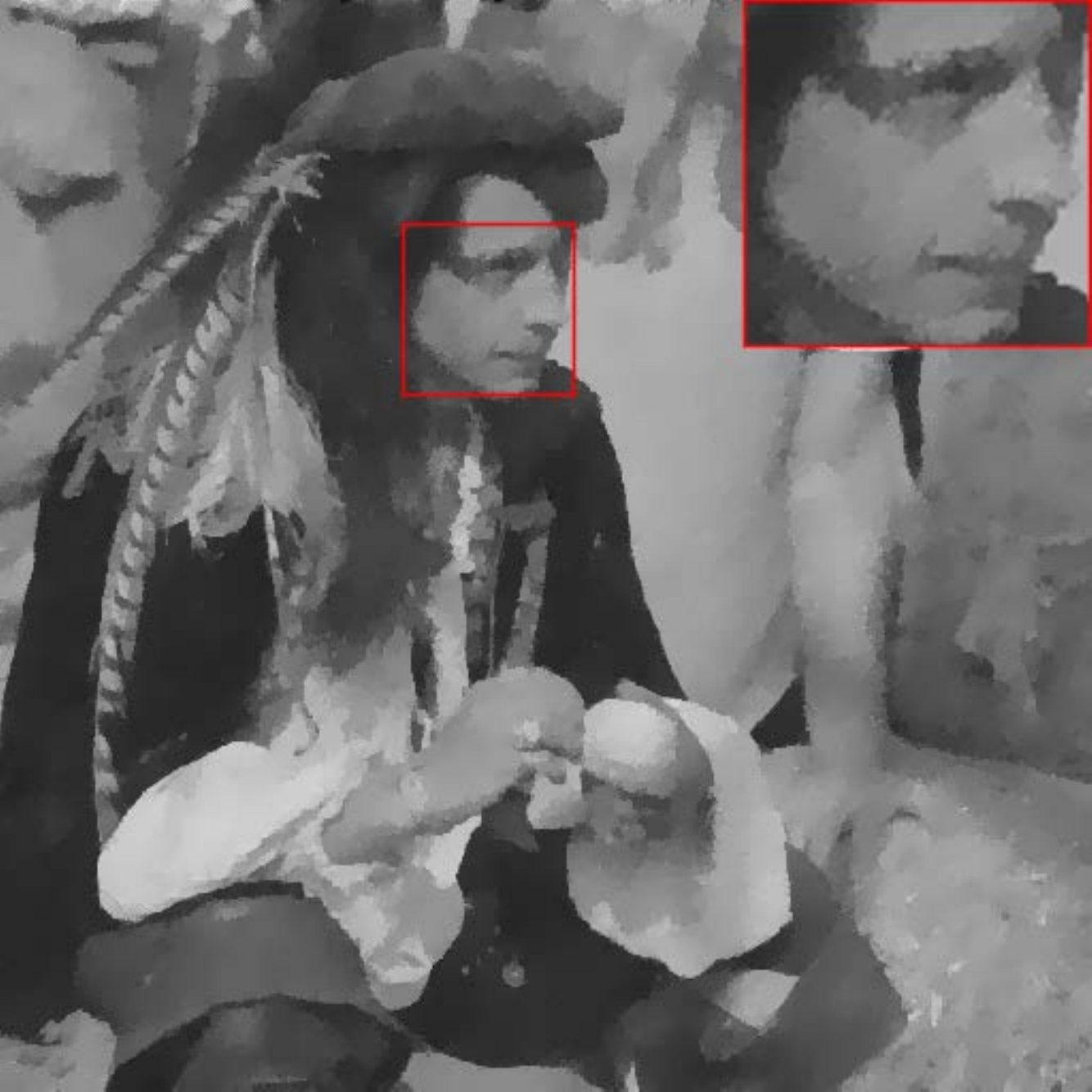}&
\includegraphics[width=3.5cm]{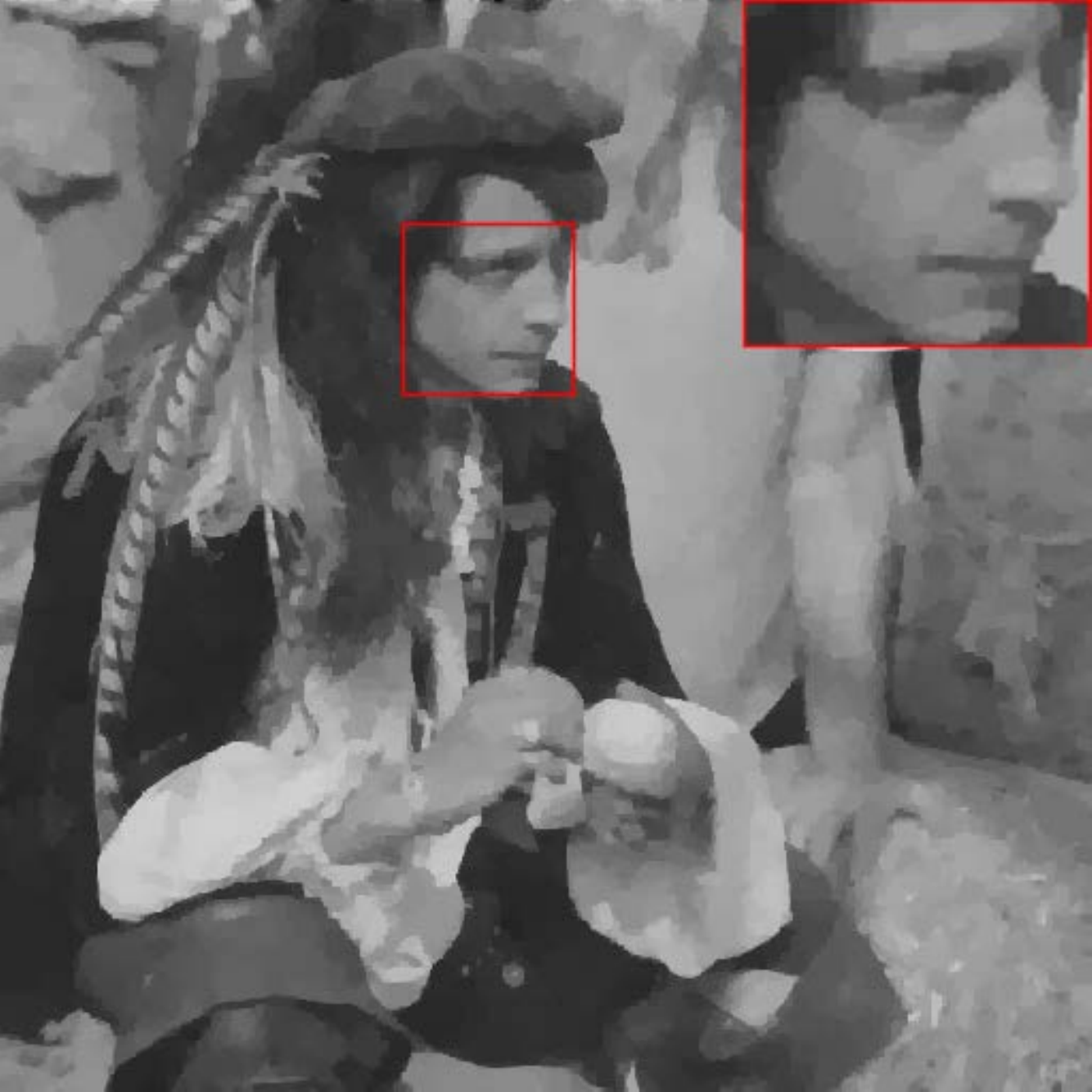}\\
\begin{sideways}\hspace{1.3cm}Boat\hspace{1.5cm}\end{sideways}&
\includegraphics[width=3.5cm]{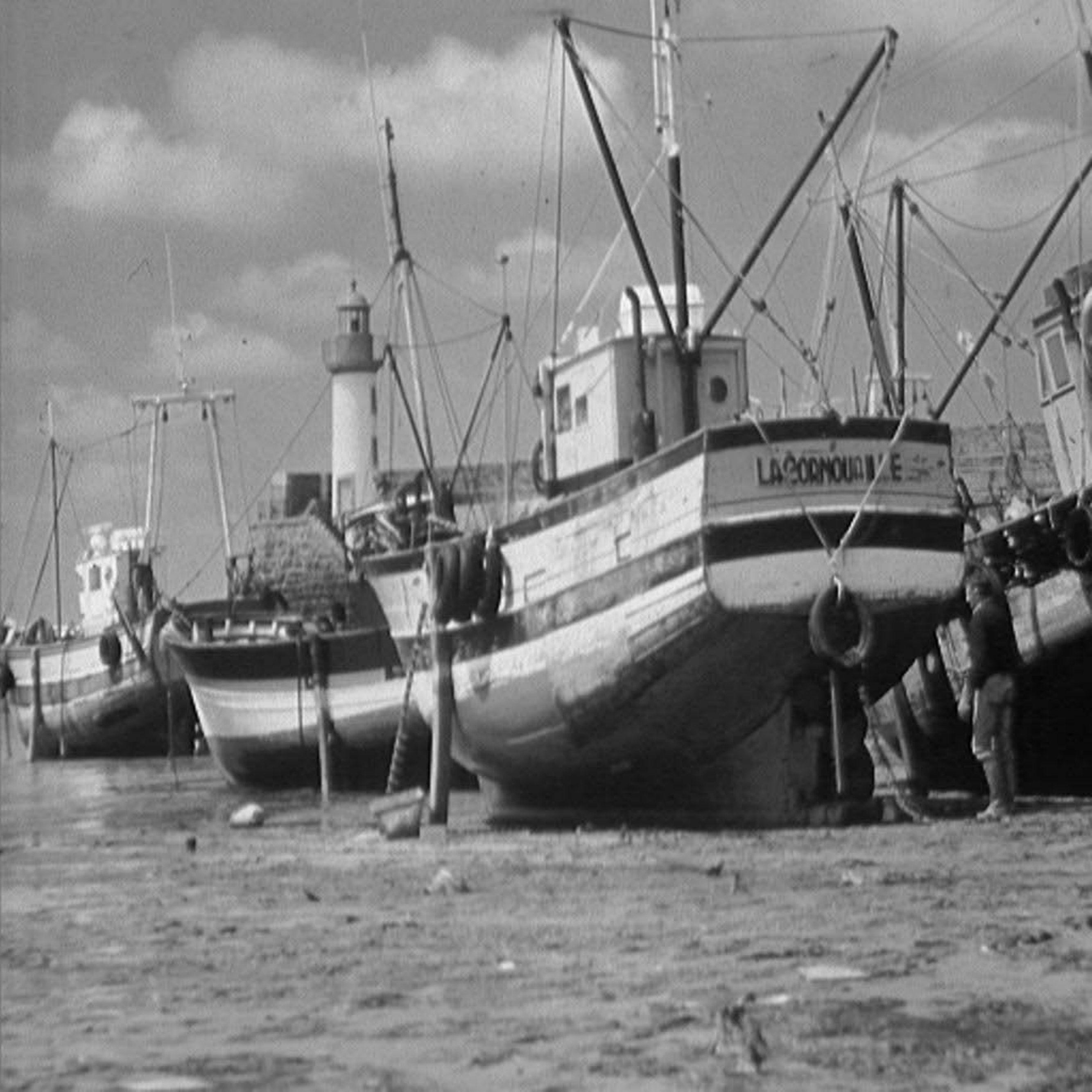}&
\includegraphics[width=3.5cm]{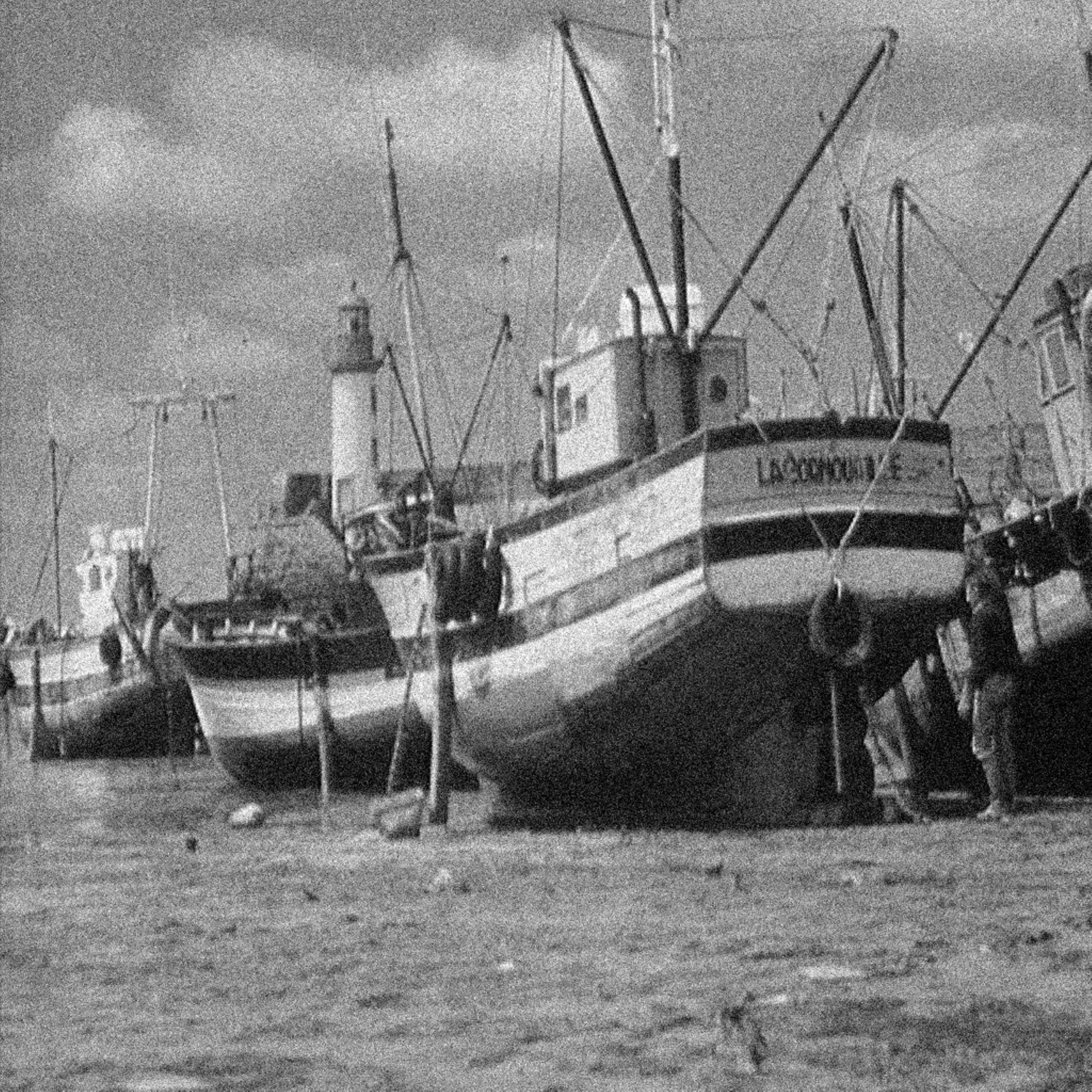}&
\includegraphics[width=3.5cm]{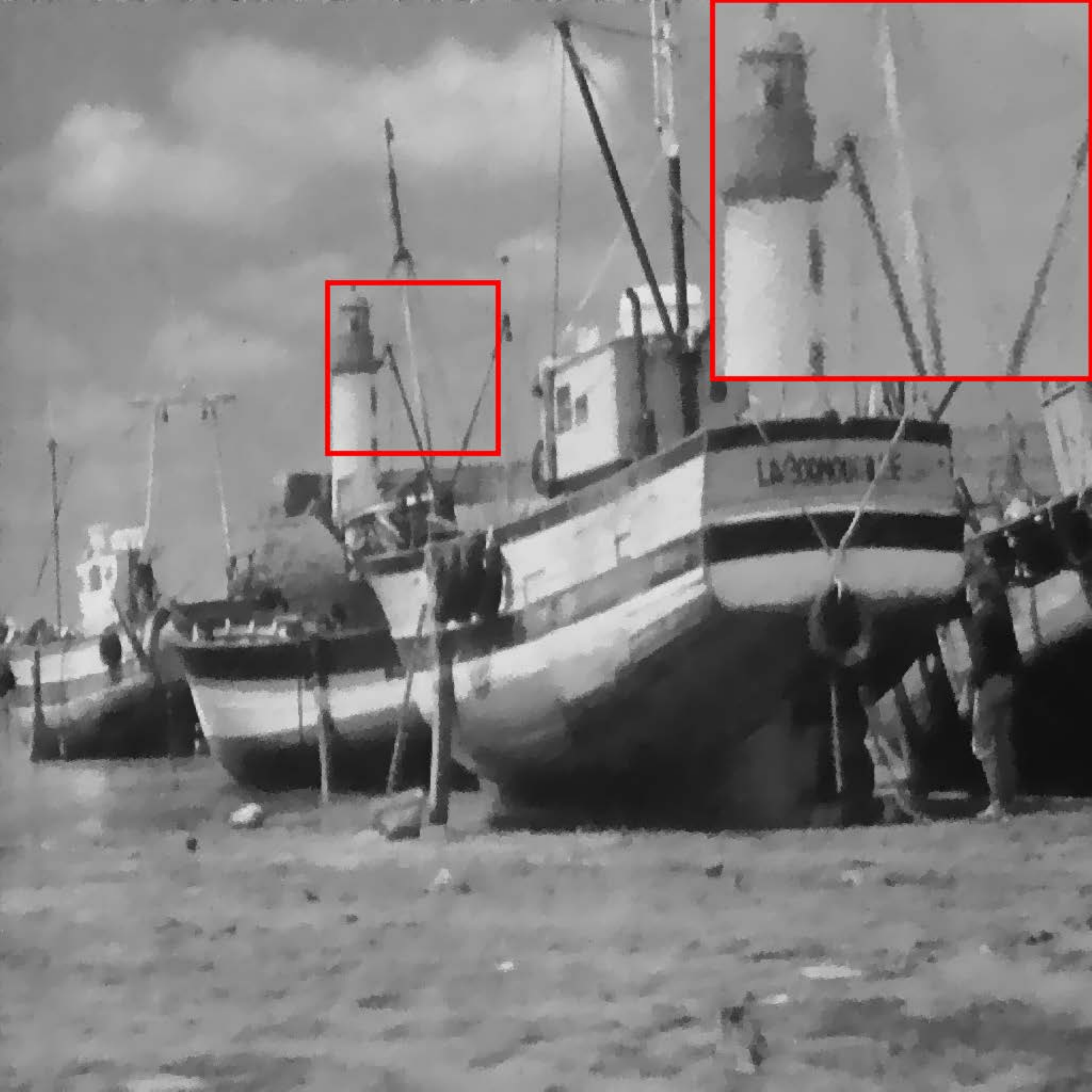}&
\includegraphics[width=3.5cm]{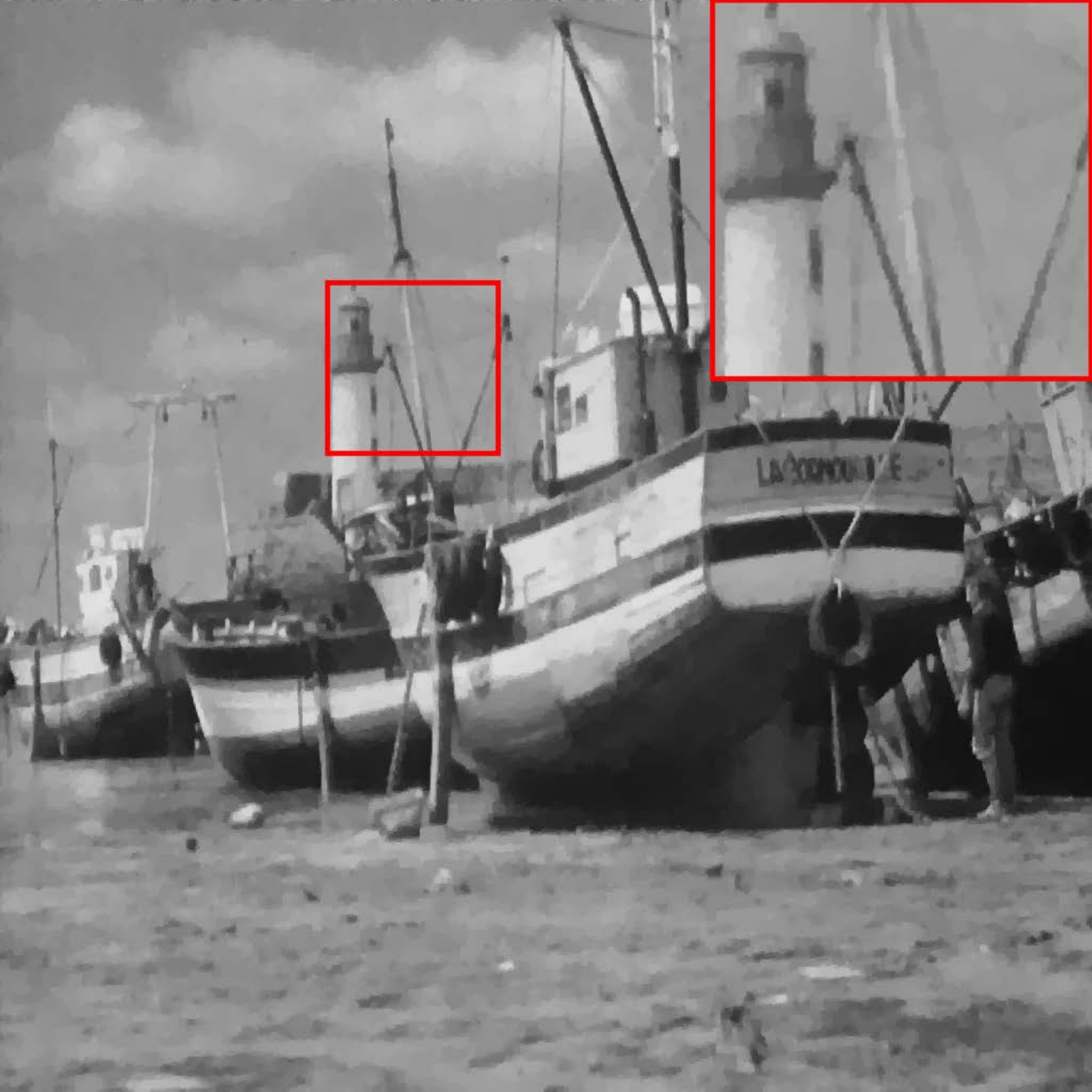}\\
\begin{sideways}\hspace{1.0cm}Montage\hspace{1.5cm}\end{sideways}&
\includegraphics[width=3.5cm]{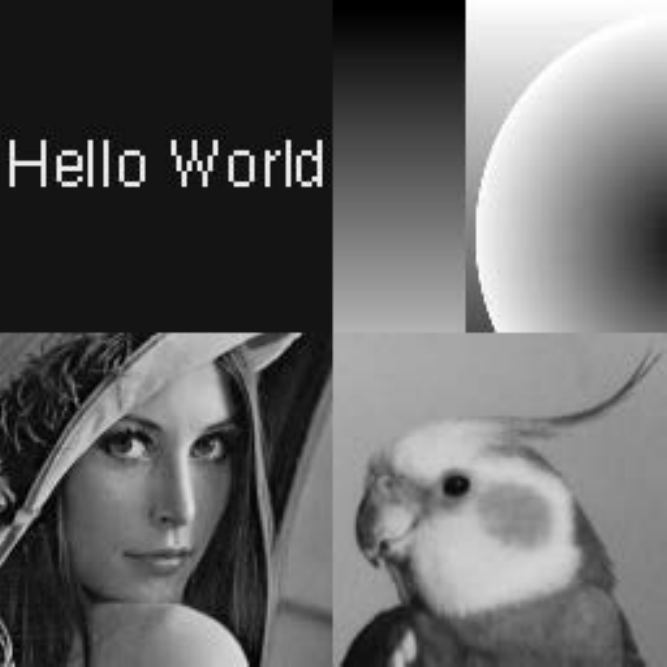}&
\includegraphics[width=3.5cm]{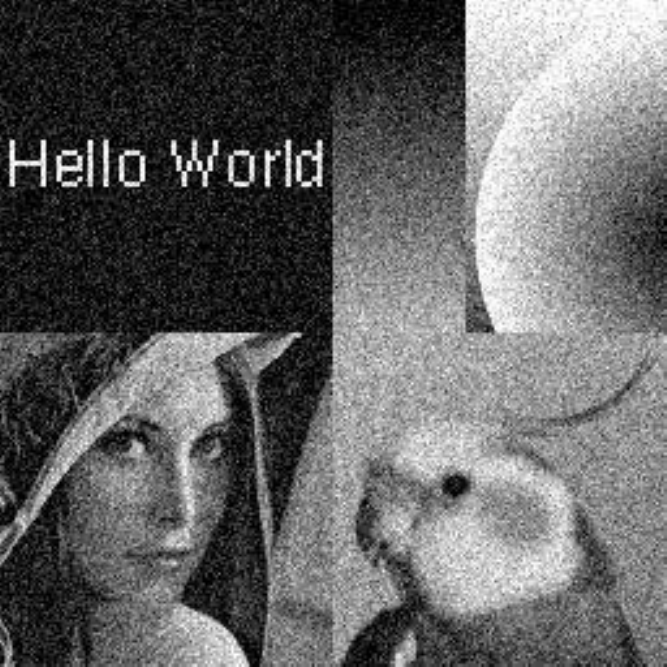}&
\includegraphics[width=3.5cm]{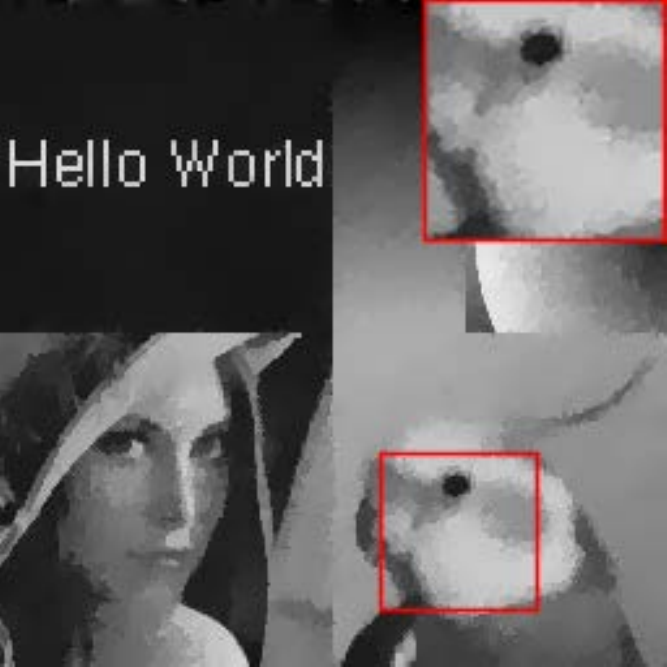}&
\includegraphics[width=3.5cm]{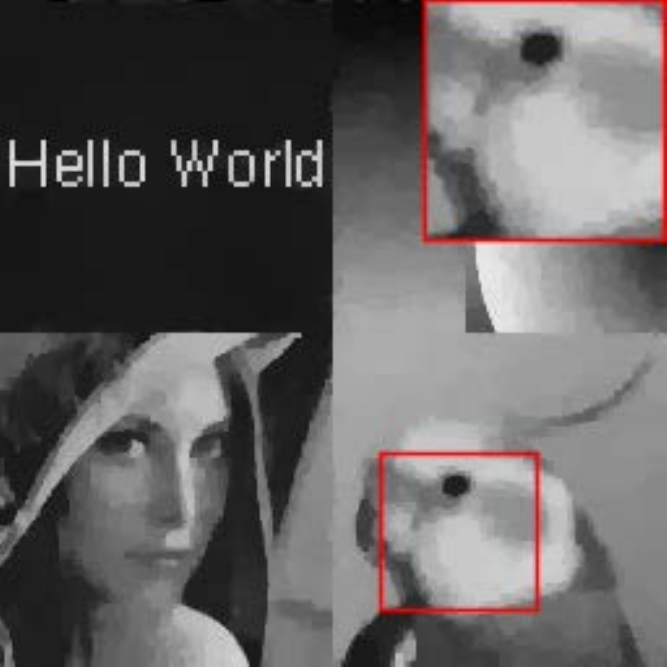}\\
\end{tabular}~
    \caption{Parameter sensitivity tests. The parameters $a=1, b=0.1,$ and $ \gamma = 1e-5$ are specified for both methods. For the remaining parameters: $\lambda = 13.5, r_1 = 60, r_2 = 2, r_3 = 1.5, \delta_1 = 2e-2, \delta_2 = 1e-2$ are specified for LALM, while $\lambda = 11, r_1 = 50, r_2 = 3, r_3 = 2, \delta_1 = 1e-2, \delta_2 = 1e-2$ are specified for RALM.}
    \label{exp1test-images} 
\end{figure*}

The Euler's elastica model (\ref{REE}) involves three parameters: $a, b$, and $\lambda$. Due to the use of the linearization technique, six more parameters: $r_1, r_2, r_3, \gamma, \delta_1, \delta_2$ are introduced for both  LALM and RALM. While the LALM can achieve promising results with properly chosen parameters, these parameters are tricky to tune. Thanks to the cutting-off strategy, the dependence between variables $\boldsymbol{p}$ and $\boldsymbol{n}$ is restricted to one direction, such that RALM has a low parameter sensibility. 
So, RALM should enjoy easier parameter tuning. To demonstrate this property, 
a  sensitivity test is performed as follows.  The two competing algorithms LALM and RALM are firstly tested on a  synthesized image size of $512\times512$ shown in the first row of Fig.\ref{exp1test-images}. For this image, the parameters for both the LALM and RALM are fine-tuned to achieve similar high quality restorations. As shown in  Fig.\ref{exp1test-images}(a), both methods achieve  quite similar PSNR values. Then the  parameters are fixed, and the two methods are applied to three other images, namely  pirate ($512\times512$), boat ($1024\times1024$) and montage ($256\times256$), to see how the performances of the two methods adapt to different images. 

\begin{figure}
\centering
\subfigure[Synthesized]{
\includegraphics[width=1.69in]{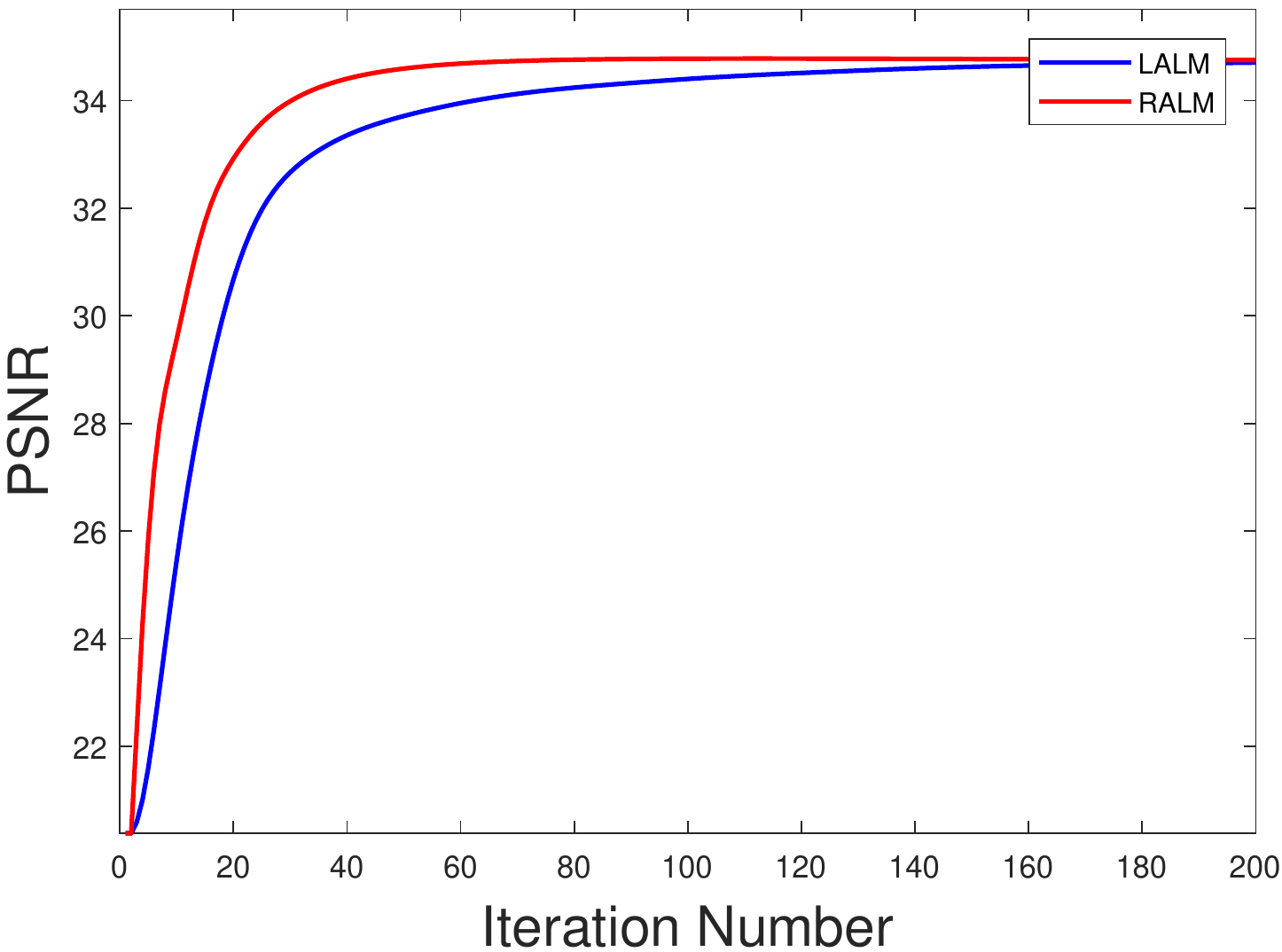}}\hfill
\subfigure[Pirate]{
\includegraphics[width=1.69in]{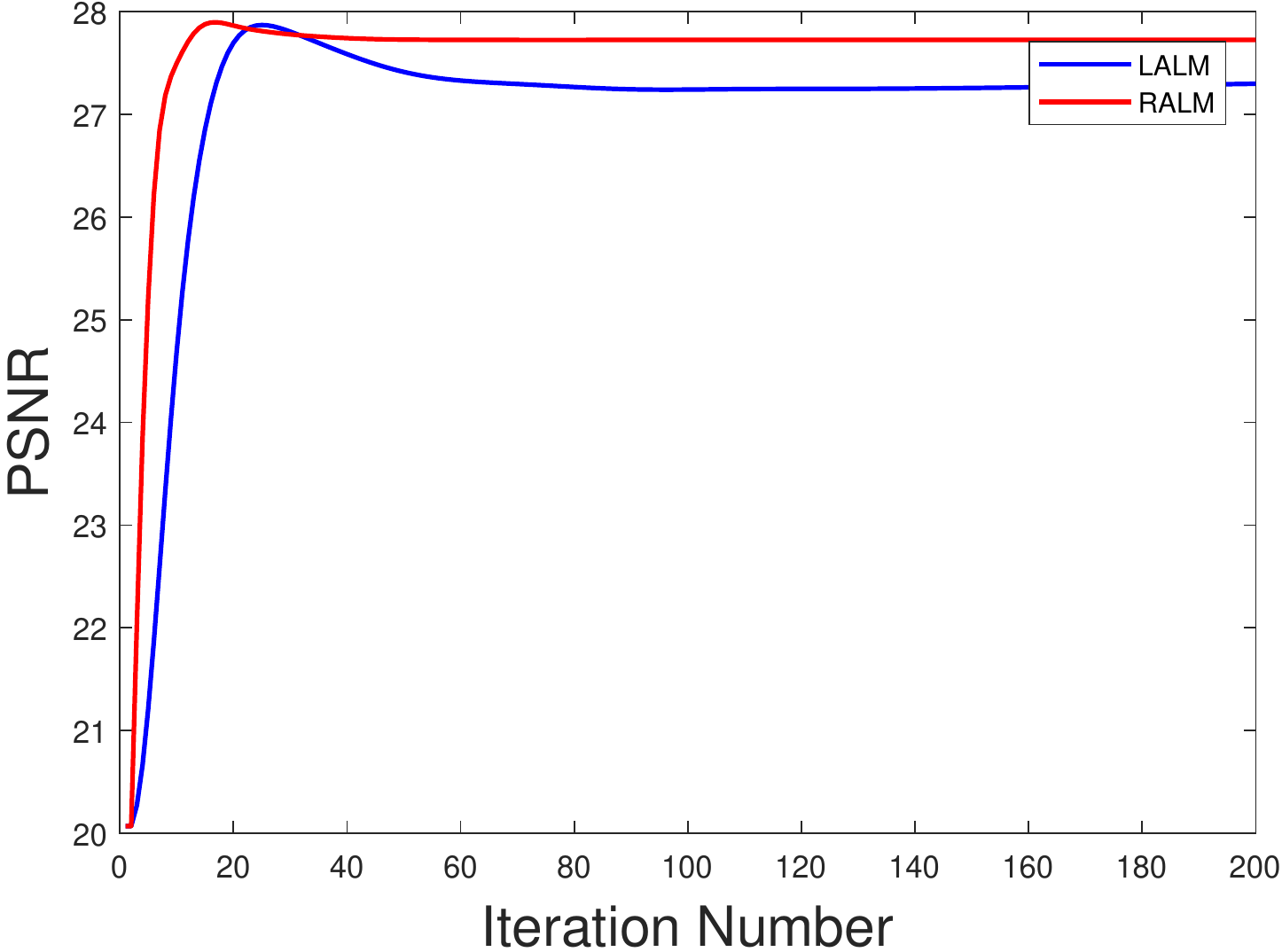}}\hfill
\subfigure[Boat]{
\includegraphics[width=1.69in]{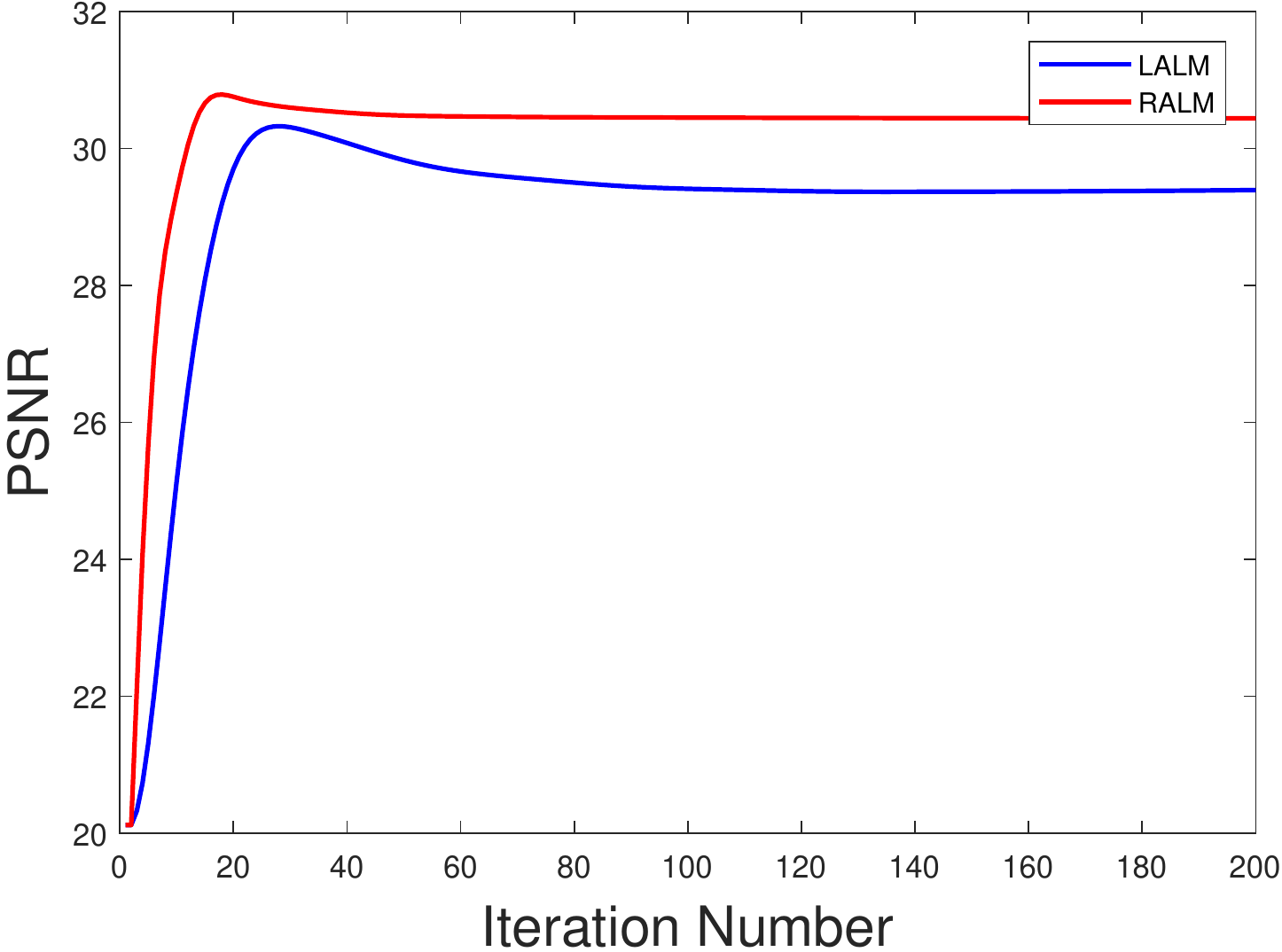}}\hfill
\subfigure[Montage]{
\includegraphics[width=1.69in]{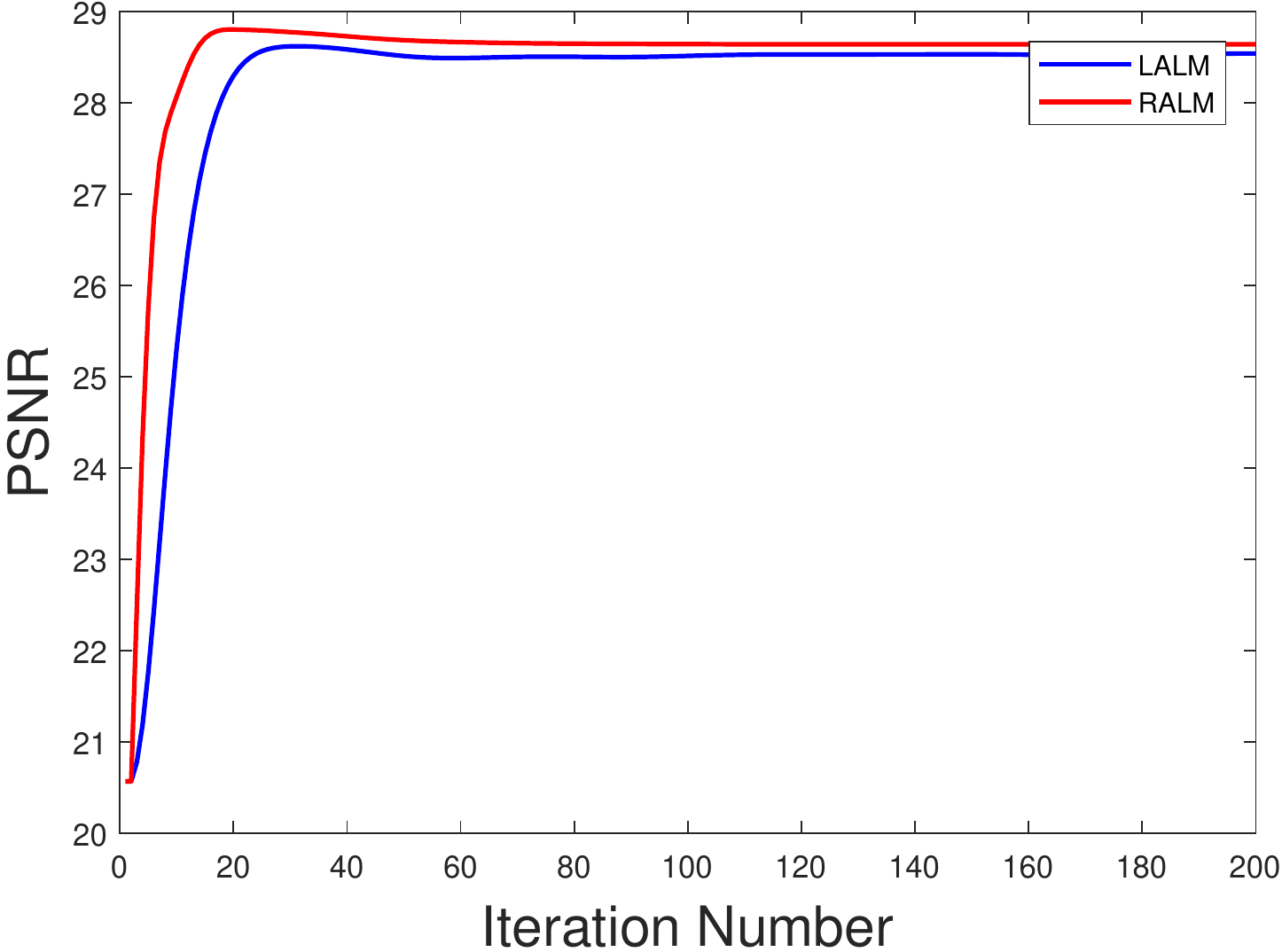}}
\caption{The PSNR plots for LALM and RALM. }
\label{exp1_psnr} 
\end{figure}

The restored images after 200 iterations are shown in the last two columns of Fig.\ref{exp1test-images}, while the PSNR curves are demonstrated in Fig.\ref{exp1_psnr}. At the first glance, both methods recover quite similar images after 200 iterations. When zooming in the resulting images and checking more closely, however, the higher quality shall be revealed for the results of RALM. For example, in the zoomed-in subregions shown at the upper right corners of the resulting images, RALM produces smoother local edges, while LALM produces small noisy structures that look like artifacts. The higher quality can also be verified by the PSNR plots shown in Fig.\ref{exp1_psnr}. An apparent observation is that, after growing to the maximum, a sharp drop of the PSNR plots for LALM occurs for the tests with the pirate and boat  images, which indicates that the optimal parameters of LALM are sensitive to different images. On the other hand, the PSNR plots of RALM  suffer from much smaller drops, which indicates that the parameters of RALM are robust across different images, i.e. they are less  sensitive than those of LALM.  Note that for the montage image, both LALM and RALM produce rather similar results. We think this might be because the montage image shares more similar properties with the synthesized image, i.e. both images present few textures, and can be well approximated by piecewise smooth functions.

\subsection{Advantage Validation of the constraint $\boldsymbol{n} ={\boldsymbol{p}}/{|\boldsymbol{p}|}$ over $\boldsymbol{p} = |\boldsymbol{p}|\boldsymbol{n}$}
As stated previously, even the two constraints  $\boldsymbol{n} =\boldsymbol{p}/|\boldsymbol{p}|$ and $\boldsymbol{p} = |\boldsymbol{p}|\boldsymbol{n}$ are mathematically equivalent to each other (when $\boldsymbol{p}\neq 0$), they might behave quite differently when being used as penalty terms.  This can be illustrated by the following comparison experiment with LALMn and LALM. The former adopts the constraint $\boldsymbol{n} ={\boldsymbol{p}}/{|\boldsymbol{p}|}$, just as described in \textbf{Algorithm \ref{alg:LALM_N}}, while the latter adopts the constraint $\boldsymbol{p} = |\boldsymbol{p}|\boldsymbol{n}$.

\begin{figure}[h]
	\centering
	\subfigure[ Original image]{
		\label{originalc} 
		\includegraphics[width=1.69in]{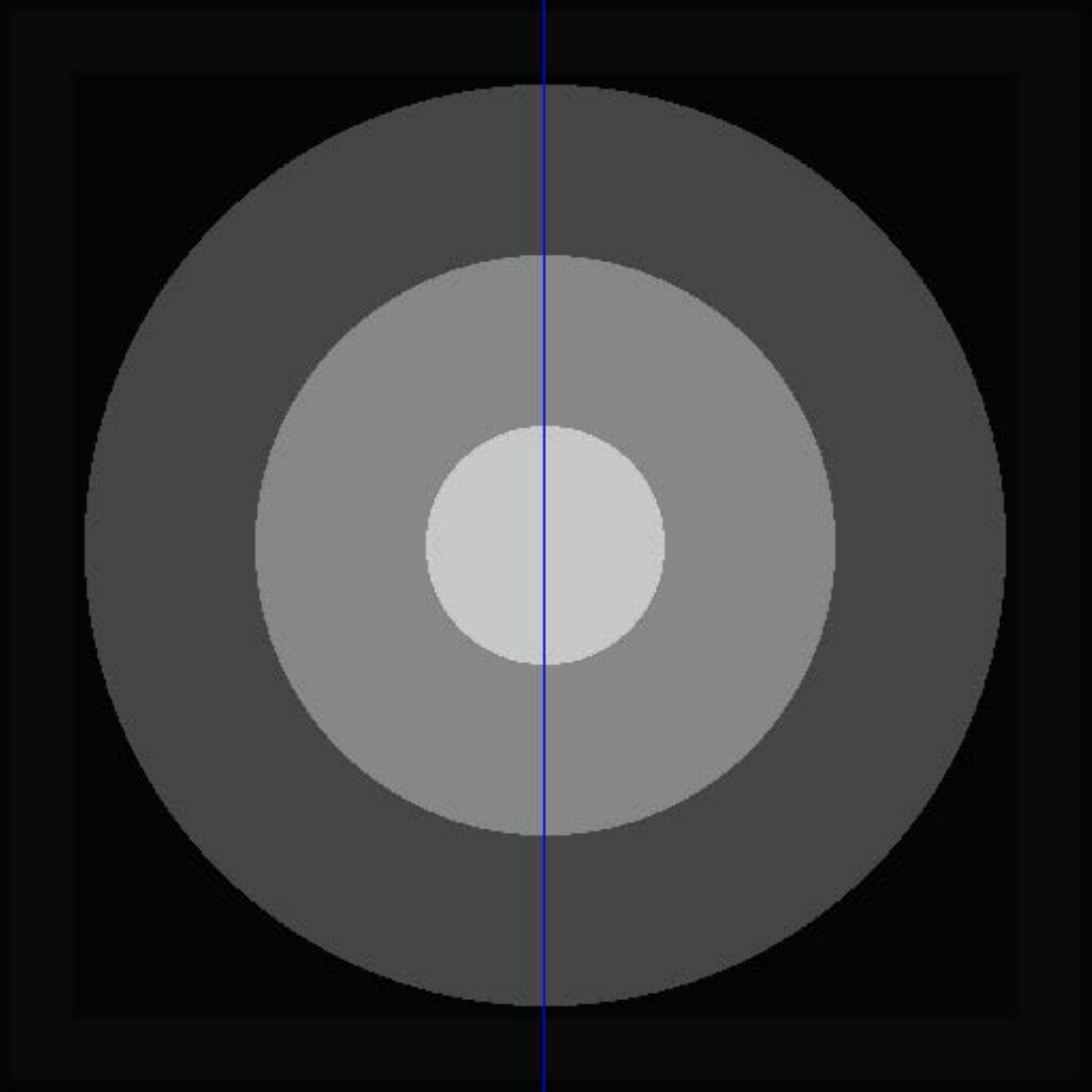}}\hfill
	\subfigure[Noisy image]{
		\label{noisyc} 
		\includegraphics[width=1.69in]{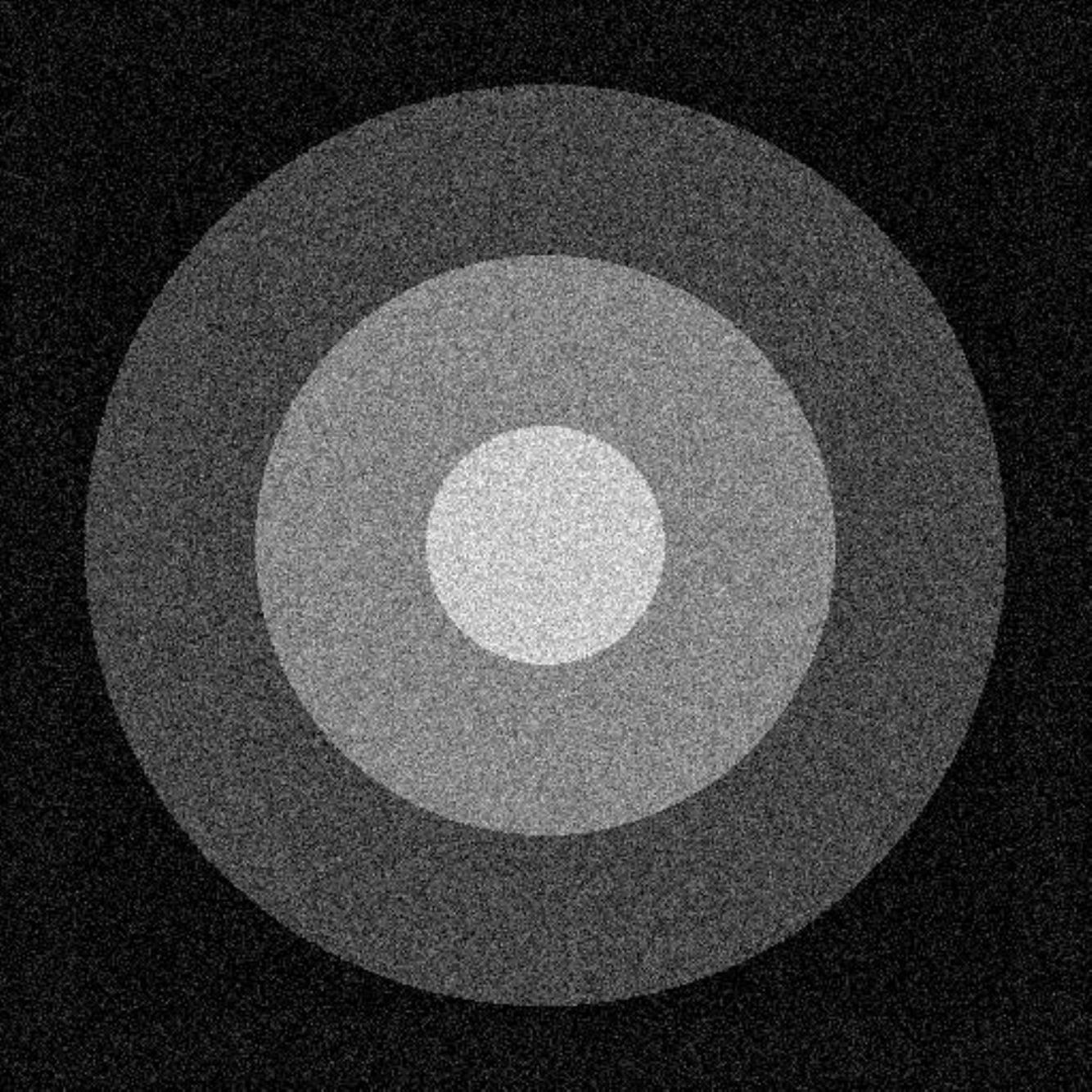}}\hfill
	\subfigure[Denoised by LALM]{
		\label{LALM_c} 
		\includegraphics[width=1.69in]{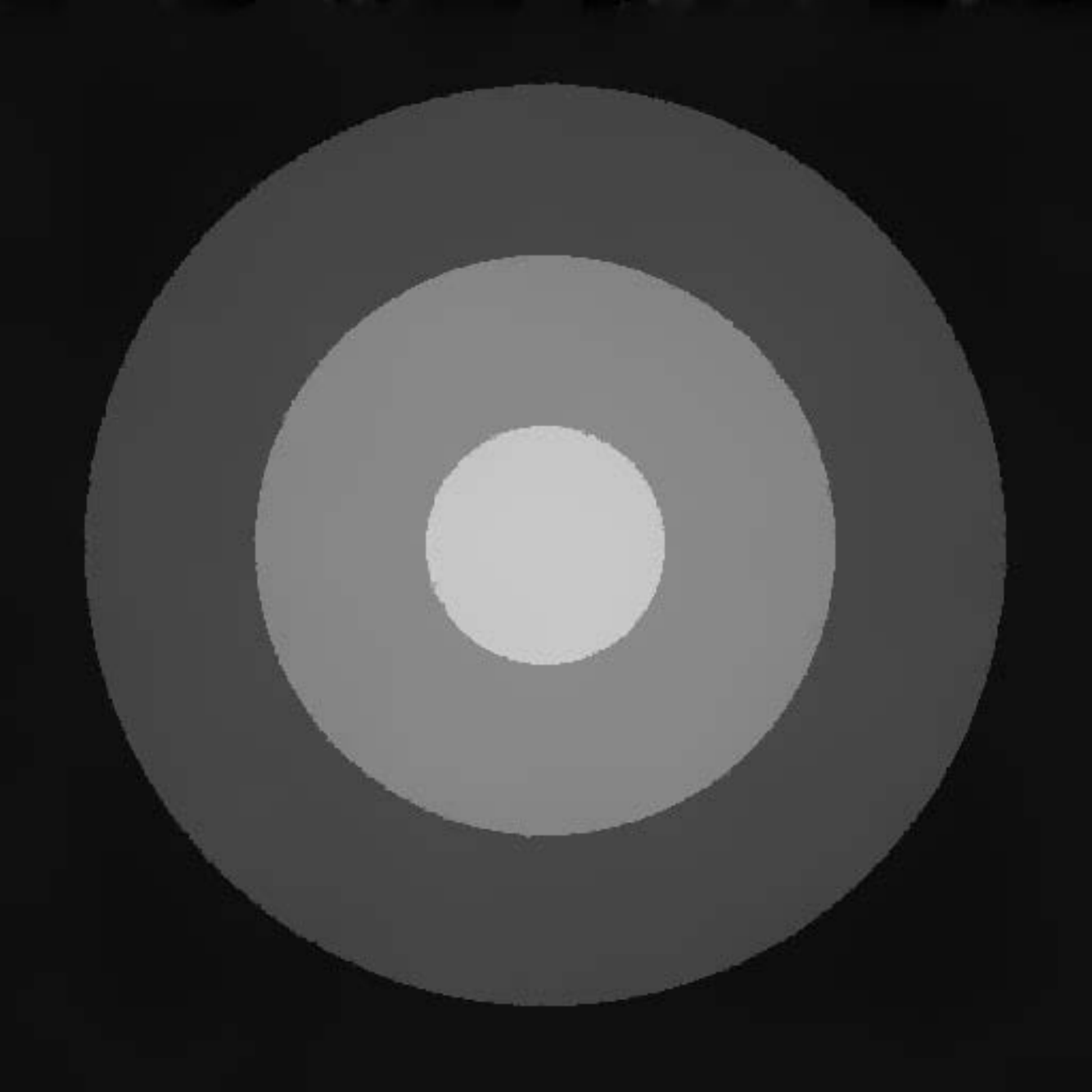}}\hfill
	\subfigure[ Denoised by LALMn ]{
		\label{LALMn_c} 
		\includegraphics[width=1.69in]{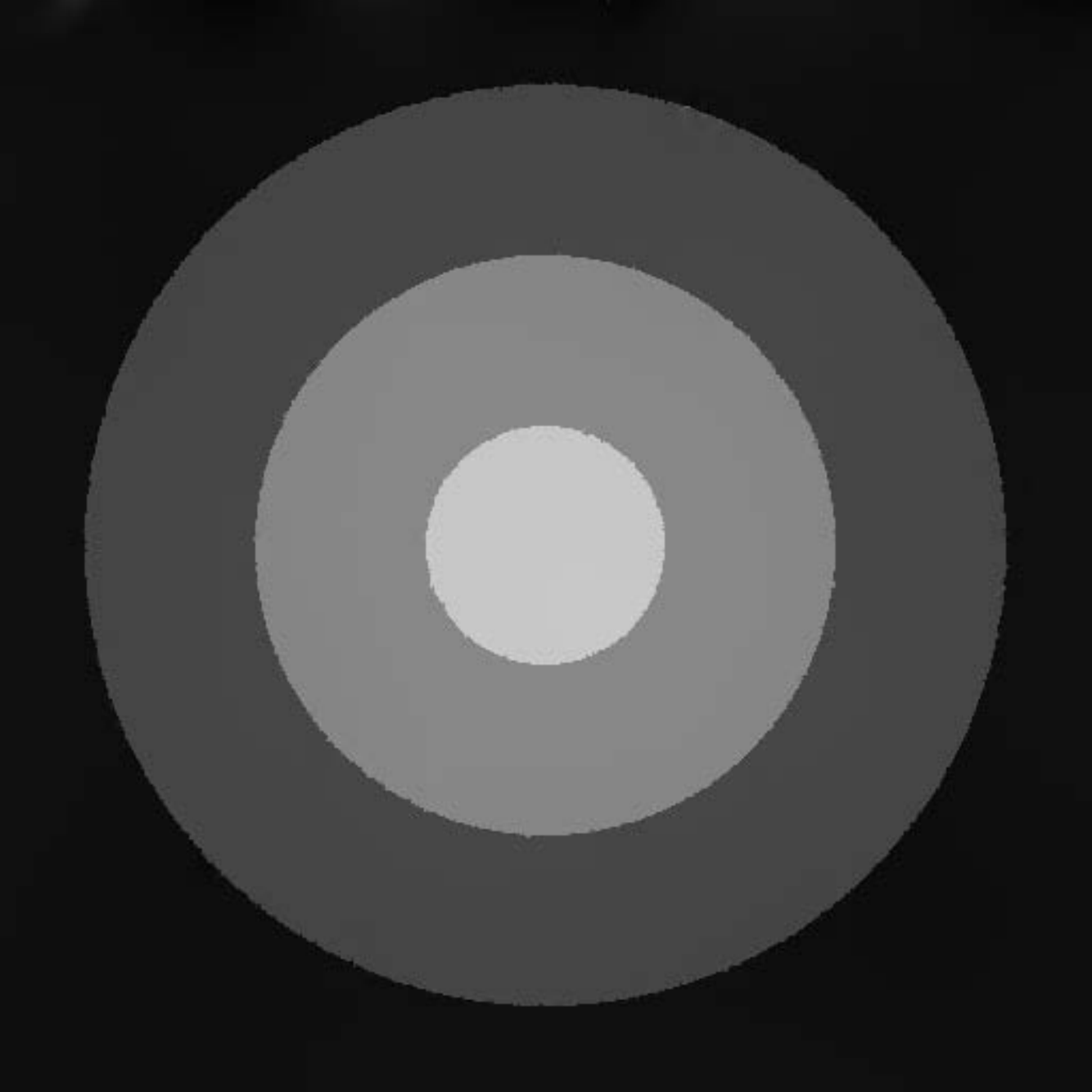}}\hfill
	\caption{(a) Original image. (b) Noisy image. (c) Denoised by LALM with $a = 2, b = 0.02, \lambda = 15.5, r_1 = 100, r_2 = 1, r_3 = 1.2, \gamma = 1e-5, \delta_1 = 0.01, \delta_2 = 3e-4$. (d) Denoised by LALMn with $a = 1,b = 0.01,\lambda = 10, r_1 = 10, r_2 = 5, r_3 = 1.2, \gamma = 1e-5, \delta_1 = 1e-3, \delta_2 = 2e-3$. }
	\label{e1circles} 
\end{figure}

\begin{figure}[!htbp]
\centering
 	\subfigure[LALM]{
		\label{LALM_256th} 
		\includegraphics[height=1.376in]{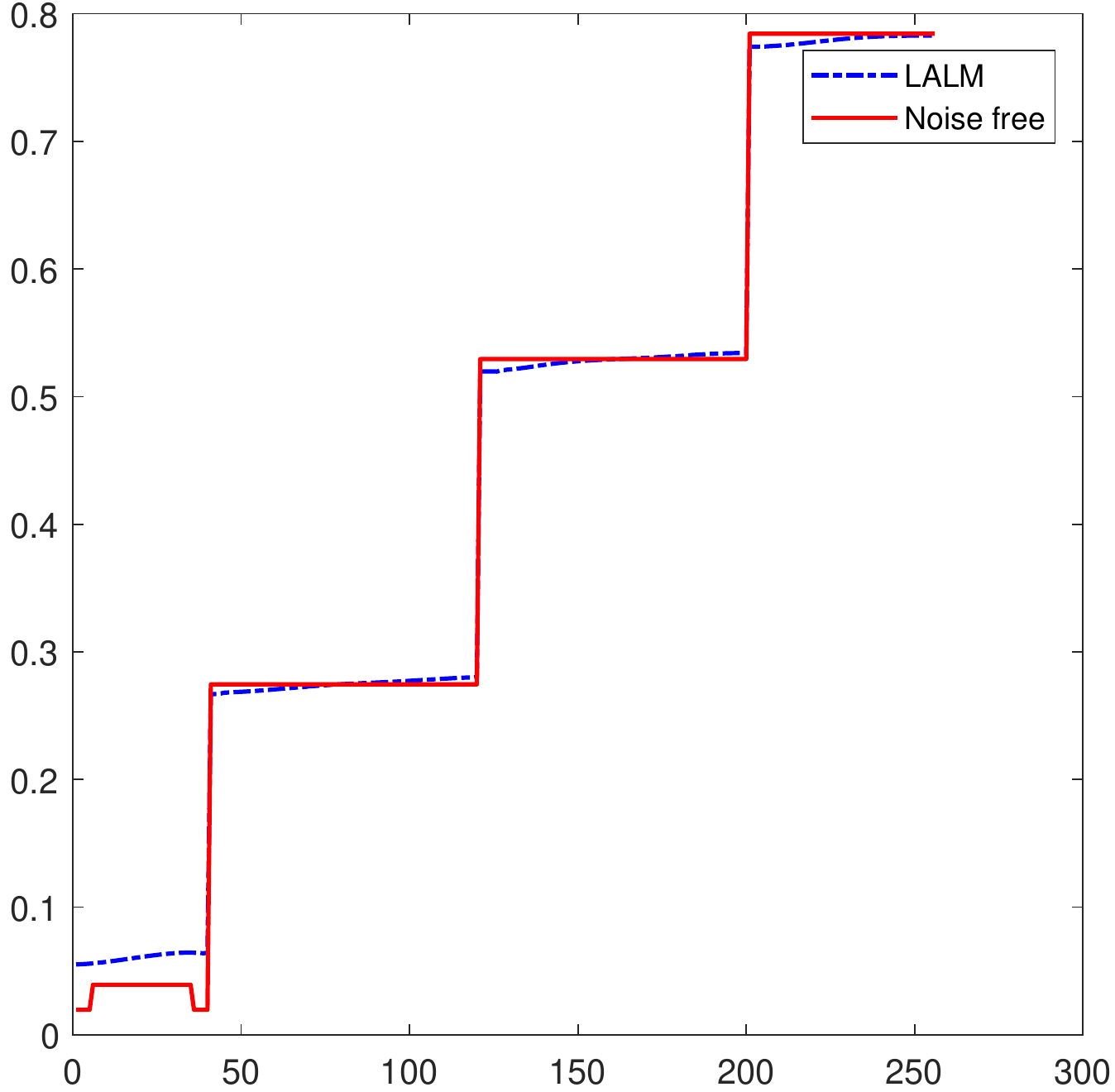}}
			\subfigure[LALMn]{
	    \label{LALMn_256th} 
	    \includegraphics[height=1.376in]{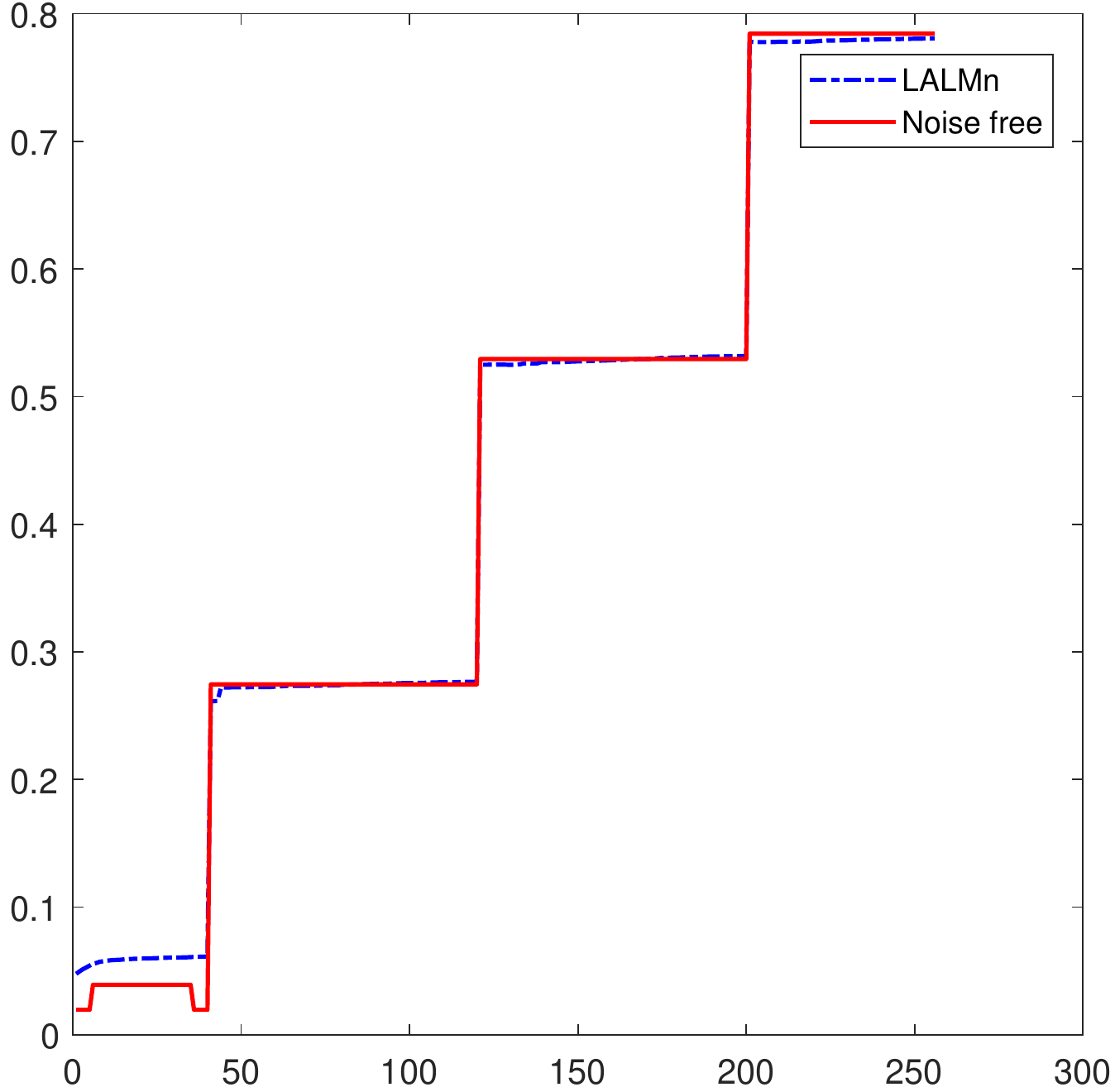}}
	\subfigure[Norm comparasion]{
		\label{normn} 
		\includegraphics[width=1.7in]{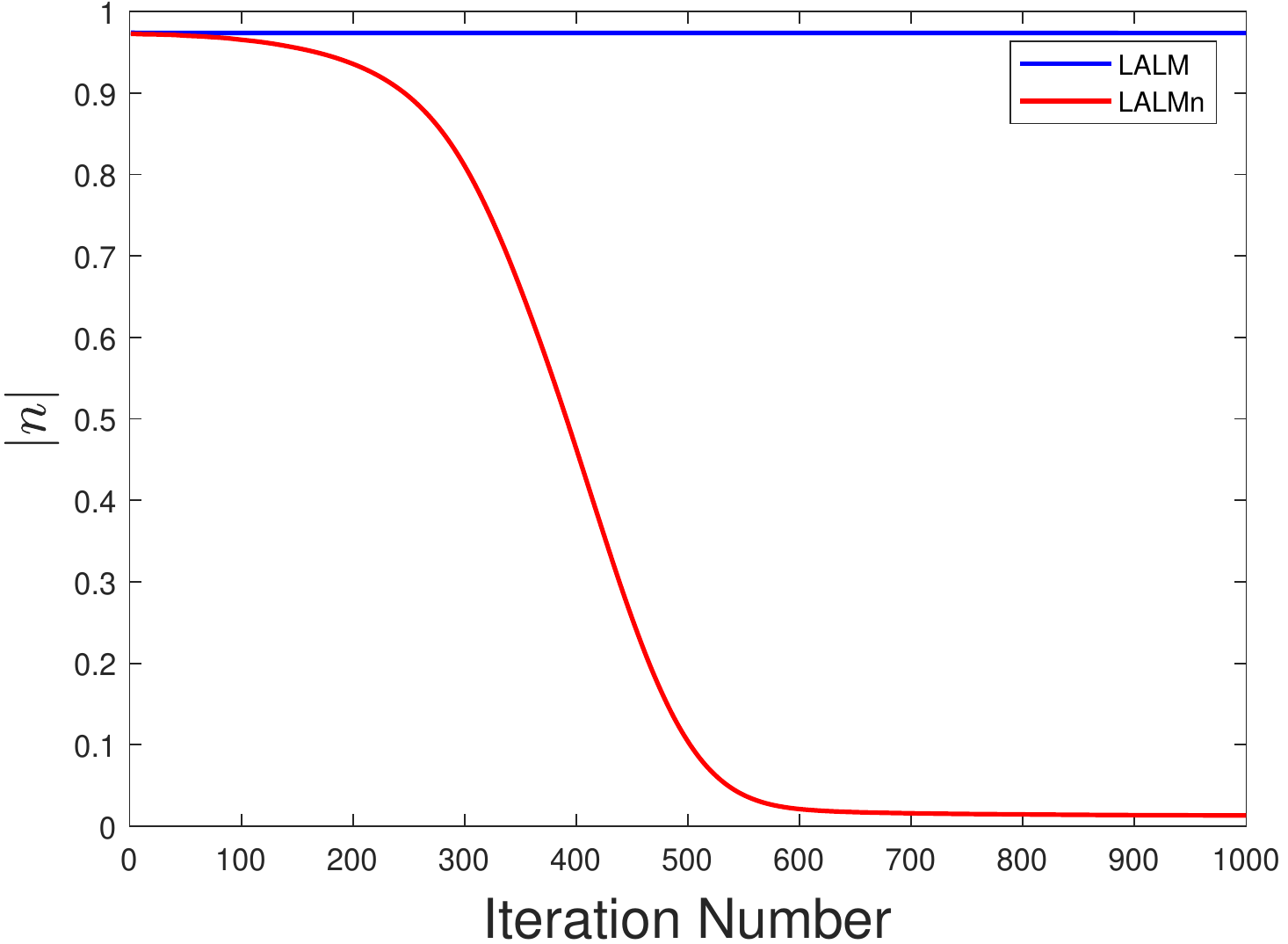}}

	\caption{The line profiles of 256th column of the denoised results, the norm of $\boldsymbol{n}$ for LALM and  LALMn respectively.}
	\label{e1_lines} 
\end{figure}

The two algorithms are tested on a synthesized image consisting of several concentric circle rings, size of $512\times 512$. Fig.\ref{e1circles}(a) and (b) show the synthesized image and its noisy version, while  Fig.\ref{e1circles}(c) and (d) show the restored images by LALM and LALMn, respectively. Please note that the vertical blue line shown in Fig.\ref{e1circles}(a) corresponds to the 256th column of the image, which shall be used for profile plots. From  Fig.\ref{e1circles}(c) and (d), one can see that both algorithms could recover the ideal image well, and little difference could be visually identified.  To examine the results more carefully, in Fig.\ref{LALM_256th}-Fig.\ref{normn}, the line profiles of  256th columns and  the norm of $\boldsymbol{n}$, which is defined by (\ref{norm_n}), are shown for the results of LALM and LALMn, respectively. All the plots are drawn up to 1000 iterations.
From the line profile plots, observable differences, though still quite weak, can be identified,
which shows that the LALMn plot agrees better with the ideal image than the LALM plot does.  We have tried our best to fine tune the parameters of LALM, however, the difference persists and can not be diminished further.

A big difference occurs when we check the plots of the norm  $\|\boldsymbol{n}\|$ produced from the two algorithms. As shown in Fig.\ref{normn},  the norm curve for LALMn converges to somewhere almost 0, while the norm curve for LALM converges to somewhere near 1. 
In fact, without counting the values near the edges, the norm for RALM should have converged to zero. We think that  
  $\|\boldsymbol{n}\| \to 0$ is more reasonable. This can be justified by the following arguments. As noted in \cite{zhu2012image},
instead of regarding an image as a function $f$ defined on a domain $\Omega \subset \mathbb{R}^2$, we can also consider the image $f$ as defining a surface in $\mathbb{R}^3$: $z = f(x,y)$, $(x,y) \in \Omega$. So the normal vector of the image surface takes the form of $ \boldsymbol{\widetilde{n}} = (f_x,f_y, -1)$. For a piecewise constant region of $f$, the surface normal $ \boldsymbol{\widetilde{n}}$ should be $(0,0,-1)$. In this regard, for the surface defined by the synthesized image, we should have $\boldsymbol{\widetilde{n}} = (0,0, -1)$ almost everywhere, except near the edges. Since $\boldsymbol{n}$ represents $\frac{(f_x,f_y)}{\sqrt{f_x^2+f_y^2+\epsilon}}$, it's norm should be $0$ almost everywhere.

\subsection{Inconsistency test for the conventional augmented Lagrangian formulation}
As stated in the introduction section, there exists an internal inconsistency in the augmented Lagrangian formulation for the Euler's elastica model. When the weighting parameter $b\to 0$, the Euler's elastica model boils down to the ROF model. So, a natural requirement is that its augmented Lagrangian formulation should correspondingly boil down to the augmented Lagrangian formulation for the ROF model. This natural requirement, however, is not fulfilled by the conventional formulation, which we think is an internal inconsistency of the formulation. Since LALM is faithful to the conventional formulation, this inconsistency lives inside it. On the other hand, by introducing a cutting-off strategy, the proposed method RALM has successfully removed such an inconsistency. This can be easily seen from the pseudo-code  described in \textbf{Algorithm \ref{alg:RALM}}.

\begin{figure}[!htb]
	\centering
	\subfigure[Numerical energy, LALM]{
		\label{e2_energy1} 
		\includegraphics[width=1.69in]{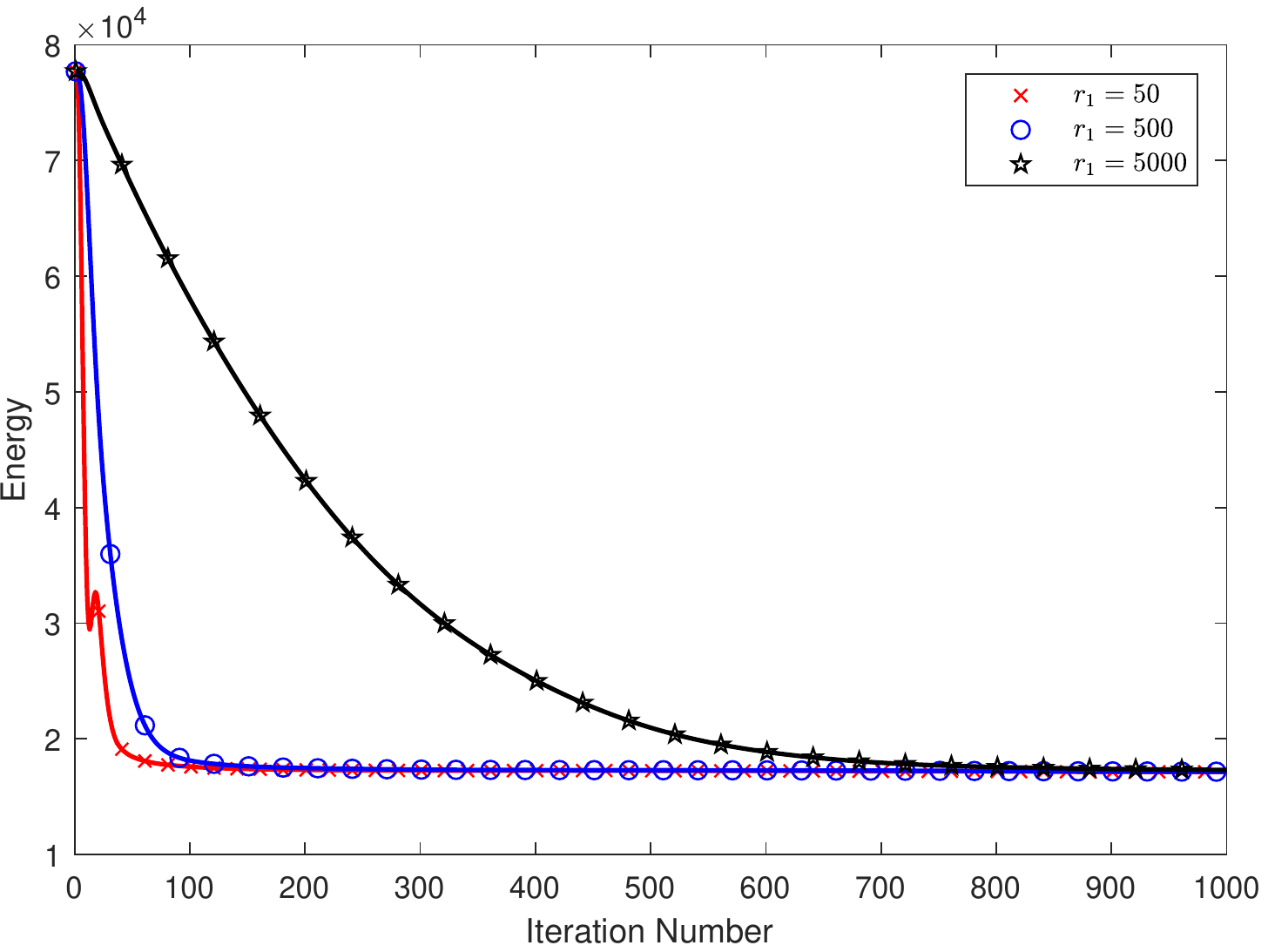}}\hfill
	\subfigure[Numerical energy, RALM]{
     	       \label{e2_energy2} 
	       \includegraphics[width=1.69in]{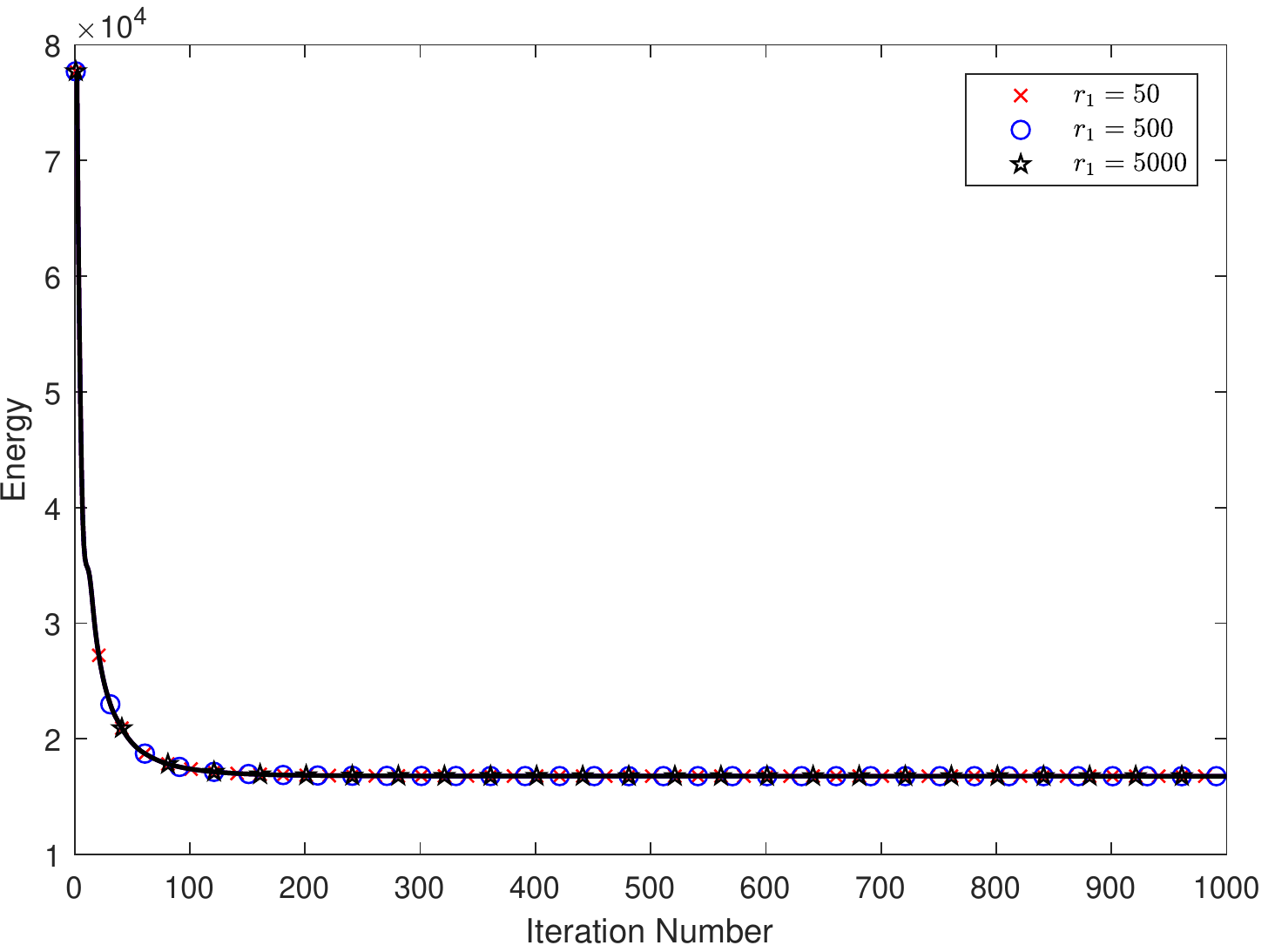}}\hfill
	\subfigure[PSNR, LALM]{
		\label{e2_psnr1} 
		\includegraphics[width=1.69in]{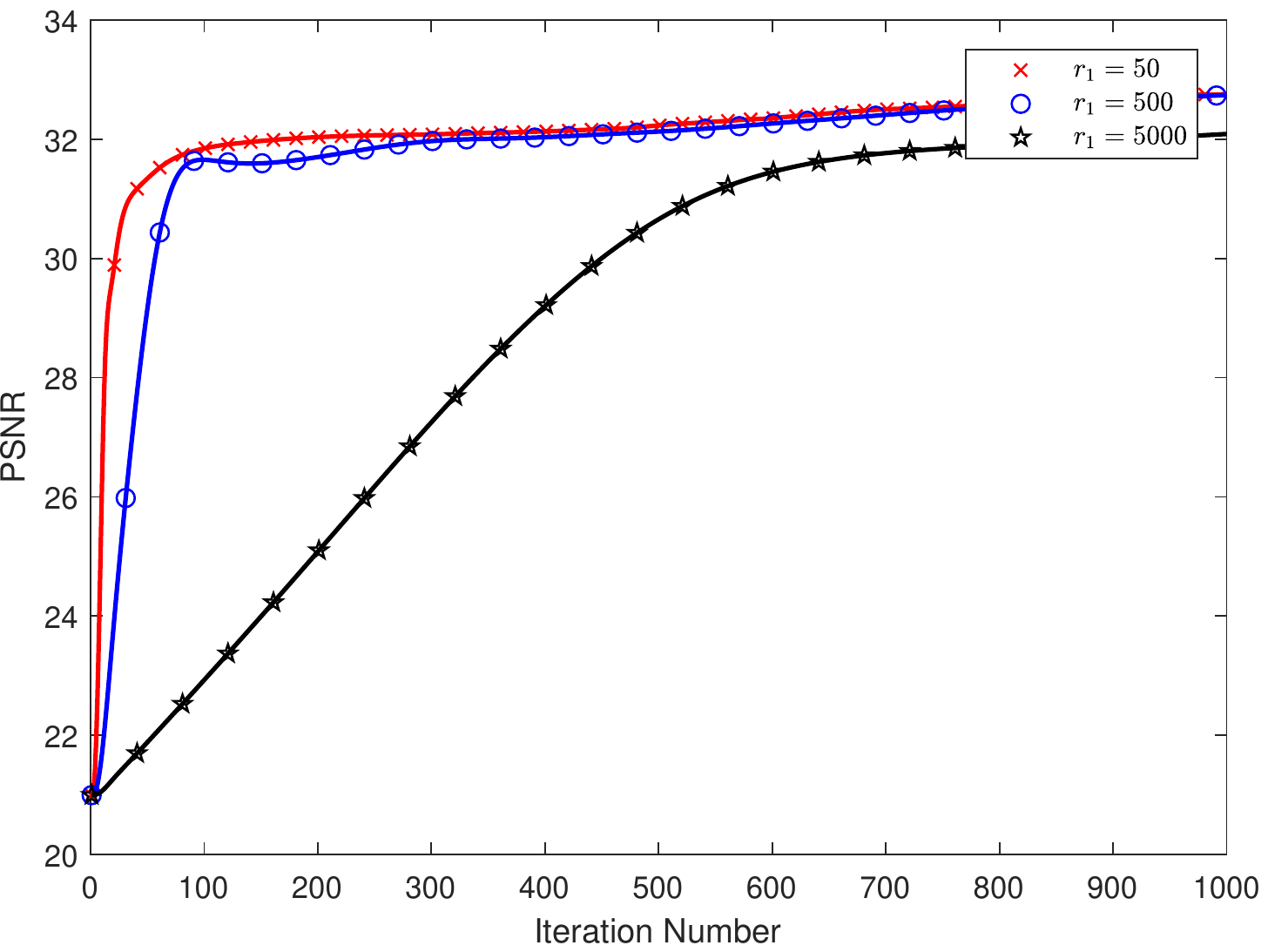}}\hfill
	\subfigure[PSNR, RALM]{
		\label{e2_psnr2} 
		\includegraphics[width=1.69in]{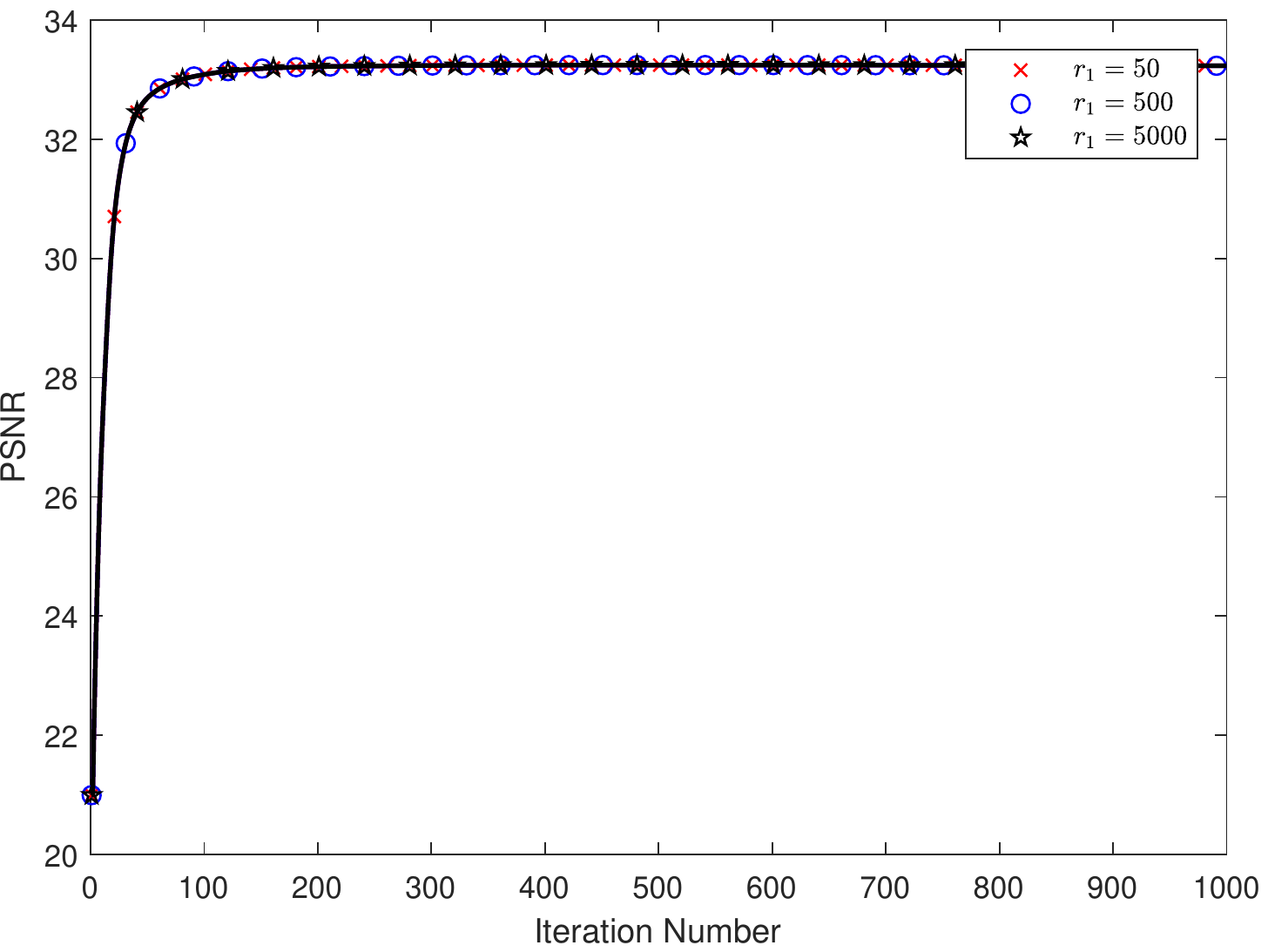}}\hfill
      \subfigure[Relative error, LALM]{
		\label{e2_error_u1} 
		\includegraphics[width=1.69in]{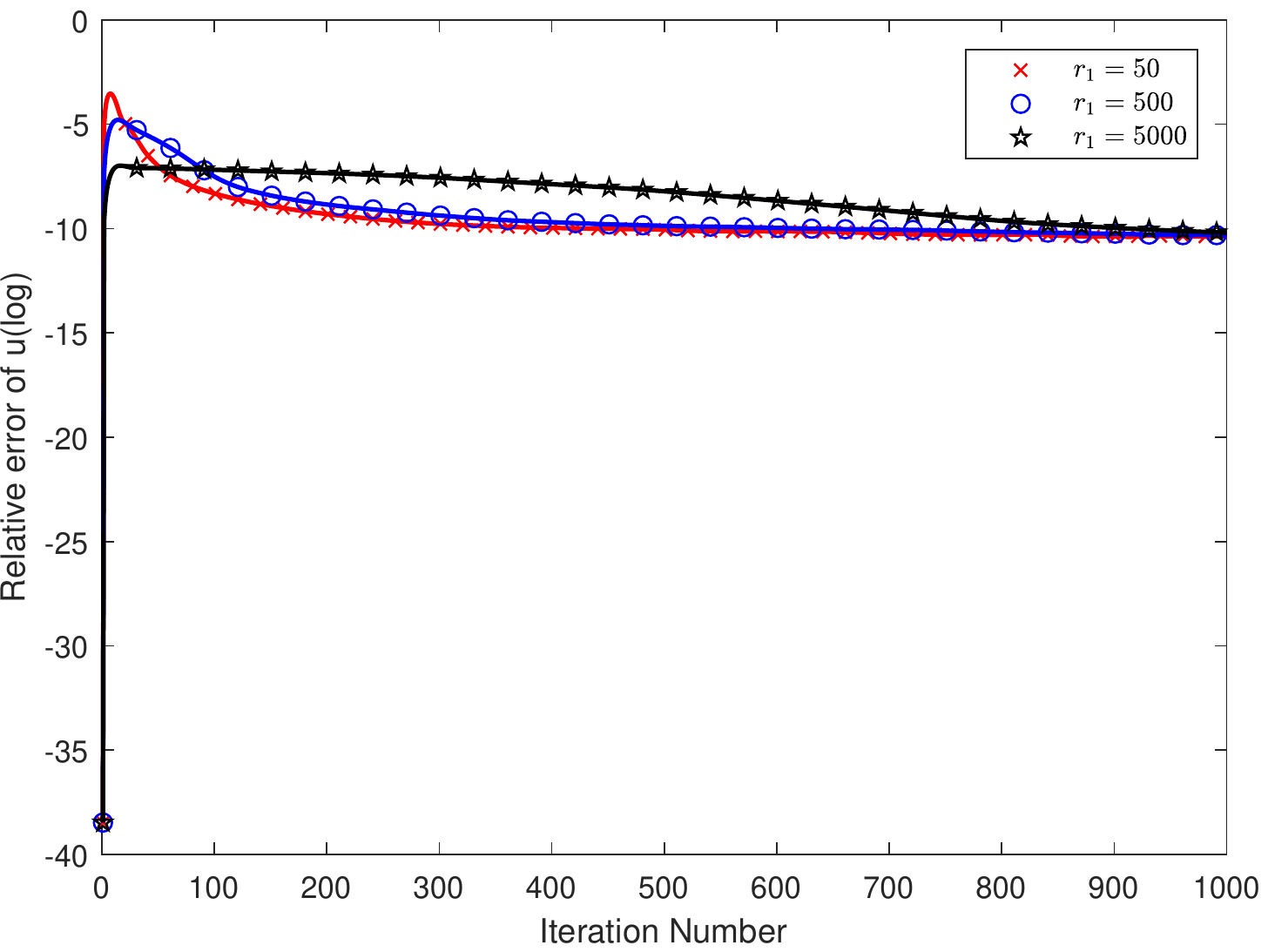}}\hfill
	\subfigure[Relative error, RALM]{
		\label{e2_error_u2} 
		\includegraphics[width=1.69in]{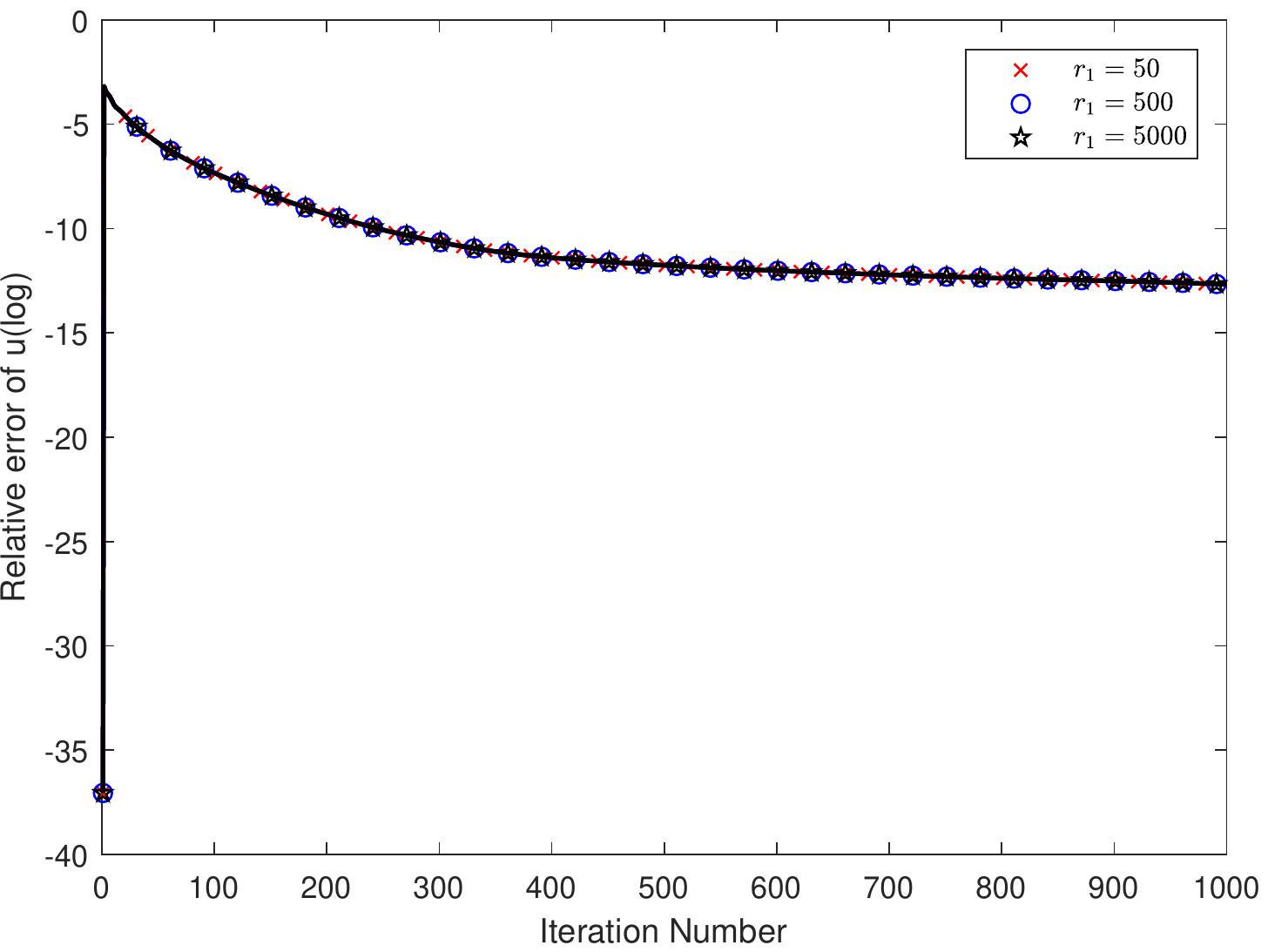}}\hfill
	\caption{Plots of numerical energy, PSNR and relative errors. In each row, the left one is from LALM and the right one is from RALM.}
	\label{e2_LALM} 
\end{figure}

In this subsection, we will test how the inconsistency will influence the convergence behavior of LALM as well as the quality of the restored images.  The synthesized image shown in Fig.\ref{originalc} is utilized again to test both LALM and RALM. With $b=0$, the Euler's elastica  model reduces to  
\begin{equation*}
\label{d_rof}
\min\limits_{u}\int_\mathrm{\Omega} a |\nabla u| d\mathrm{\Omega}+
{\frac{\lambda}{2}}\int_\mathrm{\Omega} (u-f)^2d\mathrm{\Omega},
\end{equation*} 
and the augmented Lagrangian functional associated with this optimization problem is 
\begin{equation}
\begin{aligned}
\label{dl_rof}
\mathcal{L}(u,\boldsymbol{p};\boldsymbol{\lambda}_{2})= &
\int_{\mathrm{\mathrm{\Omega}}}a|\boldsymbol{p}| + {\frac{\lambda}{2}}\int_{\mathrm{\Omega}} (u-f)^2
+\int_{\mathrm{\Omega}}\boldsymbol{\lambda}_{2}\cdot(\boldsymbol{p}-\nabla u)\\
&+\frac{r_2}{2}\int_{\mathrm{\Omega}}|\boldsymbol{p}-\nabla u|^2d\mathrm{\Omega},
\end{aligned}
\end{equation} 
which has been discussed in \cite{wu2010augmented}. 
Since only the parameters $a, \lambda, \boldsymbol{\lambda}_{2}, r_2, \delta_1$ are involved in (\ref{dl_rof}),  it's expected that the other left parameters $r_1, r_3, \delta_2$ should not influence the behavior of LALM and RALM.  
Since the connection between $\{u,p\}$ and the left auxiliary variables are established by the parameters $r_1$, we would like to test the methods with varied $r_1$.
  We monitor three quantitive indices: numerical energy, PSNR and relative errors, to see how they change with $r_1= 50, 500 \mbox{ and } 5000$, respectively. The results are shown in Fig.\ref{e2_LALM}.  We observe that the numerical energy, PSNR and relative errors of LALM are all influenced by the values of $r_1$, while our algorithm RALM does not. From Fig.\ref{e2_energy1} and Fig.\ref{e2_energy2}, one may conclude that even though the convergence behavior changes with varied $r_1$, it seems that both RALM and LALM converge to the same point. However, the result shown in Fig.\ref{e2_psnr1} strongly suggests that this should be just an illusion, and the final results of LALM will depend on $r_1$, at least to some extend. 
  
\subsection {Performance tests on real images}

\begin{figure}[h]
   \centering
\subfigure{
\includegraphics[width=1.113in]{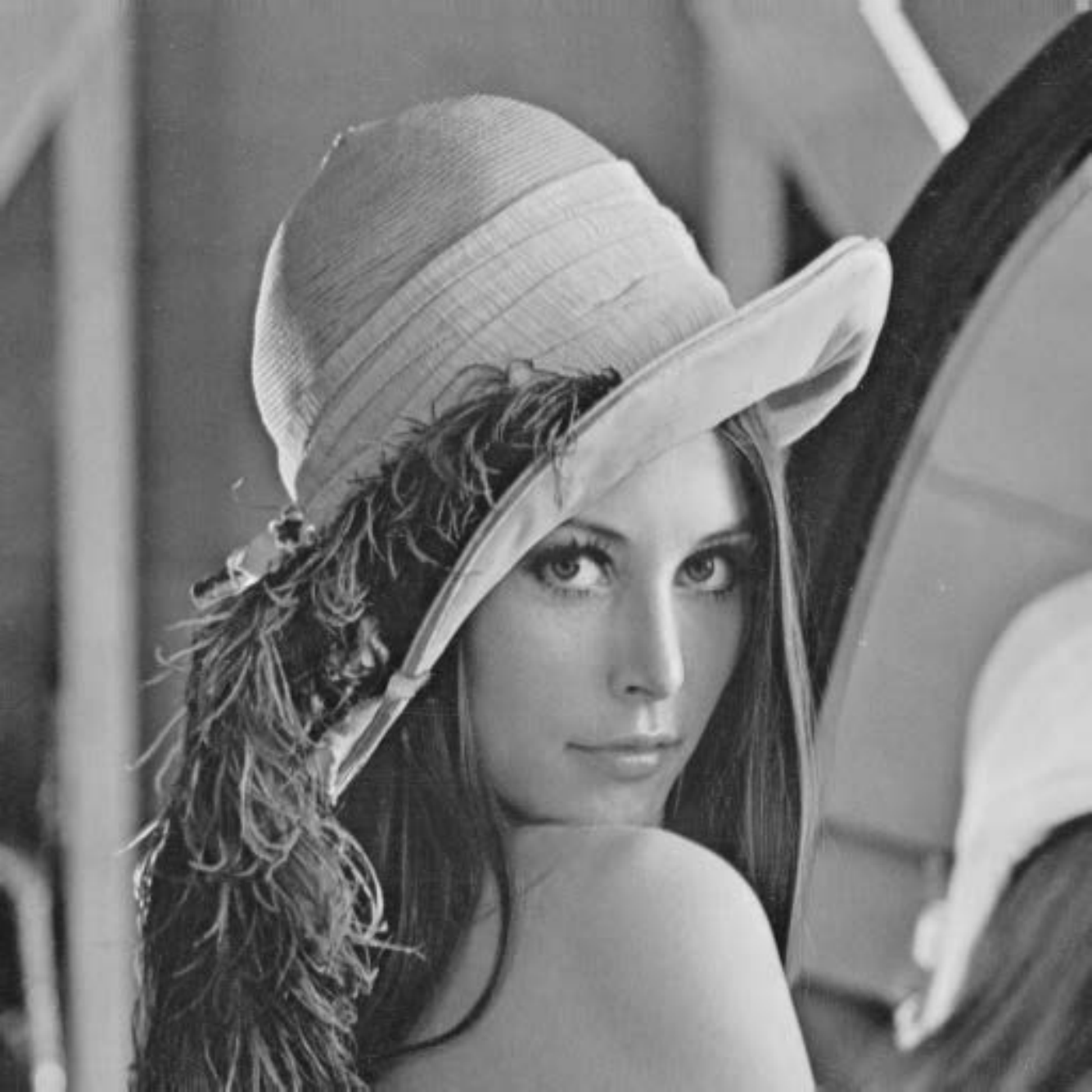}}\hfill
\subfigure{
\includegraphics[width=1.113in]{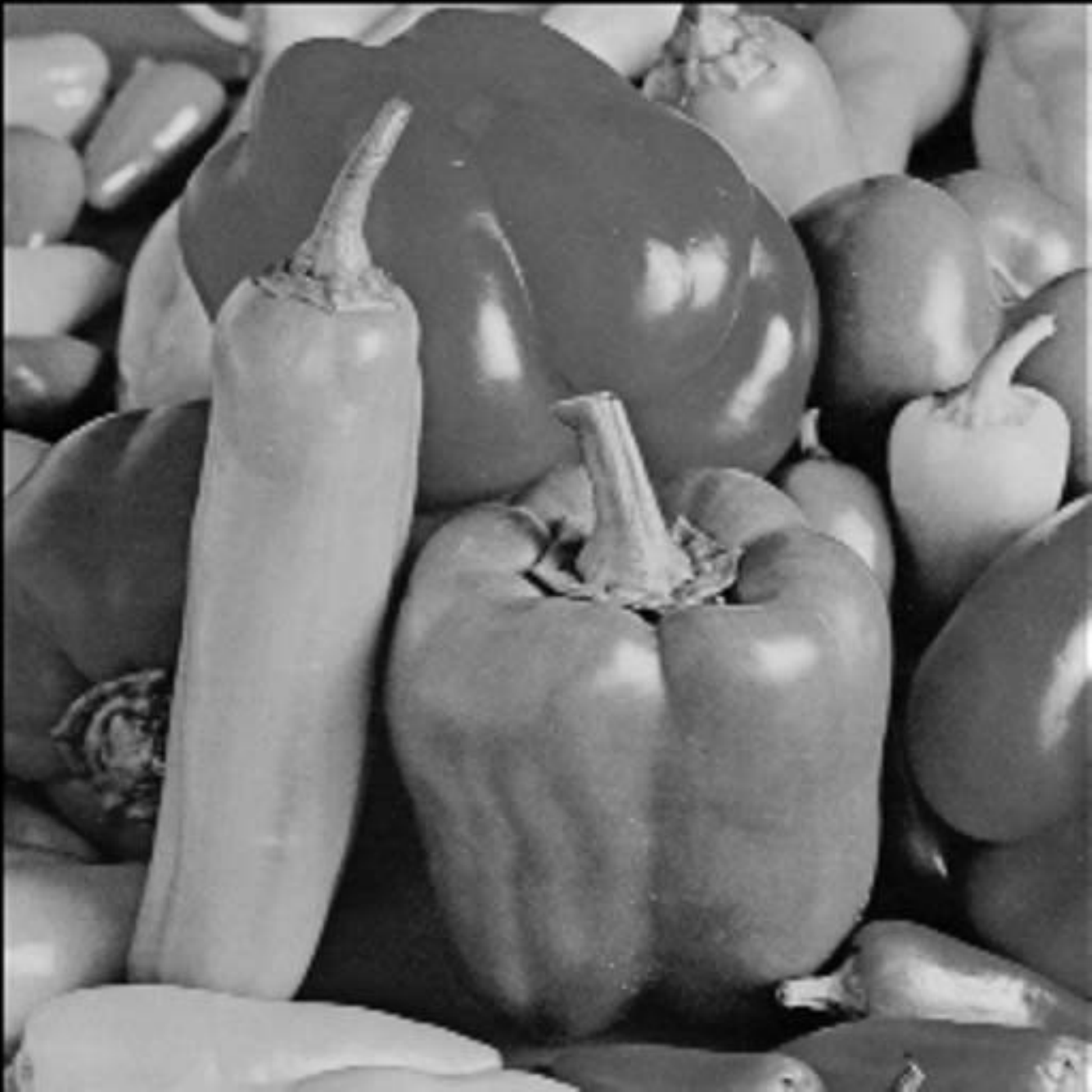}}\hfill
\subfigure{
\includegraphics[width=1.113in]{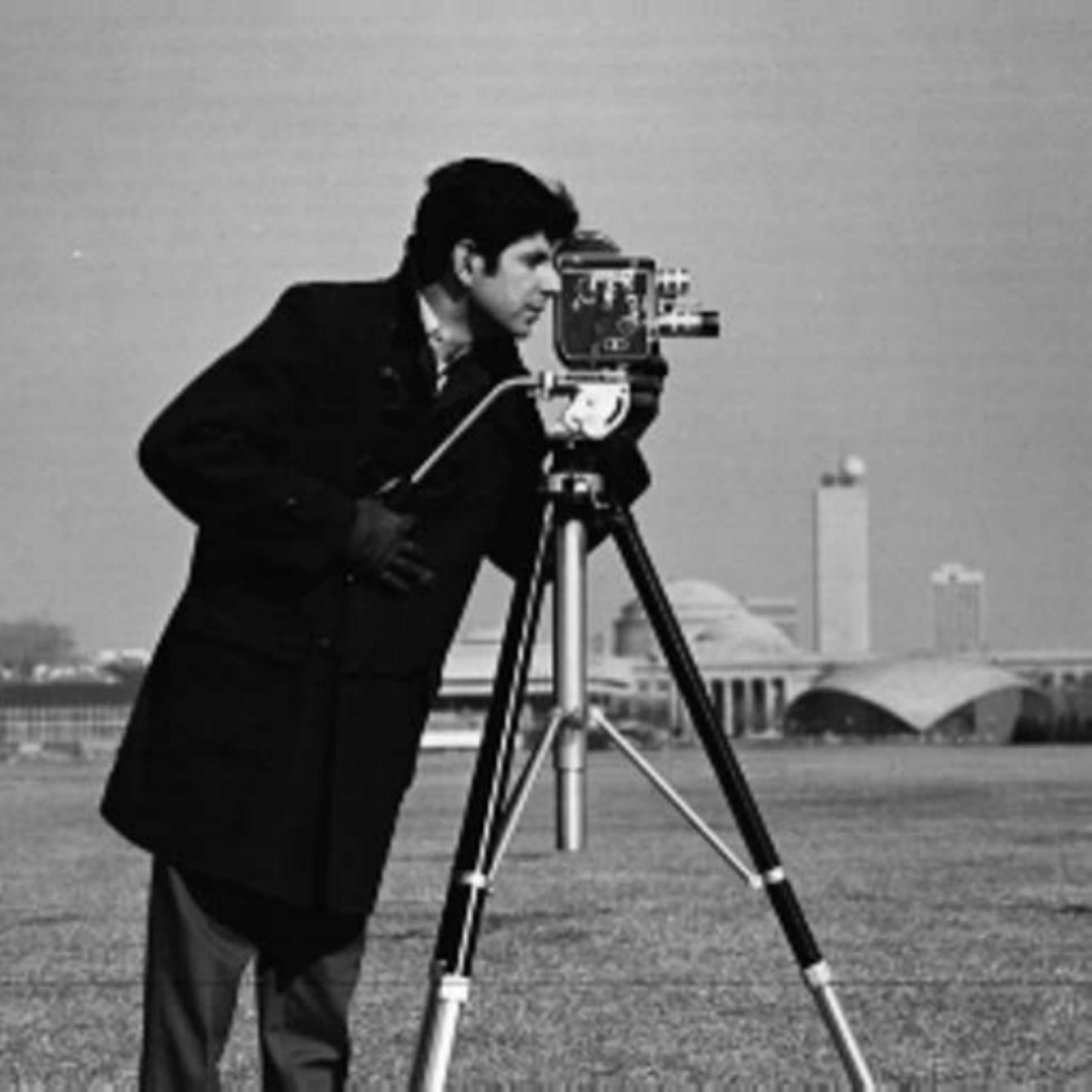}}\hfill
\subfigure{
\includegraphics[width=1.113in]{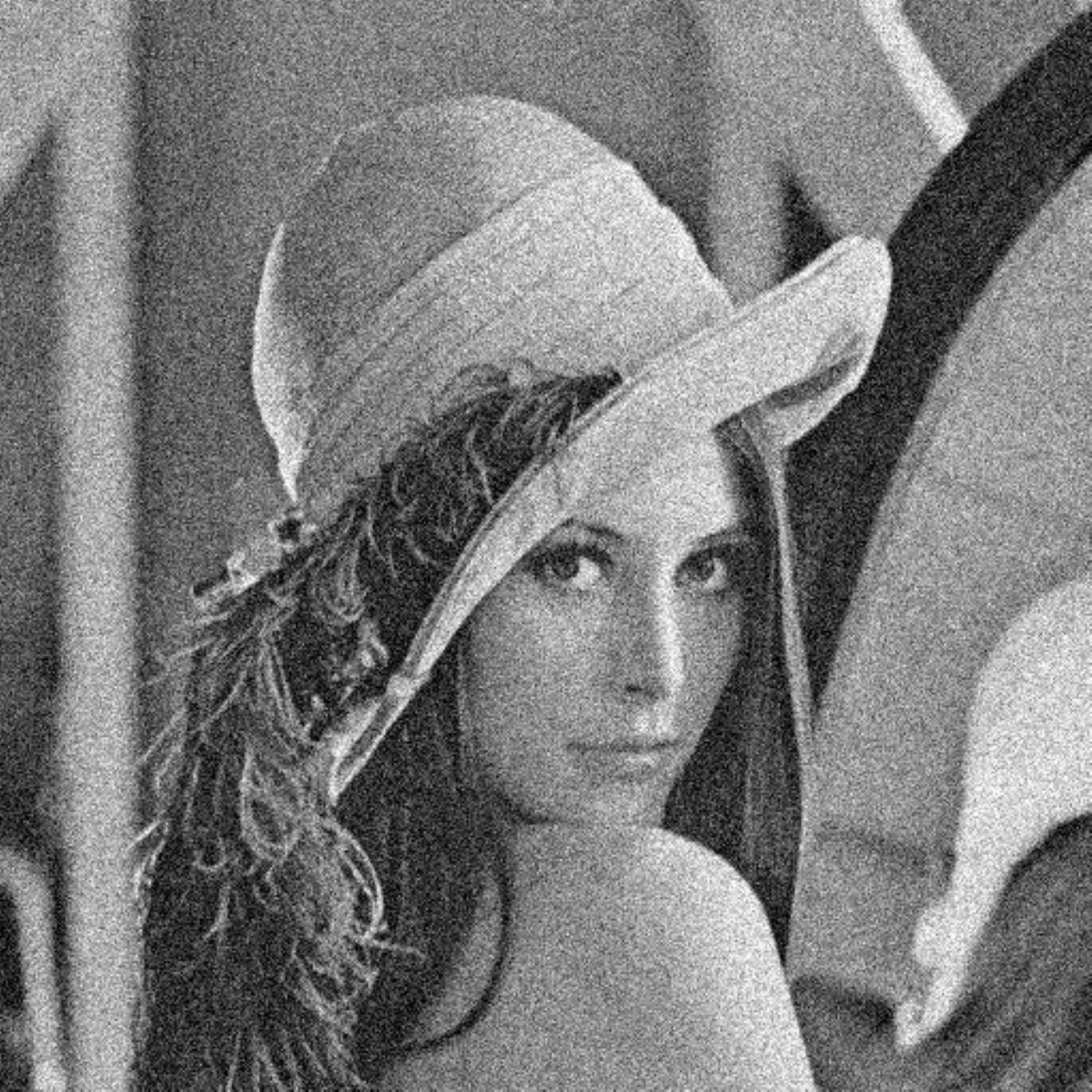}}\hfill
\subfigure{
\includegraphics[width=1.113in]{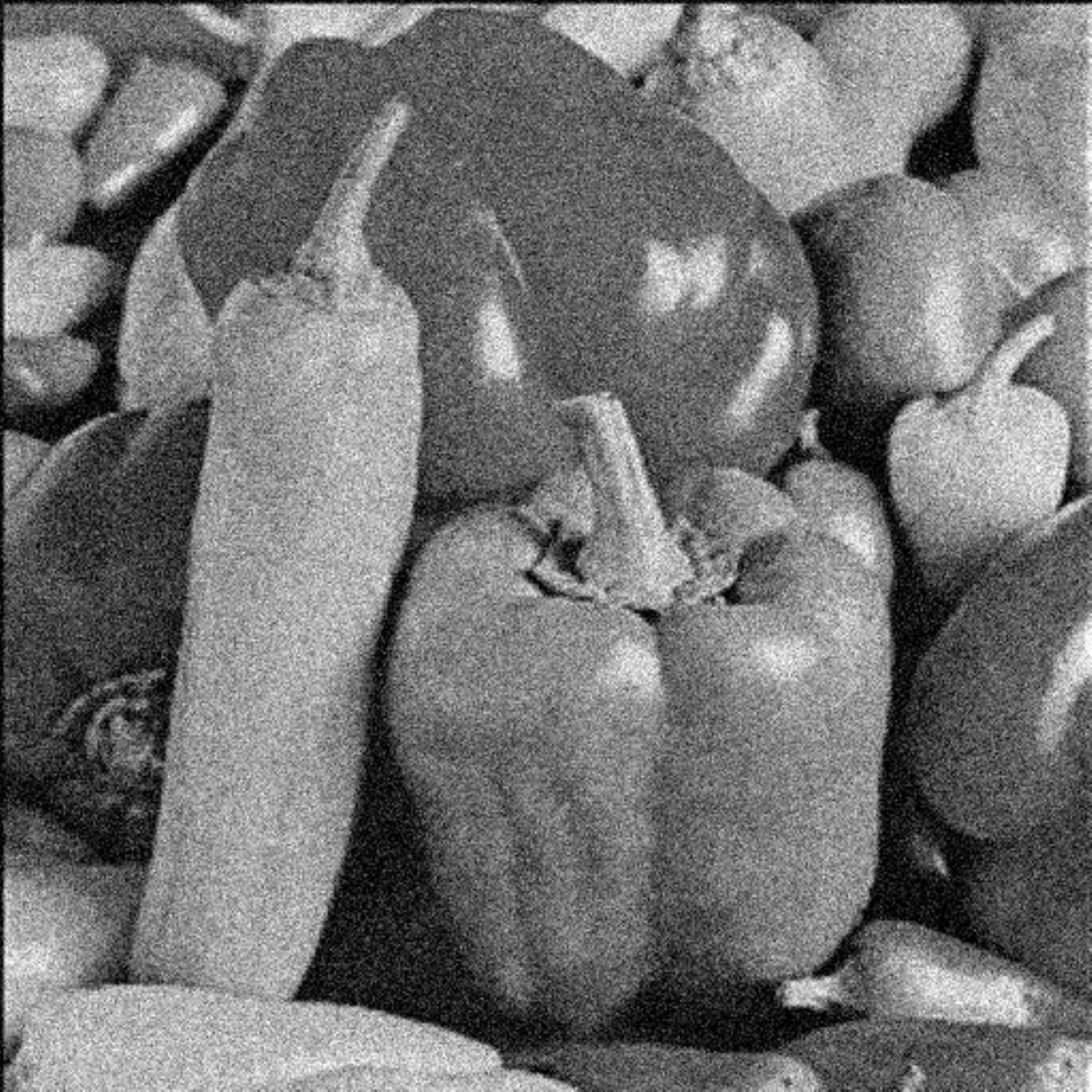}}\hfill
\subfigure{
\includegraphics[width=1.113in]{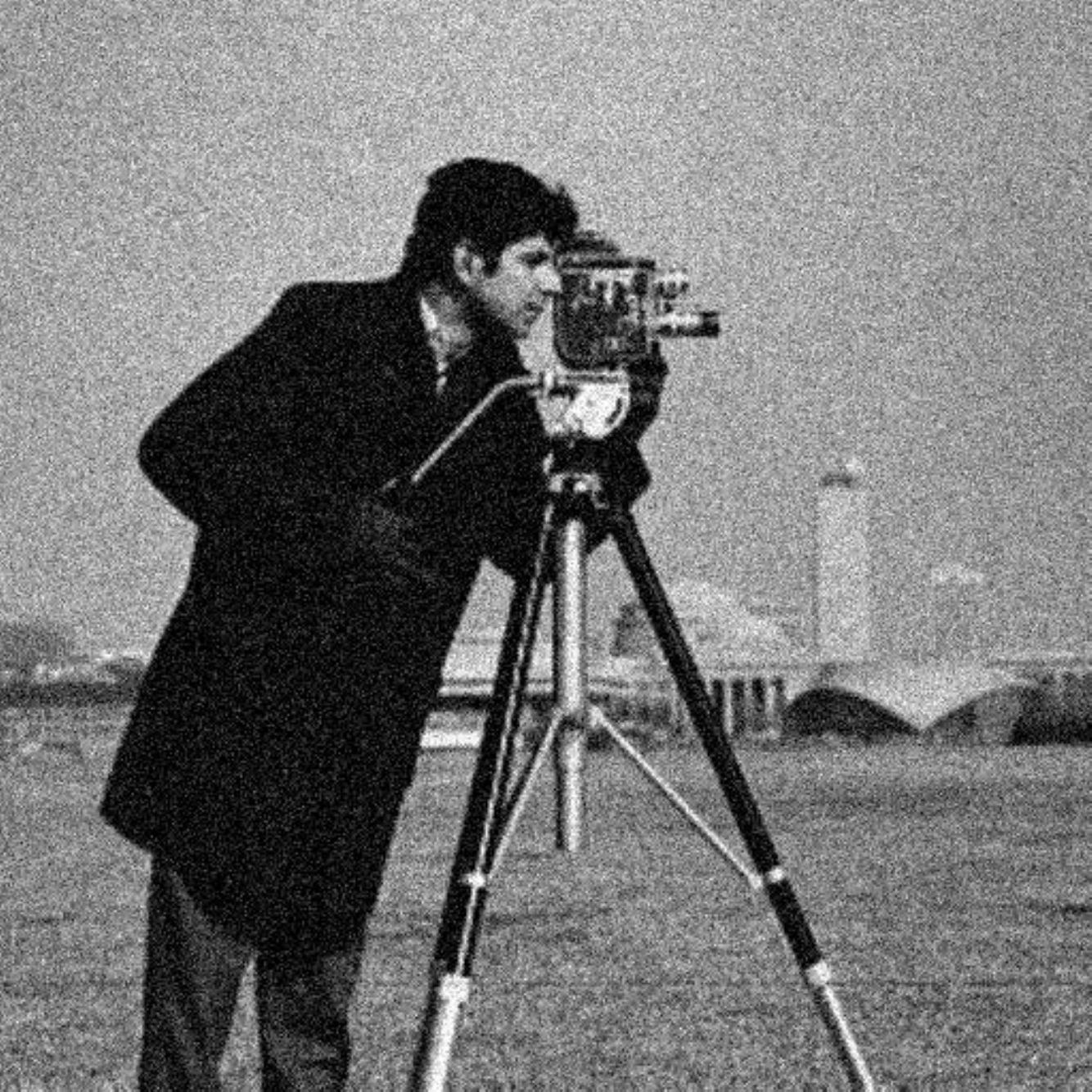}}\hfill
\caption{Test images. In each row, they are images lena, peppers and cameraman  from left to right, respectively. The upper and lower one presents the noise-free and noisy version of the corresponding image, respectively.}
 \label{exp4test-images} 
\end{figure}

In this subsection, we will test how the proposed RALM performs on real images against  LALM.
The clean images and their noisy counterparts are shown in Fig.\ref{exp4test-images}. The parameters used by the two competing methods, which are listed in Table \ref{parameters_iter}, have subjected to fine-tuning to achieve optimal performance. One can easily check that  the parameters of RALM are almost immune to the changing of images, and only slight changes are needed. As comparison the parameters of LALM  demonstrate strong image dependence, and significant adjustments are usually needed from image to image. 

 \begin{table*}[!htb]
	\centering
	\caption{The parameters used by LALM and RALM}
	\label{parameters_iter}
	\begin{tabular}{cccc cccc cccc }
		\hline\noalign{\smallskip}
		Image &Methods &$a$ &$b$ &$\lambda$ &$r_1$&$r_2$&$r_3$&$\gamma$&$\delta_1$&$\delta_2$&$tol$\\
		\noalign{\smallskip}\hline\noalign{\smallskip}
		lena              &LALM &1&0.01&15&200&5&1.2&1e-5&1e-2&2e-2&9e-5\\
		                     &RALM &1& 0.01&12.5&50&1&2&1e-5&5e-2&1e-2&9e-5\\
		peppers       &LALM &2&0.01&32&200&2&1.5&1e-5&1e-2&1e-2&2e-4\\
		                     &RALM &1& 0.01&13&50&1&2&1e-5&5e-2&1e-2&2e-4\\
		cameraman &LALM &1&0.01&15&150&1&1.5&1e-5&5e-2&3e-3&5e-5\\
		                     &RALM &1& 0.01&11.6&50&1&2&1e-5&5e-2&1e-2&5e-5\\
		\noalign{\smallskip}\hline
	\end{tabular}
\end{table*}

\begin{figure*}[!htb]
   \centering
\begin{tabular}{p{0.03cm}<{\centering}p{3.2cm}<{\centering}p{3.2cm}<{\centering}p{3.2cm}<{\centering}p{3.2cm}<{\centering}}
&LALM&RALM& LALM Contour&RALM Contour\\
 \begin{sideways}\hspace{1.2cm}Lena\end{sideways}&
\includegraphics[width=3.5cm]{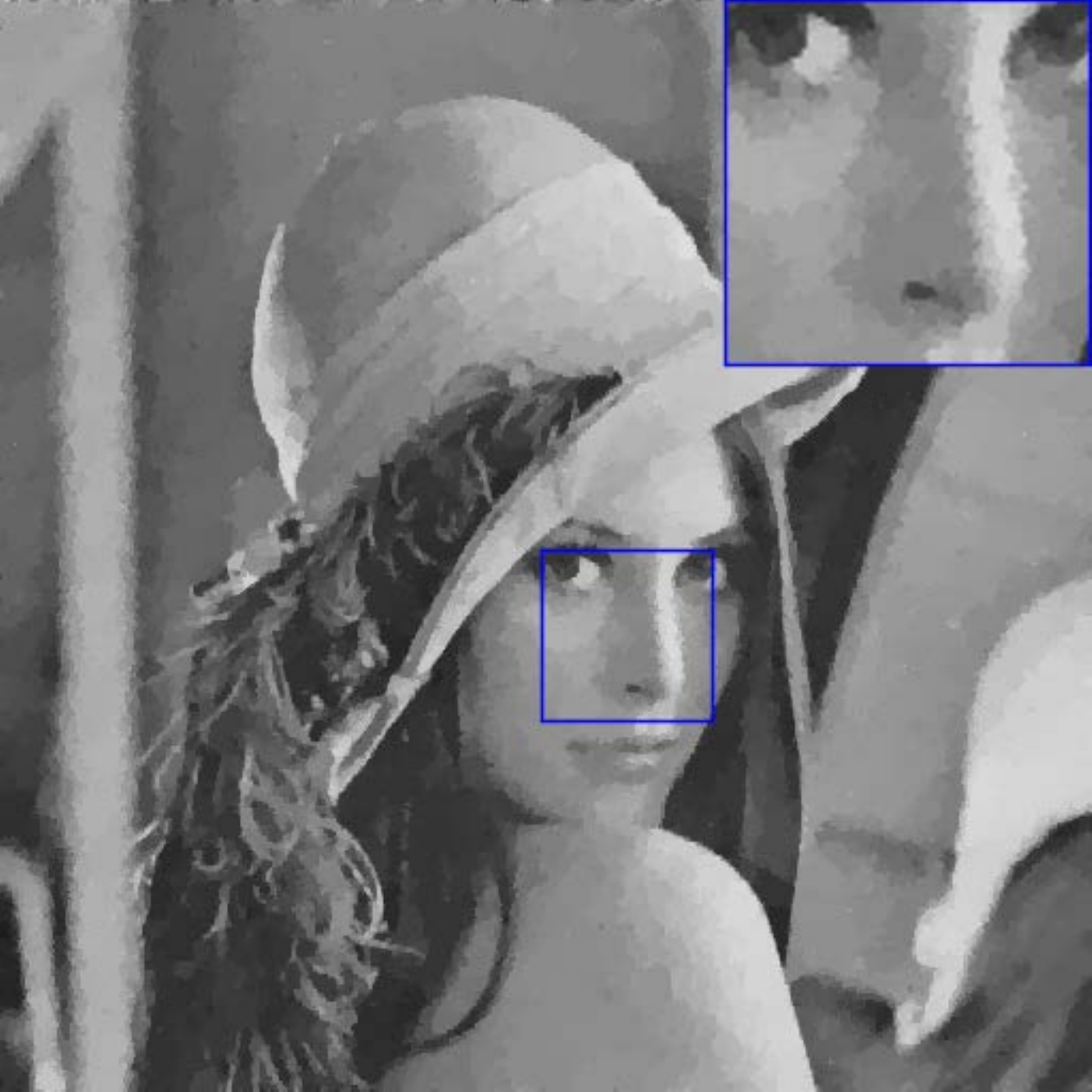}&
\includegraphics[width=3.5cm]{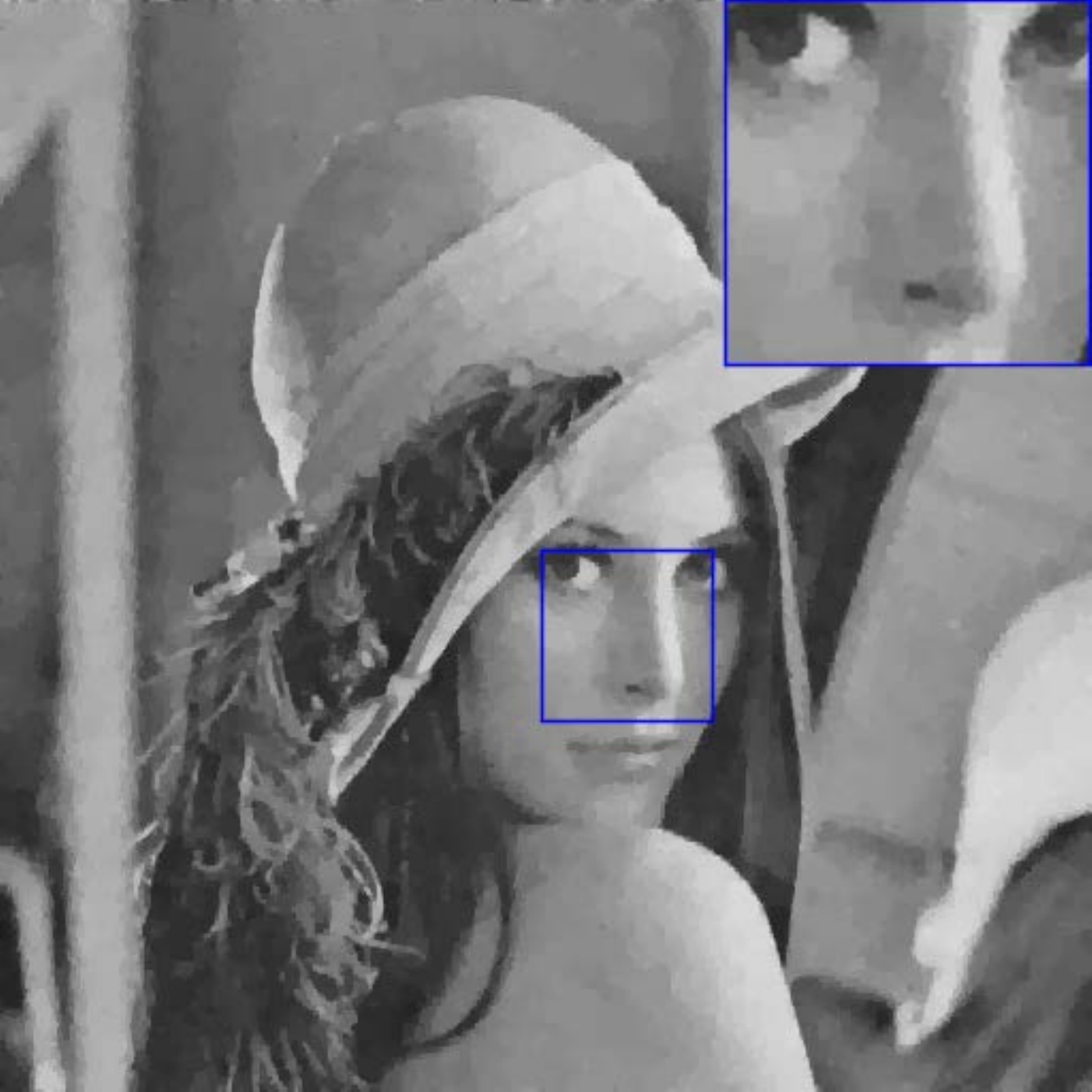}&
\includegraphics[width=3.5cm]{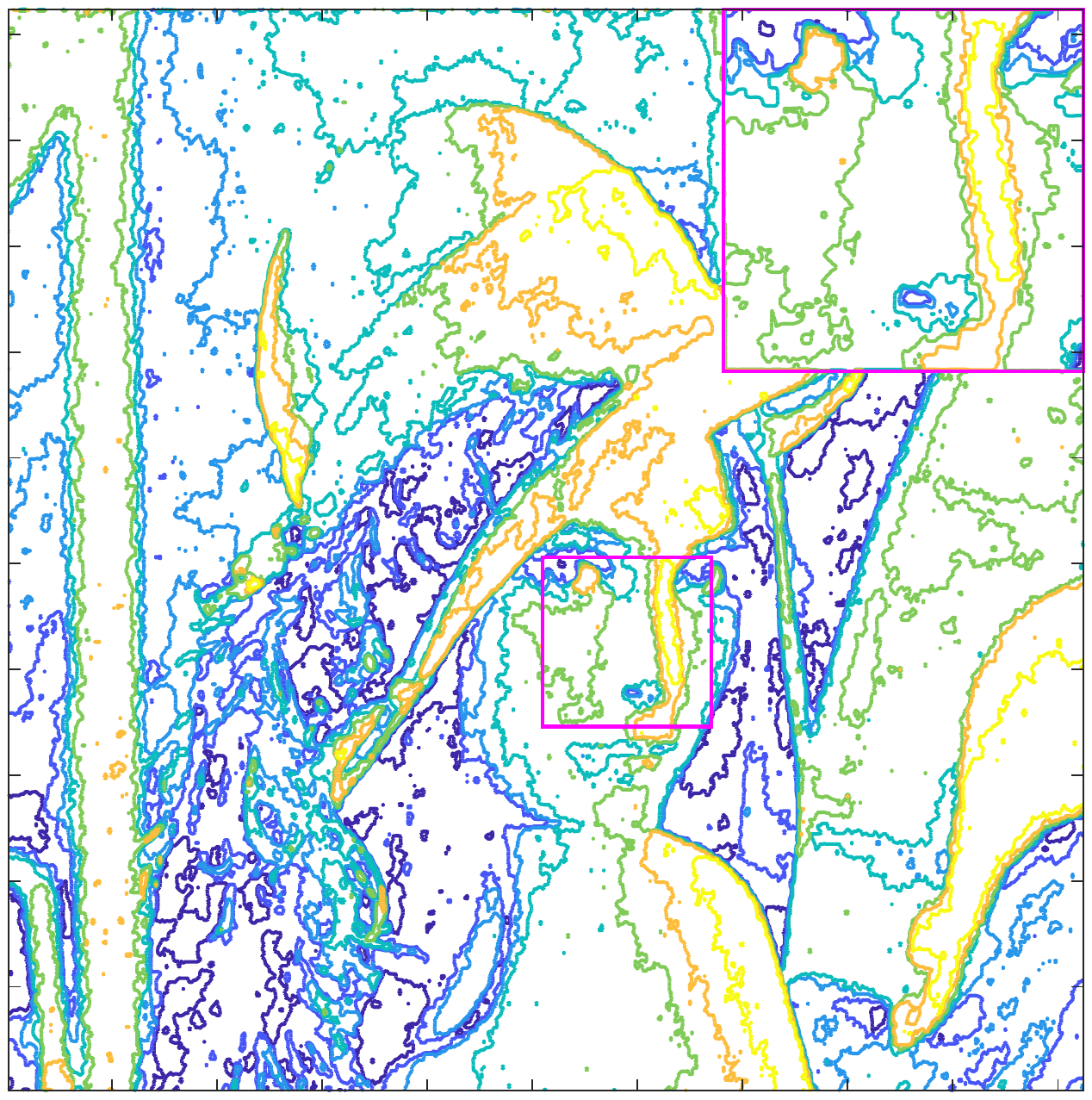}&
\includegraphics[width=3.5cm]{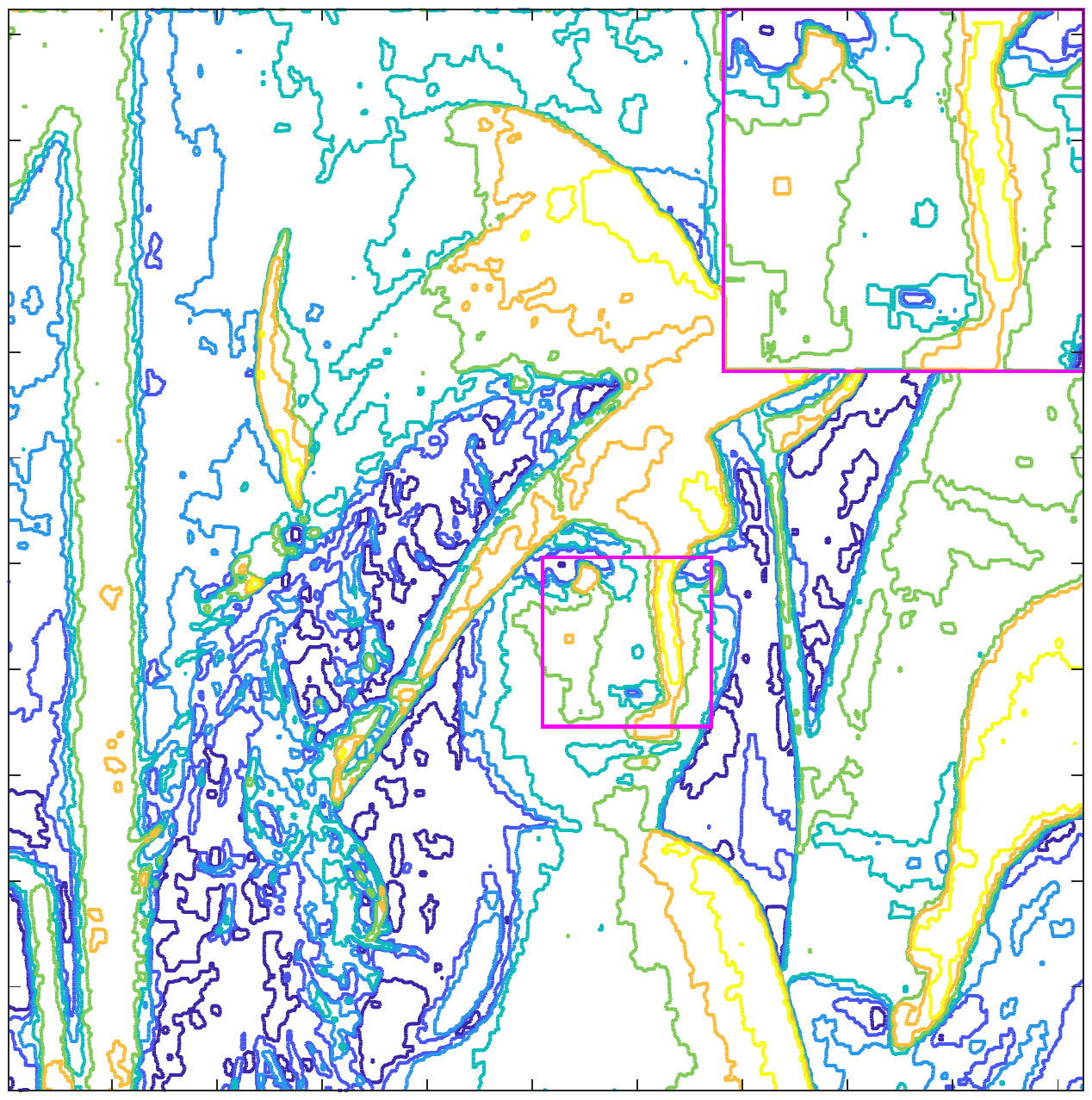}\\
\begin{sideways}\hspace{0.9cm}Peppers\end{sideways}&
\includegraphics[width=3.5cm]{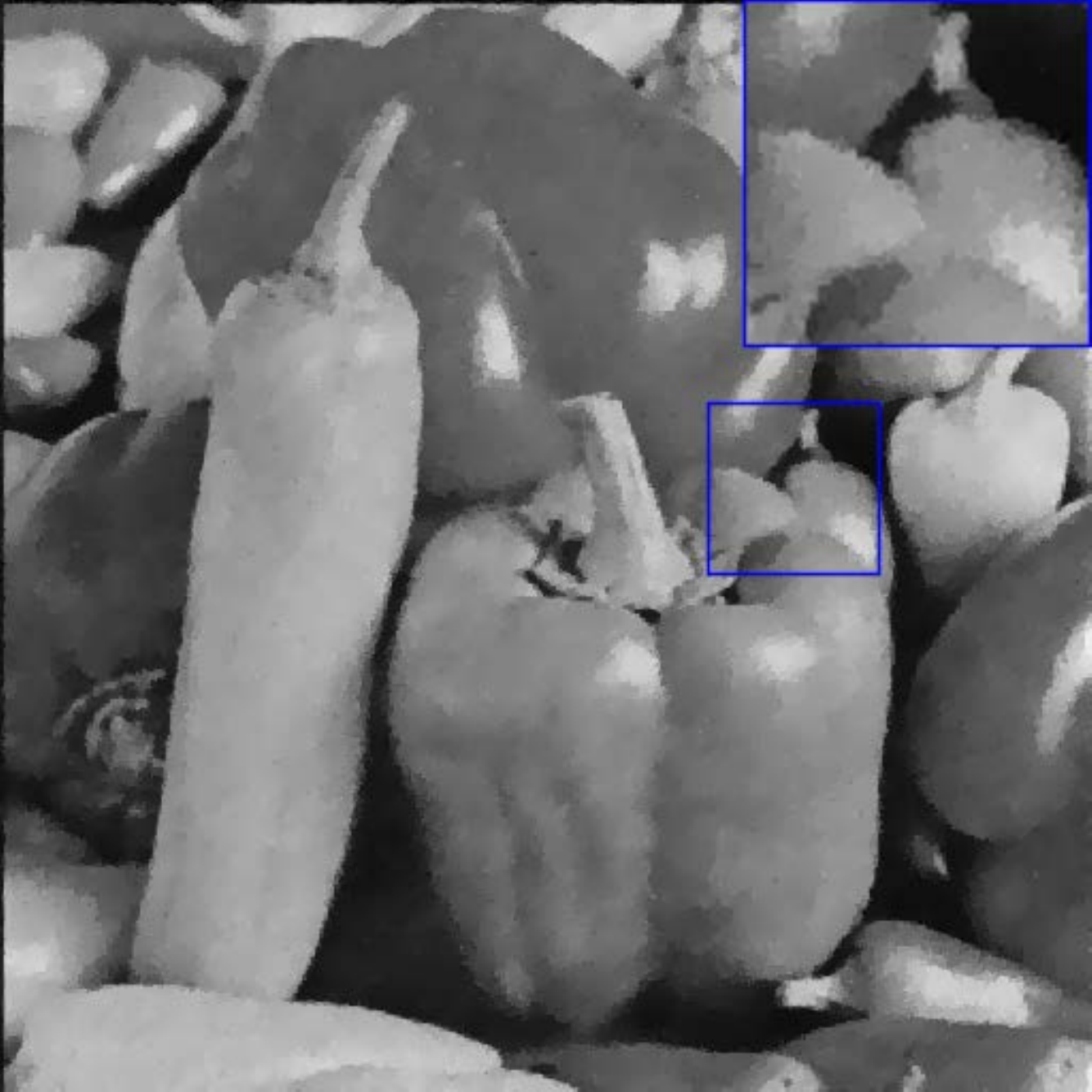}& 
\includegraphics[width=3.5cm]{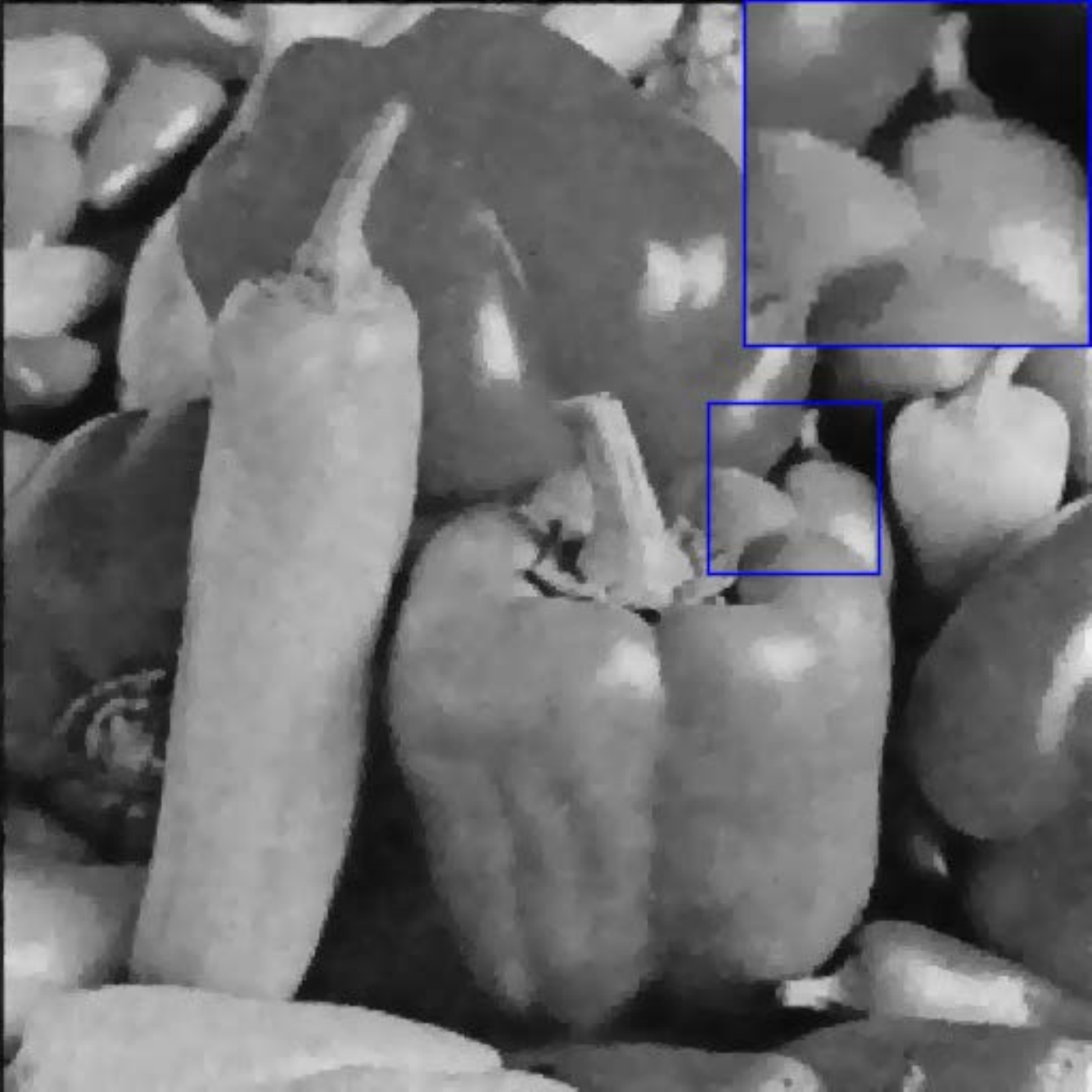}&
\includegraphics[width=3.5cm]{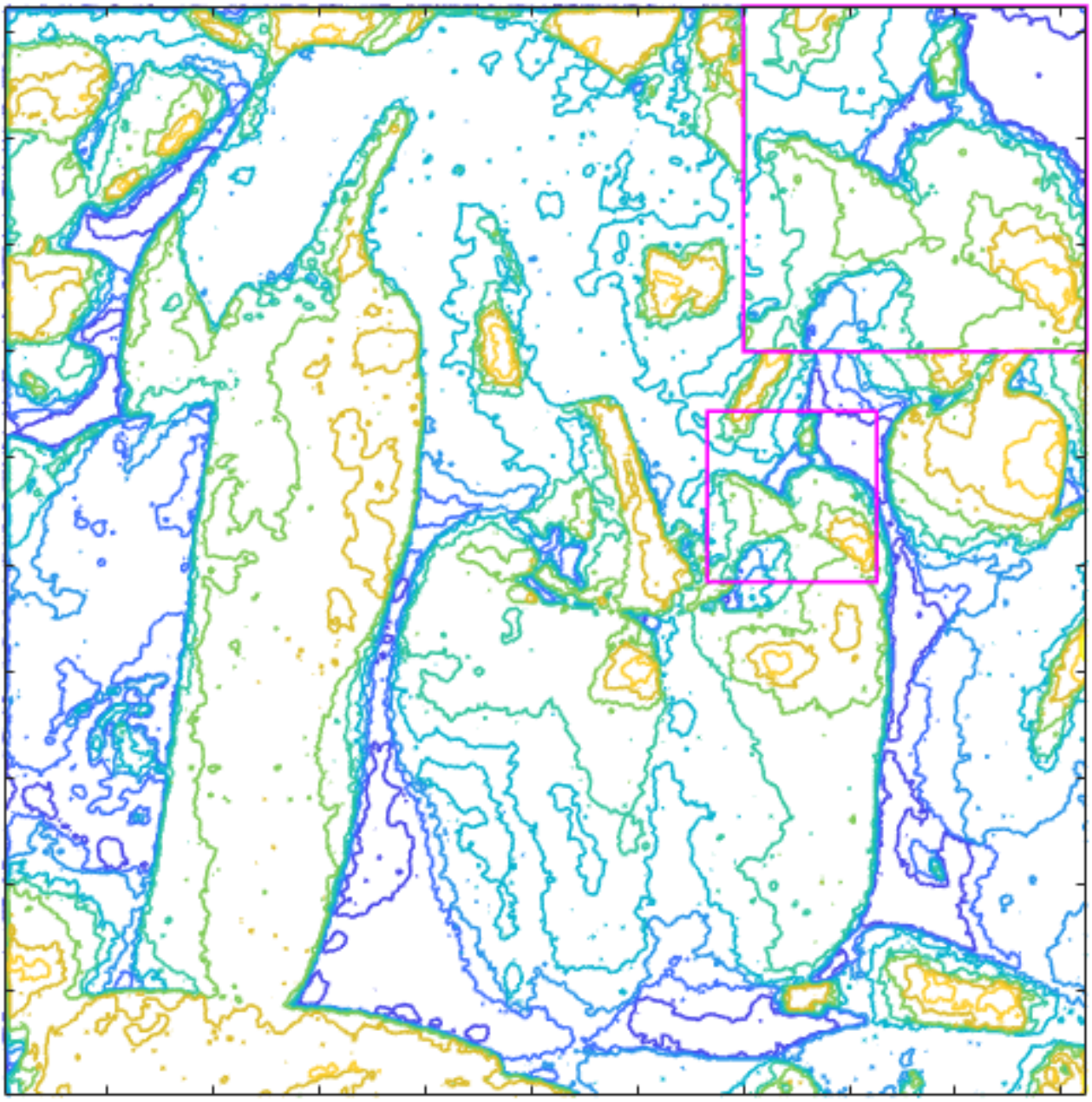}&
\includegraphics[width=3.5cm]{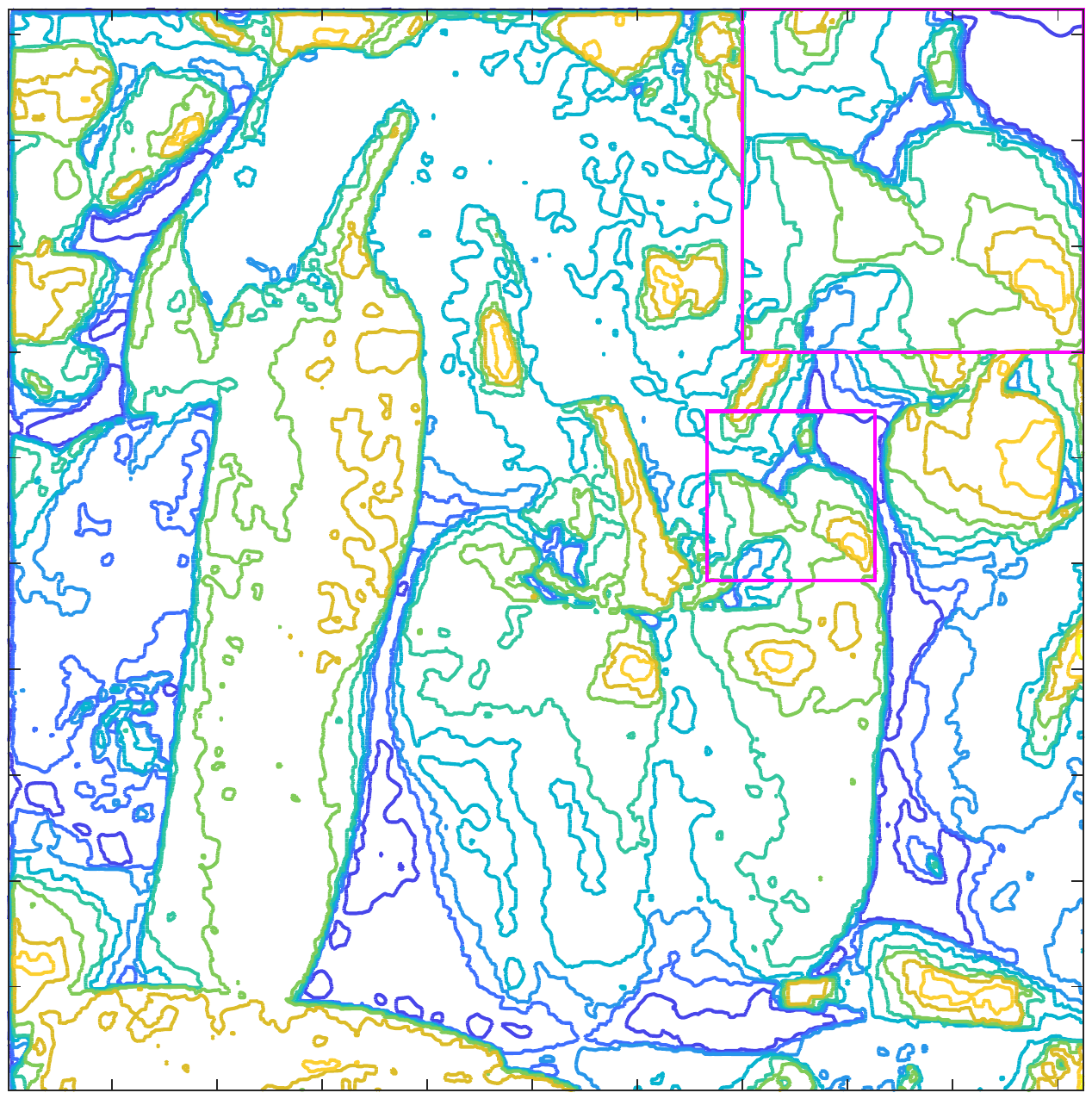}\\
\begin{sideways}\hspace{0.6cm}Cameraman\end{sideways}&
\includegraphics[width=3.5cm]{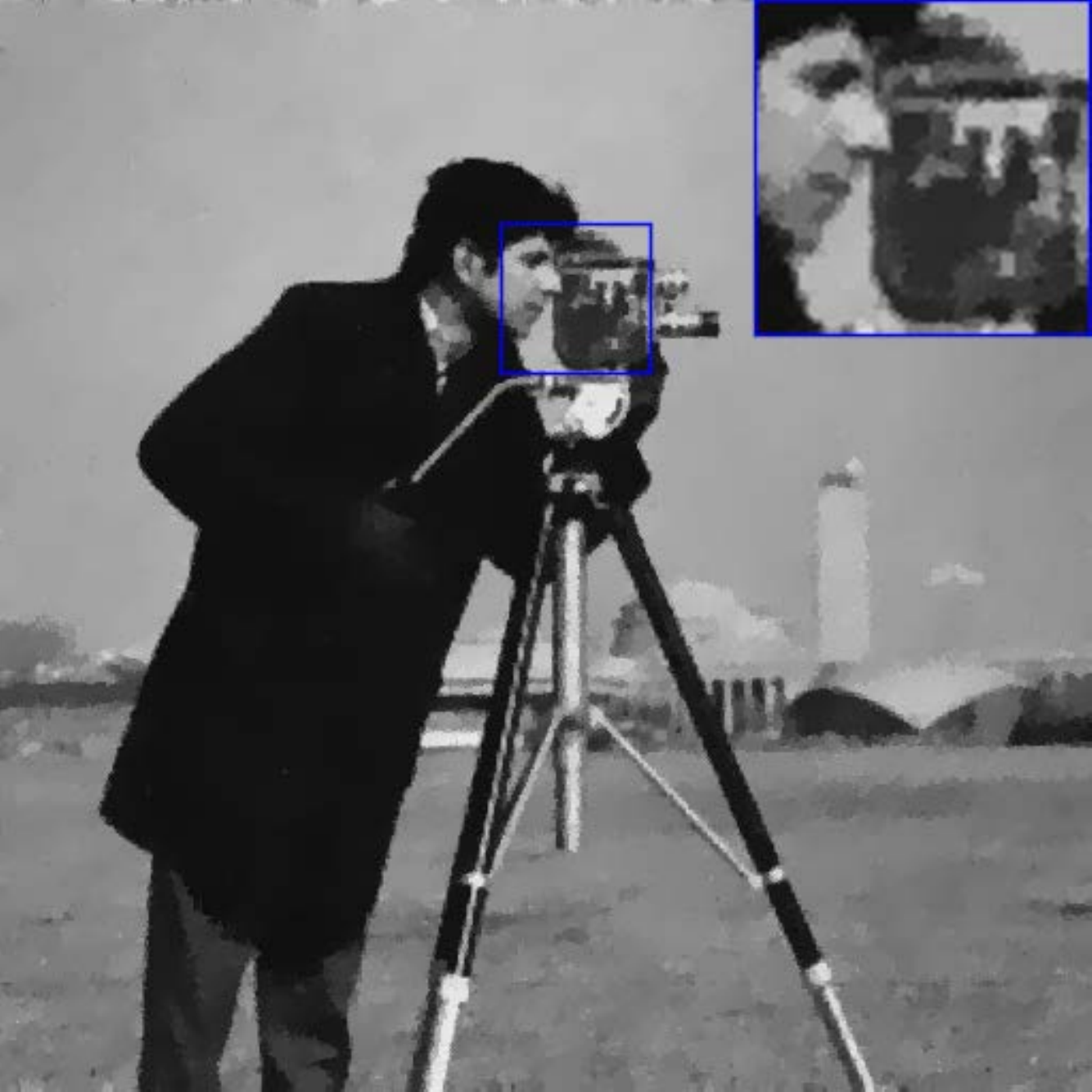}& 
\includegraphics[width=3.5cm]{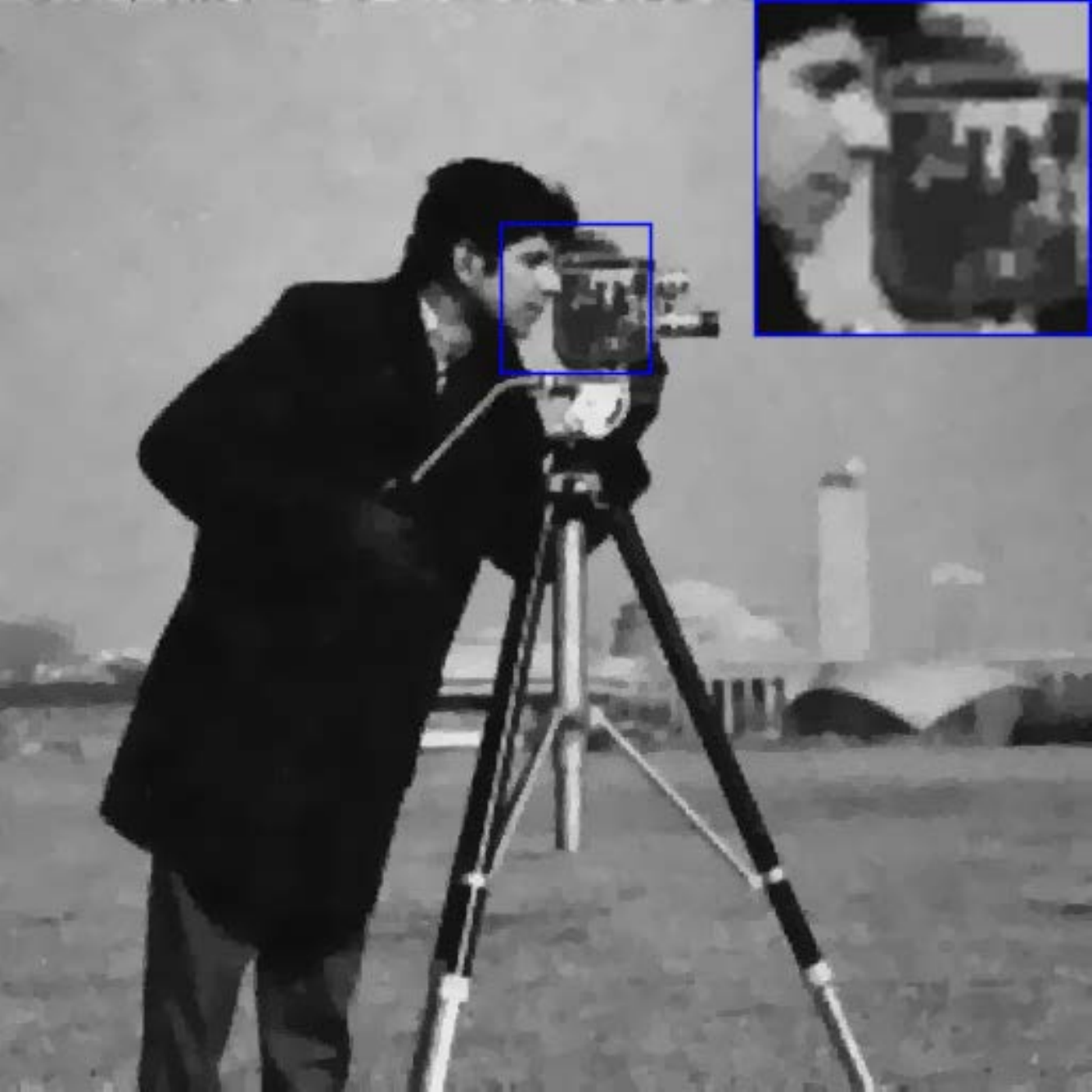}&
\includegraphics[width=3.5cm]{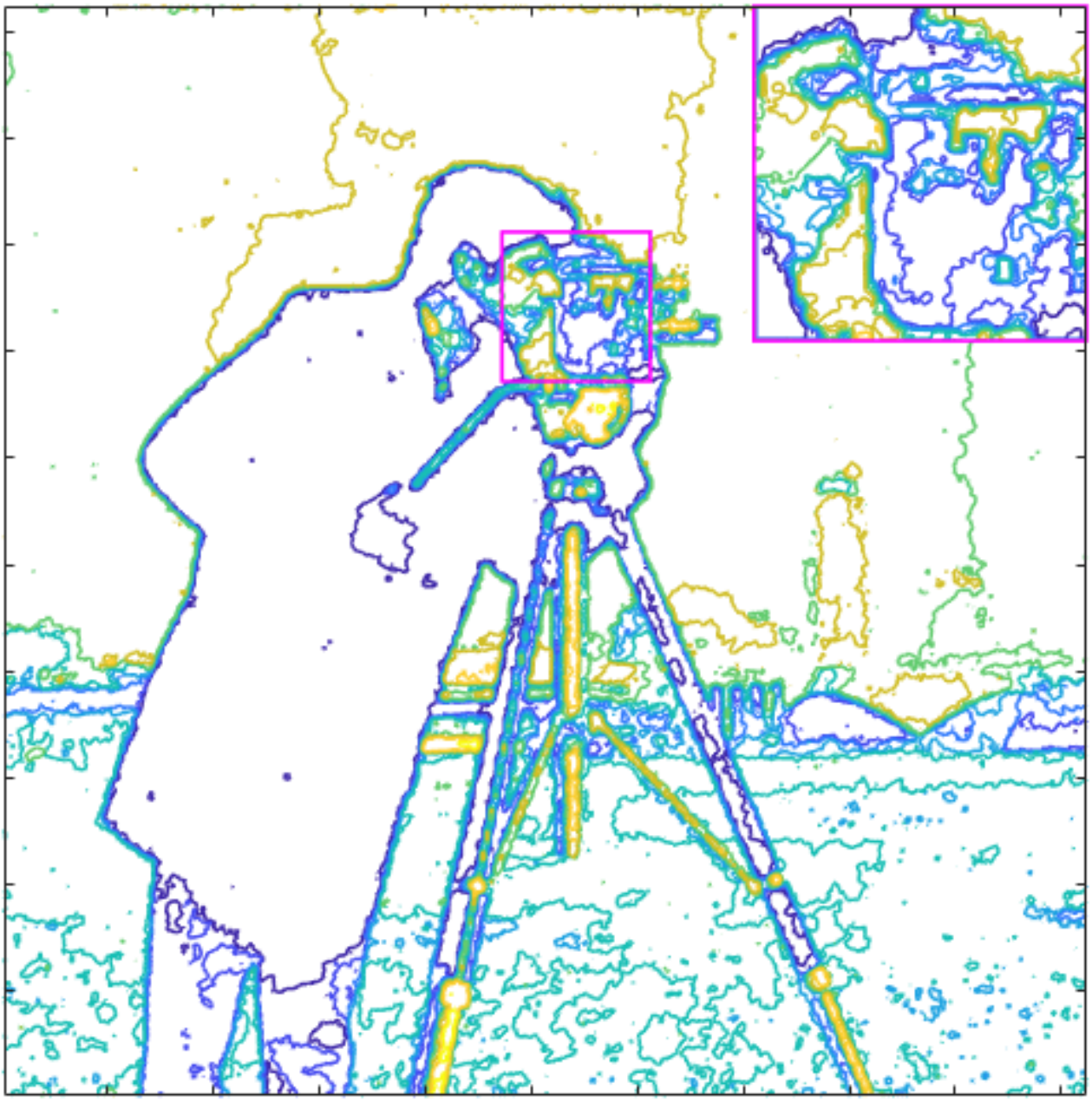}&
\includegraphics[width=3.5cm]{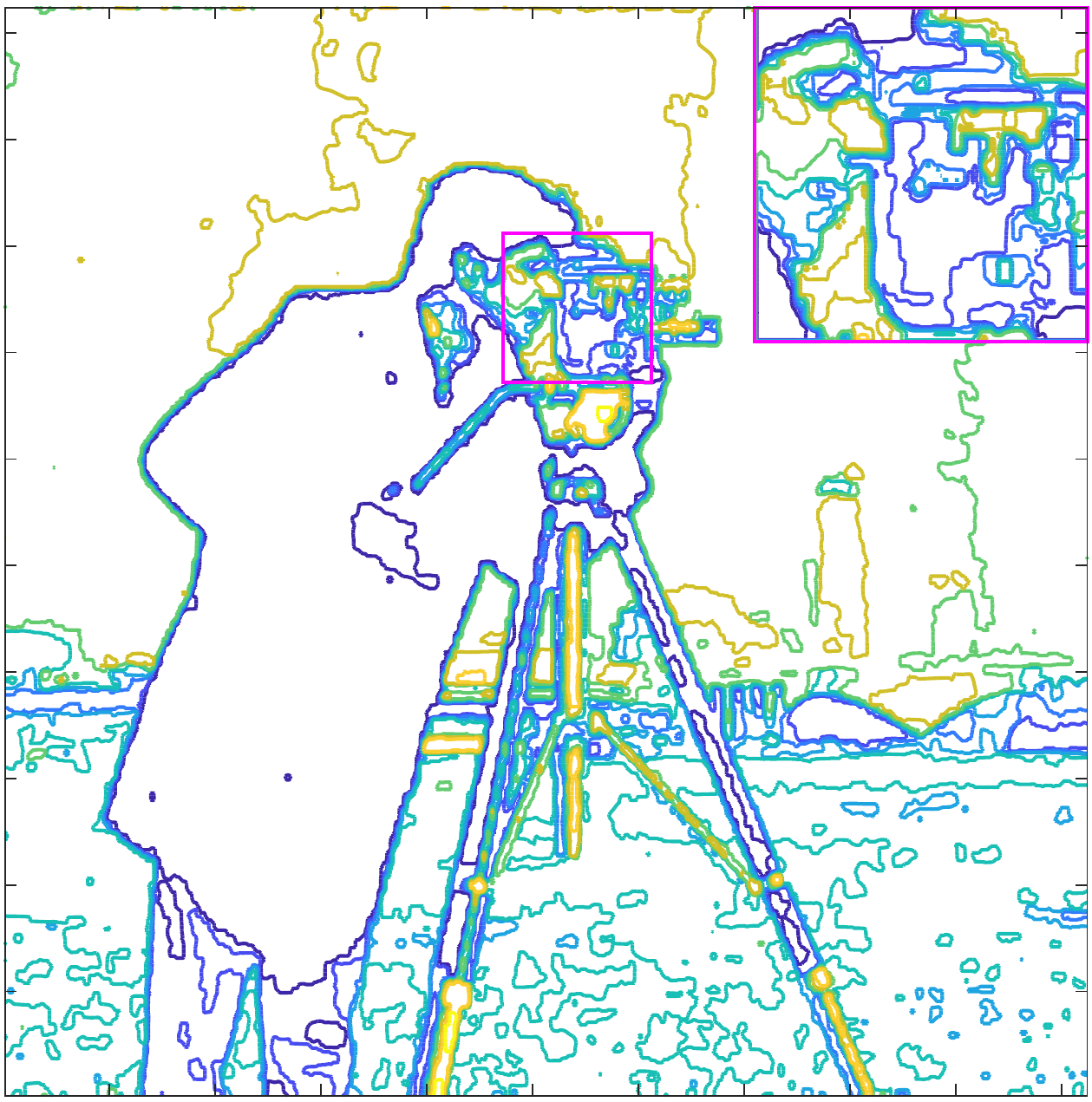}
\end{tabular}~~~~
\caption{The first and second columns show the denoised results of LALM and RALM with zoomed-in subsection of blue frame respectively.  And their contour plots with zoomed-in subsection of pink frame are shown in the third and fourth columns.}
    \label{image_contrast} 
\end{figure*}

The restored results by the two methods are shown in Fig.\ref{image_contrast}.  From the left two columns, one can conclude that the two competing methods recover quite similar images, and little difference can be visually identified. In the zoomed-in sub-images located at the upper right corners, weak but perceptible differences can be detected. When we check the isophotes shown in the last two columns, however, the difference becomes apparent. The isophotes produced from RALM are less fragmented and visually more pleasant than LALM does. From the two zoomed-in areas of the lena image, one can easily check that for the results of LALM, there are image edges collapsing in the nose part, while for the proposed RALM, the isophotes are connected. The same phenomenon occur in the isophotes of the recovered  peppers and cameraman images, shown in the last two rows of Fig.\ref{image_contrast}. In the zoomed-in junction parts of the peppers, as well as in the zoomed-in parts of the cameraman's face area, the proposed method produces smoother isophotes than LALM does. 

\begin{figure}[!htb]
   \centering
\centering
\includegraphics[width=1.69in]{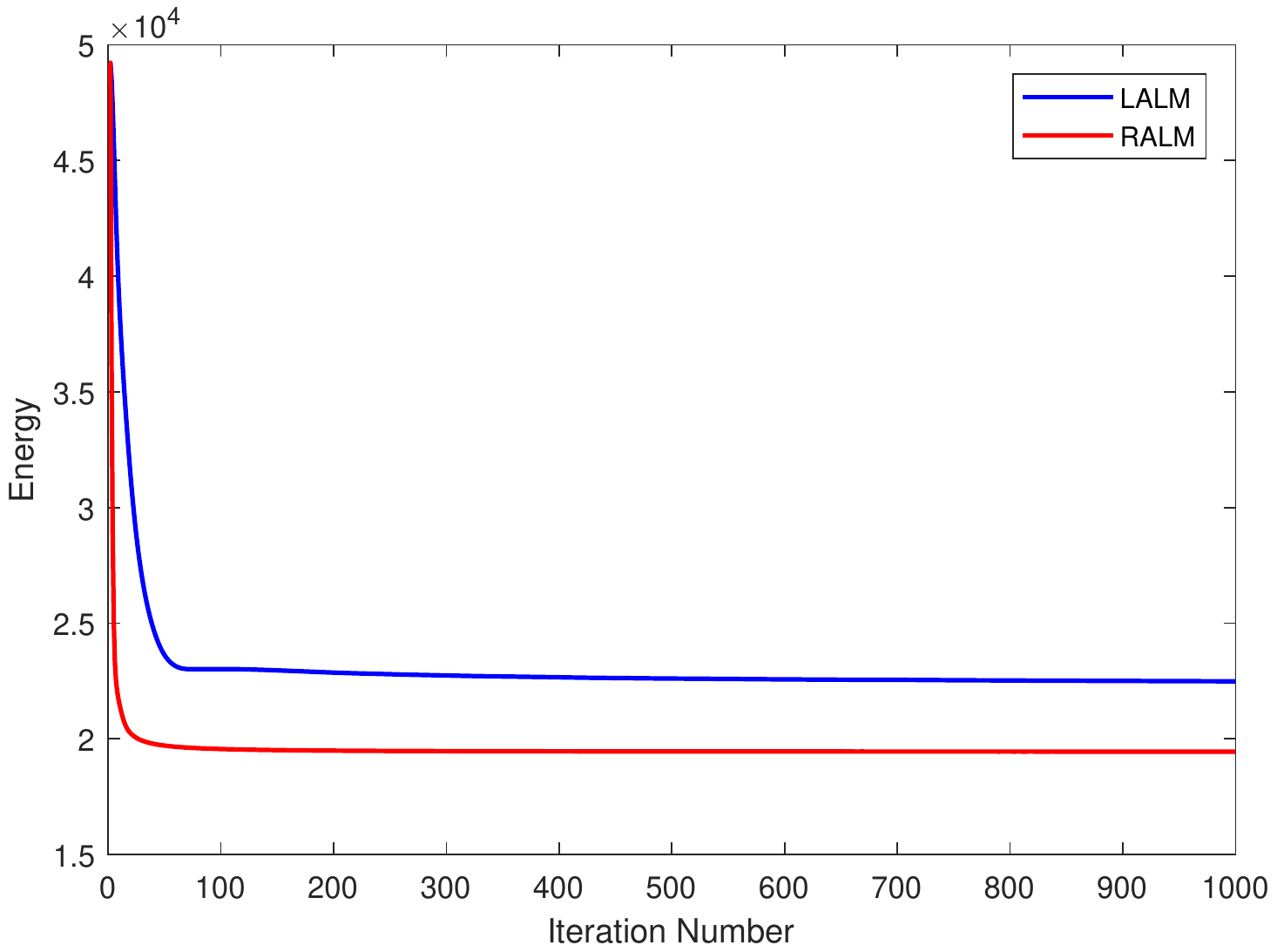}
\includegraphics[width=1.69in]{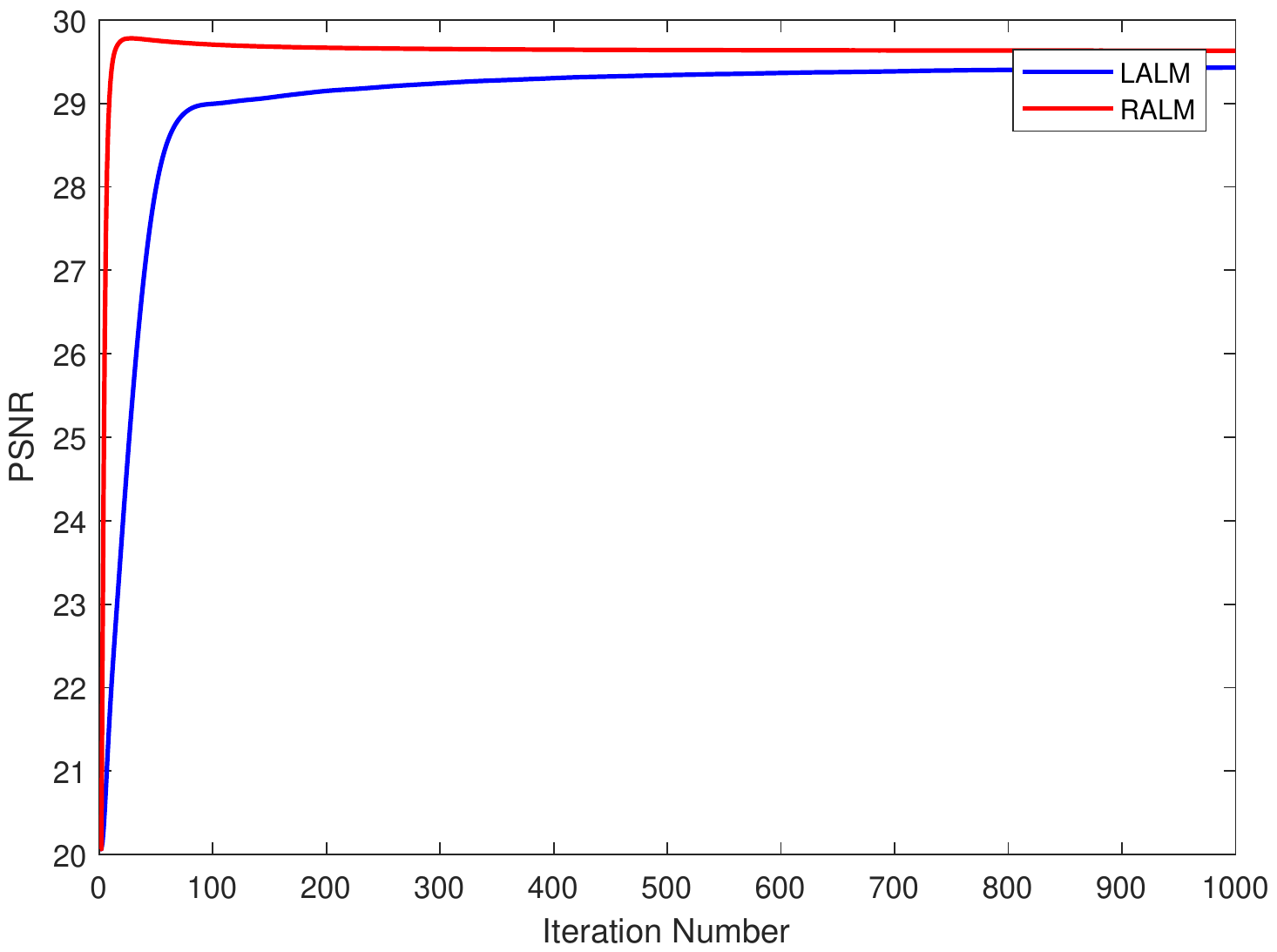}
\includegraphics[width=1.69in]{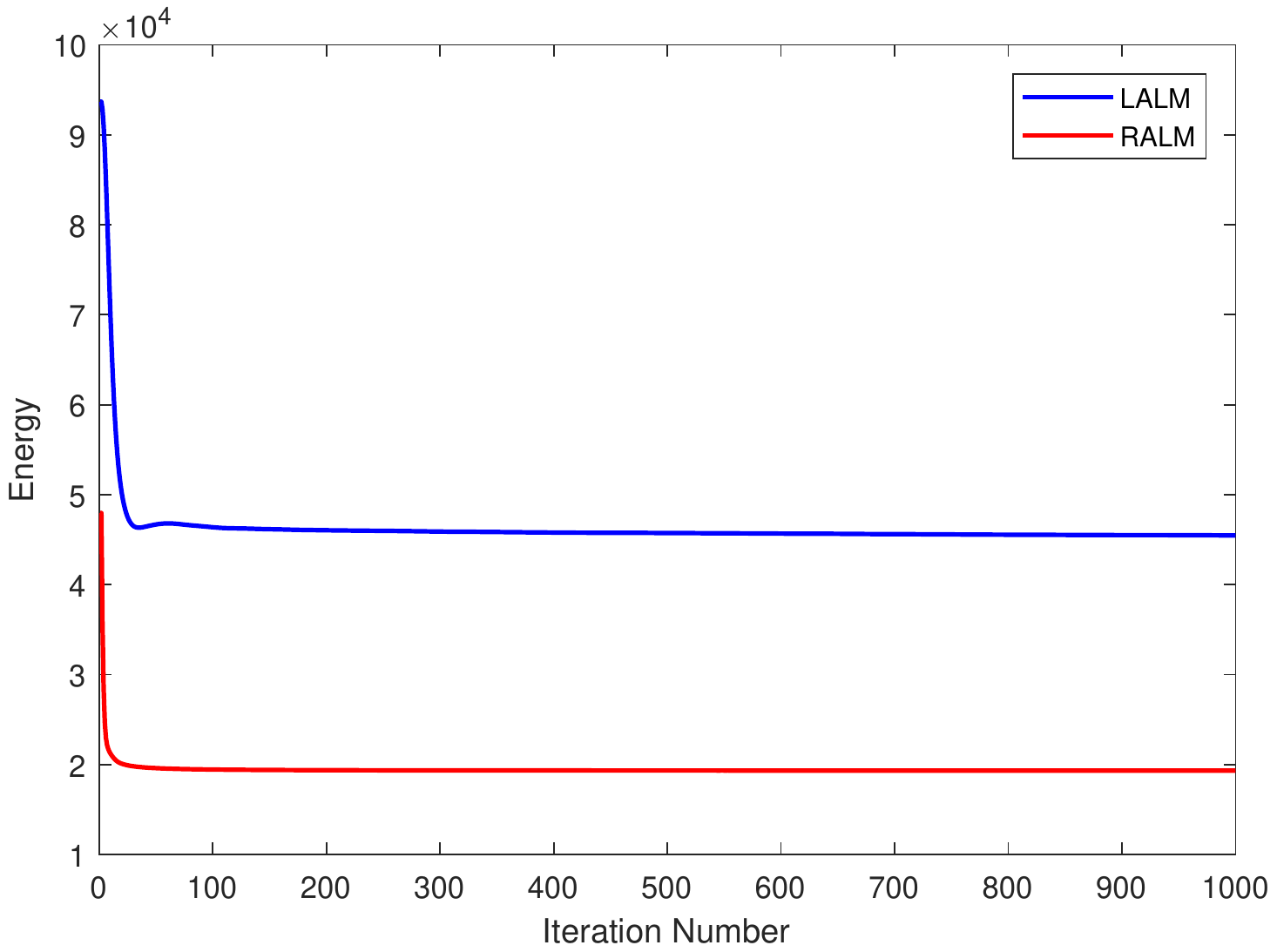}
\includegraphics[width=1.69in]{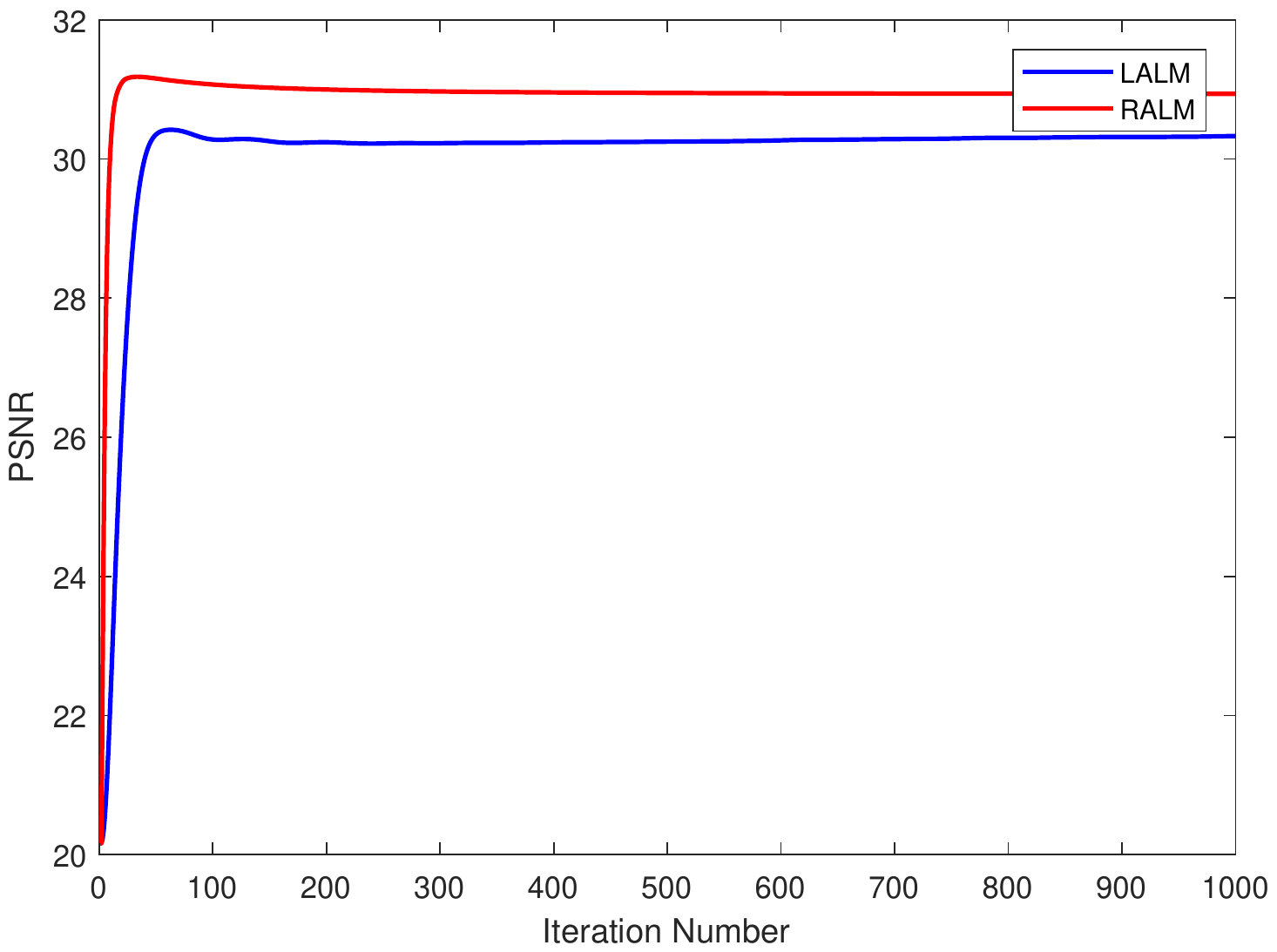}
\includegraphics[width=1.69in]{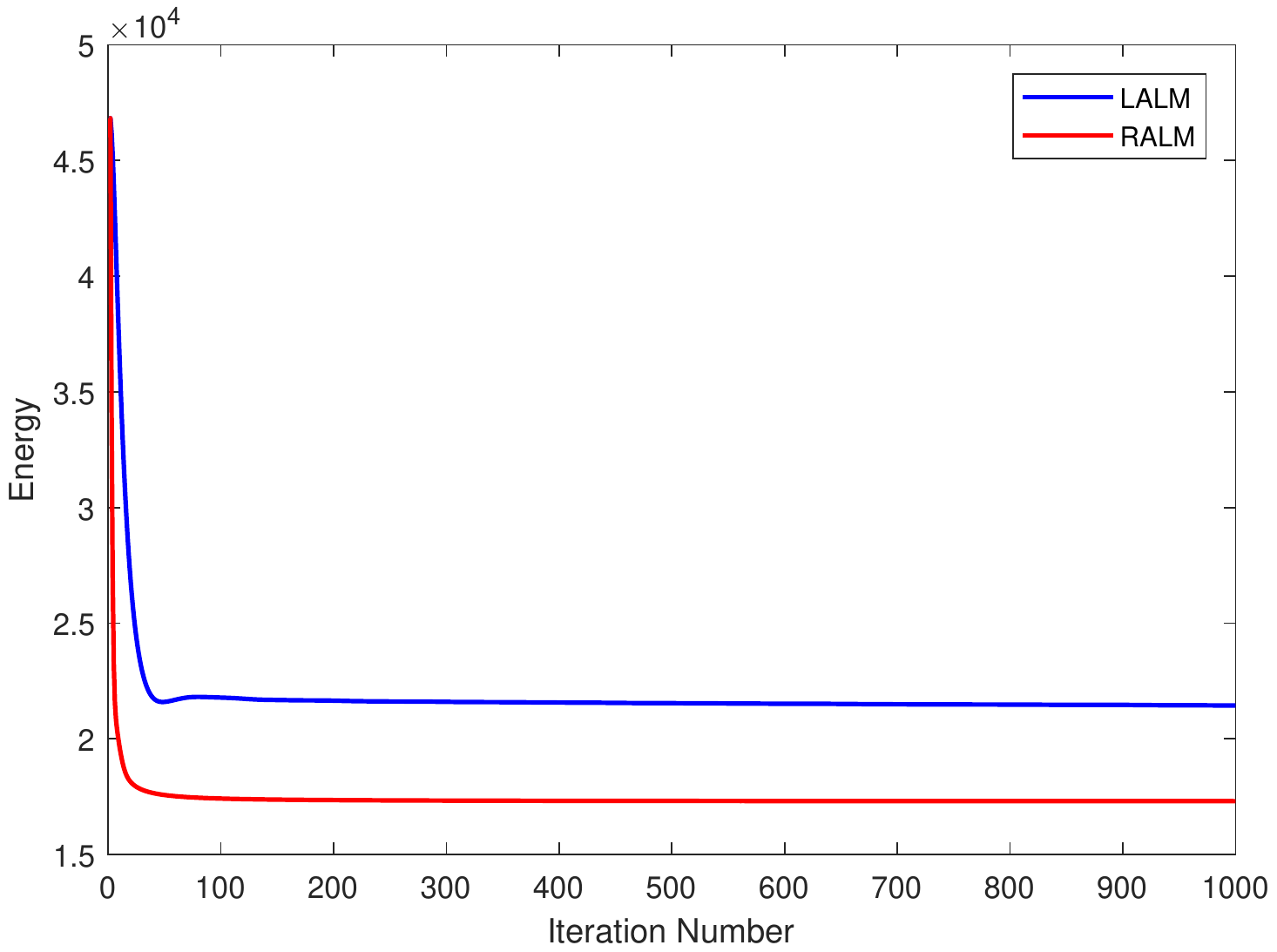}
\includegraphics[width=1.69in]{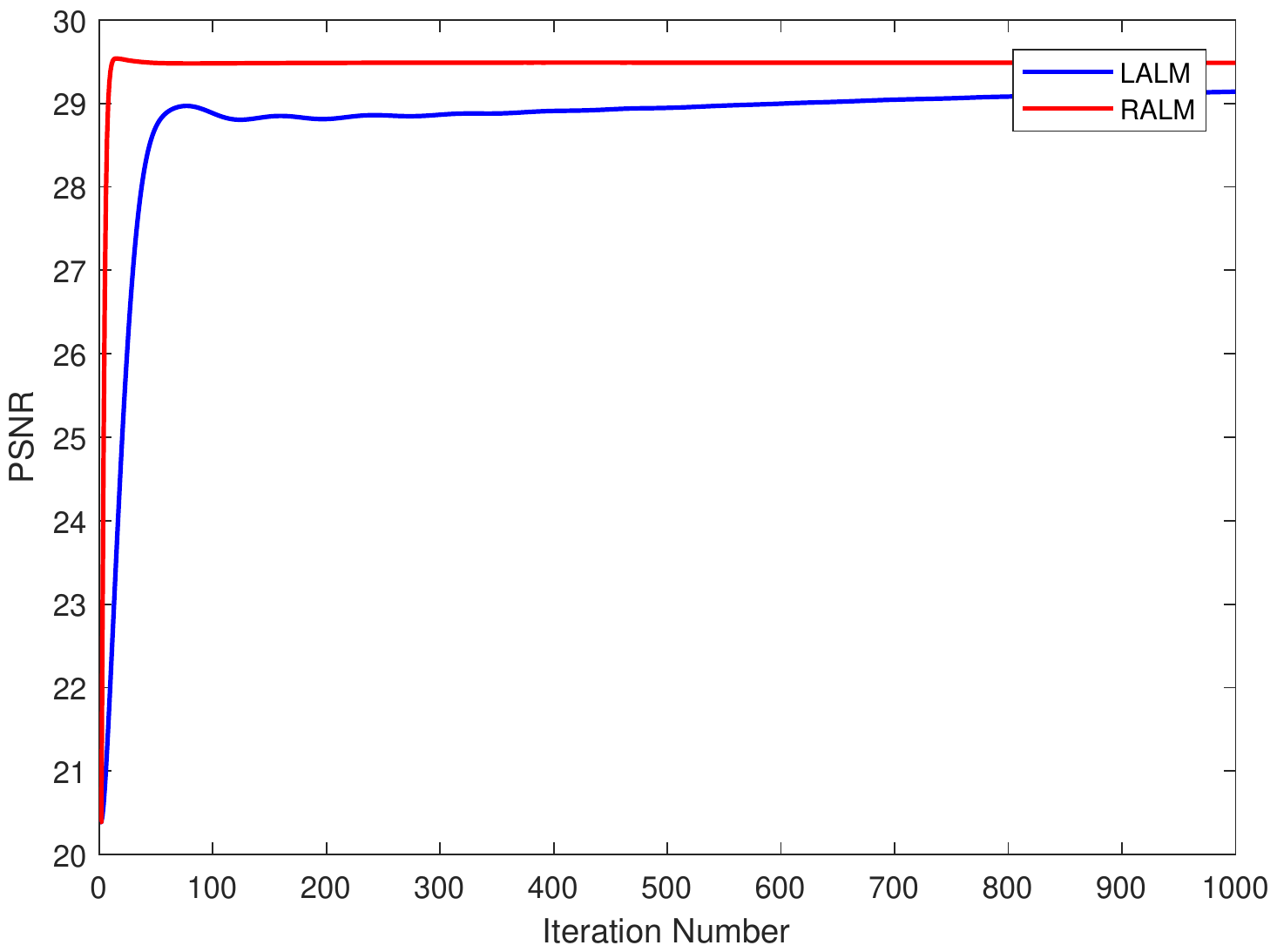}
  \caption{The efficiency and stability comparison of LALM and RALM. The first column shows energy how the energy is evolving for the two methods; The second column presents the PSNR variation of the two algorithms. The plots correspond to the lena, peppers and cameraman images from top to bottom respectively. }
	\label{ep} 
\end{figure}

To compare  in a quantitative manner,  the energy and PSNR evolutions for the two methods are monitored and shown 
in Fig.\ref{ep}. To fully demonstrate their behaviors, 1000 iterations are computed, which are beyond the need for convergence. From the plots, one can conclude that the energy and PSNR curves of RALM converge to steady state much faster than LALM does. Furthermore, compared with LALM, RALM achieves lower energy value and higher PSNR value, which suggests that the restored images by RALM possess higher quality. The quantitative indices shown in Table \ref{index} further validate the above conclusion.  one can check that  RALM recovers higher quality images in terms of NRMSE, NMAD, and PSNR, while consumes much less iterations.

\begin{table}[!htb]
	\centering
	\caption{PSNR, NMAD and NRMSE for images in our experiments }
	\label{index}
	\begin{tabular}{ccccc}
		\hline\noalign{\smallskip}
		Image &  Index &Noisy Data & LALM & RALM \\
		\noalign{\smallskip}\hline\noalign{\smallskip}
		lena     &NRMSE &0.2797&0.0329&\bf{0.0305}\\
		$512\times512$&NMAD & 0.1632 & 0.0488&\bf{0.0463}\\
		&PSNR & 20.0658& 29.3617&\bf{29.6942}\\
		&Iter.&        &590        &\bf{115}           \\
		peppers  &NRMSE &0.2473&0.0243&\bf{0.0199}\\
		$512\times512$&NMAD & 0.1725 & 0.0468&\bf{0.0433}\\
		 &PSNR & 20.1624&30.2318&\bf{31.1161}\\
		 &Iter. &       &215       &\bf{69}            \\
		cameraman  &NRMSE &0.1546&0.0211&\bf{0.0191}\\
		$512\times512$&NMAD & 0.1630 & 0.0506&\bf{0.0487}\\
		&PSNR & 20.3904& 29.0483&\bf{29.4845}\\
		  &Iter. &      & 712     &\bf{192}      \\
		\noalign{\smallskip}\hline
	\end{tabular}
\end{table}

\section{Conclusion}\label{sec:conclusion}

By considering the limiting case of $b\to 0$, an internal inconsistency is identified in the conventional augmented Lagrangian formulation for solving the Euler's elastica model. Based on this key observation, a simple cutting-off strategy is introduced to the solution procedure of the $\boldsymbol{p}$ subproblem which helps to remove the inconsistency. Besides, our analysis shows that the commonly used relaxation technique for imposing the constraint for the auxiliary variable $\boldsymbol{n}$ is actually  problematic, and its original form should be instead adopted.  The above two observations lead to the proposed restricted augmented Lagrangian method (RALM), which enjoys easier parameter tuning, faster convergence, and better image restoration ability. 

We think that the proposed {\bf RALM } provides a strategy for accelerating and improving other augmented Lagrangian based  algorithms for solving the Euler's elastica model or other models that need to introduce multiple auxiliary variables with ordered dependence.     
\bibliographystyle{IEEEtran}
\bibliography{euler_elastica}

\ifCLASSOPTIONcaptionsoff
  \newpage
\fi

%

\end{document}